\documentclass[a4paper,12pt]{report}
\usepackage{amsmath}
\usepackage{wasysym}
\begin{document}
\title{Perturbative Quantum Gravity and Yang-Mills Theories in de~Sitter Spacetime}
\author{Mir Faizal\\ A Thesis Submitted for the Degree of PhD\\  Department of Mathematics\\ University of York }
\date{October 2009}
\evensidemargin = 20 mm
\oddsidemargin = 20 mm
\textwidth= 5.5 truein
\linespread{1.4}
\maketitle

\begin{abstract} 
This thesis consists of three parts. In the first part we review the quantization of 
Yang-Mills theories and perturbative quantum gravity in curved spacetime. In the second part we calculate 
the Feynman propagators of the Faddeev-Popov ghosts for Yang-Mills theories and perturbative quantum gravity in the covariant gauge.
 In the third part we investigate the physical equivalence of covariant Wightman graviton  two-point function with the  physical
 graviton  two-point function. 

The Feynman  propagators of the Faddeev-Popov ghosts for Yang-Mills theories and perturbative
quantum gravity in the covariant gauge are infrared (IR) divergent in de~Sitter spacetime. We
point out, that if we regularize these divergences by introducing a finite mass and take 
the zero mass limit at the end, then the modes responsible for these divergences will not contribute to loop
diagrams in computations of time-ordered products in either Yang-Mills theories or perturbative
quantum gravity. We thus find effective Feynman propagators for ghosts in Yang-Mills theories and perturbative
quantum gravity by subtracting out these divergent modes. 

It is known that the covariant graviton two-point function in de~Sitter spacetime is infrared divergent for some choices 
of gauge parameters. On the other hand it is also known that there are no infrared problems for the physical 
graviton two-point function obtained by fixing all gauge degrees of freedom, in global coordinates.
We show  that the covariant Wightman  graviton two-point function  is equivalent to the physical one in the sense that 
they result in the same two-point function of any local gauge-invariant quantity.  Thus any infrared divergence in the  
Wightman graviton two-point function in de~Sitter spacetime can only be an gauge artefact.
\end{abstract}

\tableofcontents
\chapter*{Acknowledgements}
I would like to thank my supervisor A. Higuchi for all the useful discussions we have had during my PhD.
\chapter*{Author's  Declaration}
I declare that this work on effective Feynman  ghost propagator for perturbative quantum gravity  in de~Sitter spacetime 
and gauge equivalence of covariant Wightman graviton two-point functions to physical graviton two-point function
  in de~Sitter spacetime is original. It  has been done in close 
collaboration with my supervisor.  It has not been submitted for any other degree or diploma before.
\chapter{Introduction}

Quantum field theory in de~Sitter spacetime  has been actively studied recently due
to its relevance to inflationary cosmologies $\cite{qft2}$. Furthermore, the current observations
indicate that our Universe is expanding in an accelerated rate and may approach de Sitter
spacetime asymptotically $\cite{qft3}$. The original part of this thesis consists of two parts. 
In the first part  we find effective Feynman ghost 
propagators for Yang-Mills theories and perturbative quantum gravity in de~Sitter spacetime.
 In the second part  we find the 
equivalence of the covariant Wightman graviton two-point function   with arbitrary value of gauge parameters
to a infrared finite  physical Wightman graviton two-point function.

In order to study higher-order quantum effects for Yang-Mills theories or perturbative gravity, one needs to introduce
 Faddeev-Popov  ghosts
except in unwieldy gauges such as the axial gauge in Yang-Mills theories  $\cite{axial}$. In this thesis we shall   study the Feynman propagators for 
the ghosts in Yang-Mills theories and perturbative quantum gravity in de~Sitter spacetime. We shall  find that ghosts 
and anti-ghosts for Yang-Mills theories satisfy the minimally-coupled massless scalar field equation. It is known that there is no 
Feynman propagator for  minimally-coupled massless scalar fields that respects de~Sitter invariance $\cite{7ch31}$, due to 
infrared (IR) divergence of the Feynman propagator for these fields  in de~Sitter spacetime. We also show that the Feynman propagator for ghosts
in perturbative quantum gravity is also infrared divergent in de~Sitter spacetime.  Thus if the ghost fields in these theories were physical fields,
we would need to break  de~Sitter invariance of the vacuum for these fields $\cite{6cha}$. However we shall  see that this
problem can be circumvented because they are unphysical fields. The interaction between the Yang-Mills/gravitational field and 
the  ghosts is such that, if we regularize the infrared divergences by introducing a small mass term, the modes responsible for 
the infrared divergences will not contribute in the computation of time-ordered products of physical fields. For this reason, we
 propose that one should regularize the infrared divergences of the Feynman propagators  for the ghosts and anti-ghosts in these theories and then take 
the limit where the regularization is removed. This proposal is equivalent to using the effective Feynman propagators obtained by 
subtracting the  regularized modes responsible for the infrared divergence in perturbative calculations. 

In fact this effective Feynman propagator for Yang-Mills theories has been derived and used in a different context. It was used in
 calculating the  covariant graviton propagator two-function in  de~Sitter spacetime  $\cite{a}$. It may be noted
 that even though the physics there is very different from the physics we are considering, the mathematics involved is the same for both these cases.    

Infrared  divergences  in graviton two-point functions have been a matter of contention for over two decades $\cite{deb}$.
Since linearized gravity has gauge invariance, it is important to determine whether or not these IR divergences are gauge 
artefacts. The graviton two-point function obtained by fixing all the gauge degrees of freedom is called the physical graviton 
two-point function. The physical graviton two-point function in spatially flat coordinate system was analyzed  in Ref. $\cite{two}$. 
It was found that this two-point function is IR divergent. Infrared divergence of the  physical graviton two-point function in spatially flat coordinate system
was further studied in Ref. $\cite{flat}$. The physical graviton two-point function in global coordinate system was obtained in Ref. $\cite{phy}$. It was found that 
this  physical graviton two-point function is IR finite. The physical graviton two-point function in hyperbolic  coordinate system was also found to be IR finite
 $\cite{hyper}$. Thus the  IR divergence of the physical graviton two-point function in spatially flat coordinate system  occurred due 
to the coordinate system used  and thus was not a physical effect. However we will not study the IR divergences of the physical graviton two-point functions 
in this thesis.  

In this thesis we will present a work relevant to  the IR  divergences in the graviton two-point functions that occur for some choices of gauge parameters. 
Allen found that in the covariant gauge  the  graviton two-point function has IR divergences for certain values of the gauge
parameters $\cite{7cha4}$. Thus if a gauge is chosen with one of these  values of gauge parameters, we shall  get IR divergences. 
 The graviton two-point function  used by Antoniadis and Mottola $\cite{7cha5}$ was IR divergent as they used one of these values of gauge parameters. 
Covariant  graviton two-point function with  different values of gauge parameters was obtained in  Ref. $\cite{a}$.
Covariant  graviton two-point function with arbitrary values of gauge parameters was obtained in Ref. $\cite{15}$.
 In this thesis we shall show that the  covariant Wightman graviton two-point function with any choice of gauge parameters is physically  
 equivalent to 
the physical one obtained in Ref. $\cite{phy}$. 

For this purpose, we will split the covariant Wightman graviton two-point function into vector, tensor and scalar  parts $\cite{15}$. 
It will be shown that apart from a part of tensor part, that also contributes to the  physical  graviton two-point function,
all the contributions  are pure gauge in the following sense.  A contribution to a Wightman graviton two-point  function will be said 
to be pure gauge if at least in one of its two slots  it is a sum of tensors obtained by symmetrized 
derivatives acting on vectors. So a pure gauge contribution will be of the form: $\nabla_{a} K^1_{ba'b'}(x,x') + \nabla_{b} K^1_{aa'b'}(x,x') 
 +\nabla_{a'} K^2_{abb'}(x,x') +\nabla_{b'} K^2_{aba'}(x,x')$ for some $ K^1_{ba'b'}(x,x') $ and $  K^2_{aa'b'}(x,x')$.
  The two-point function of a local gauge-invariant
 tensor field will be the same for two Wightman two-point
 functions if they differ from each other by a pure gauge contribution. The scalar part has in fact  already been shown to be
 pure gauge in a  restricted  gauge  in Ref. $\cite{15a}$. In this thesis we   show  that this is true for any value of the gauge parameters. 
We  also show that the contributions coming from the vector part and a part of the  tensor part are also pure gauge. 
Thus  the only non-gauge contribution will come from the other part of the tensor part.  This contribution will be shown to be exactly equal to
the physical graviton two-point function obtained in Ref. $\cite{phy}$.

This physical graviton two-point function  obtained  in Ref. $\cite{phy}$  suffers from no IR divergences. Now if any graviton Wightman two-point function
 in linearized gravity in  de~Sitter is physically equivalent to that obtained in Ref. $\cite{phy}$, as we show in this thesis,
 then any IR divergences  in the covariant graviton Wightman two-point
function in linearized gravity in  de~Sitter spacetime for example as noticed in Ref. $\cite{7cha5}$, has to be  a gauge artefact. Thus  the IR divergences
will not show up in the graviton Wightman two-point function of any local gauge-invariant tensor field, e.g., the linearized Weyl tensor $\cite{b}$ .

The rest of the thesis is organised as follows: 
We shall  review Yang-Mills theories and perturbative quantum gravity in Chapters $2$ and $3$, respectively. Then after reviewing some 
basic properties of de~Sitter spacetime  in Chapter $4$  we shall  go on to study ghost fields in de~Sitter spacetime. We shall  
derive the  explicit expression  for the effective  ghost propagator for Yang-Mills theories in Chapter $5$ as a 
warm up for the derivation of the explicit expression for the effective ghost propagator for perturbative quantum gravity in Chapter $6$. 
Then we shall  describe a general method for calculating the Wightman two-point  function for free field theories in Chapter $7$.
 We shall  apply this formalism to linearized quantum gravity in Chapter $8$. 
Finally we shall  show explicitly that the   covariant graviton Wightman two-point function with any choice of gauge parameters is
 physically equivalent to the physical one obtained in Ref. $\cite{phy}$ in Chapter $9$. We shall  conclude the thesis in Chapter $10$.

We shall  do our calculations for Feynman ghost propagators on $S^4$ and then analytically continue the result to de~Sitter spacetime in Chapters $5$ and $6$.
 However we shall  do our calculations directly in de~Sitter spacetime while calculating the Wightman two-function 
in Chapters $7$ and $9$. So in Chapter $5$ and $6$ propagator  means a Feynman propagator obtained by analytical continuation from $S^4$ 
and in Chapters $7$ and $9$ a two-point function  means a Wightman function calculated directly in de~Sitter spacetime. 

The signature we adopt is $(-,+,+,+)$ and we denote a spacetime point as $x$, the spatial component of $x$ as ${\bf{x}}$ and its temporal component as $t$. 
 We define the Riemann tensor, Ricci tensor and scalar curvature by 
\begin{eqnarray}
 {R^d}_{cab}A_d &=&[ \nabla_b \nabla_a - \nabla_a \nabla_b ]A_c,\\
R_{bd} &=& {R^a}_{bad}, \\
R &=& g^{ab}R_{ab},
\end{eqnarray}
where 
\begin{eqnarray}
 \nabla_b A_c &=& \partial_b A_c - \Gamma^d_{bc}A_d,\\
\Gamma^d_{bc} &=& \frac{1}{2}g^{da}[\partial_b g_{ac} +\partial_c g_{ab} -\partial_a g_{bc}]. 
\end{eqnarray}

\chapter{Yang-Mills Theories}
In this chapter we shall  discuss the quantization of  Yang-Mills theories  in general curved spacetime. After reviewing the quantization of 
 Yang-Mills theories we shall also discuss the BRST symmetry for Yang-Mills theories. 
\section{Basic Formalism}Yang-Mills theories were originally proposed to explain occurrence of isospin $\cite{2ch1}$. 
They have been  used to unify the electromagnetism and the weak force into a single electro-weak force $\cite{2ch2}$.
 Quantum Chromodynamics, which is an asymptotically free  non-abelian gauge theory,  is now thought be the theory of strong interaction $\cite{2ch4}$. 
Thus except gravity all the forces of nature can be described by Yang-Mills theories. 
Apart from electromagnetism which is invariant under  $U(1)$ symmetry all the other gauge theories
 describing forces in nature are invariant under $SU(N)$ symmetry. So  we shall  first review $SU(N)$ Lie groups and some of their properties (see Chapter 
$15$ of Ref. $\cite{1chaaa234}$). 
From now on we shall  use the convention that the repeated group indices are summed over. An element $u$ of $SU(N)$ can be written as 
\begin{equation}
 u = \exp [i g\Lambda^A T_A].
\end{equation}
Here $T_A$ are $N \times N$ traceless Hermitian matrices, which satisfy
\begin{equation}
 Tr (T_A T_B) = \frac{1}{2} \delta_{AB}, \label{2de}
\end{equation}
The matrices $T_A$ are the generators of $SU(N)$ Lie algebra, which is given by 
\begin{equation}
 [T_A, T_B ] = i f_{AB}^C T_C,  \label{lie}
\end{equation}
where $f^C_{AB}$ are called the structure constants.
  Strictly speaking $T_A$ form a basis of the defining representation  of the $SU(N)$ Lie algebra.

These structure constants  are antisymmetric in their lower indices: as 
\begin{equation}
 [T_A, T_B] = - [T_B, T_A], 
\end{equation}
 we have 
\begin{equation}
f^C_{AB} = - f^C_{BA}. 
\end{equation}
They also satisfy the Jacobi identity: as
\begin{equation}
 [T_A ,[T_B , T_C]] +  [T_B ,[T_C , T_A]] +  [T_C ,[T_A , T_B ]] =0,
\end{equation}
 we have 
\begin{equation}
 f^E_{AD} f^D_{BC} +  f^E_{BD} f^D_{CA} +  f^E_{CD} f^D_{AB}=0. \label{jacobi}
\end{equation}
We can now define matrices $\tau_A$ which form a basis for the adjoint representation of this Lie algebra: as
\begin{equation}
 (\tau_A)^B_C = i f_{AC}^B.
\end{equation}
Now from the Jacobi identity given in Eq. $(\ref{jacobi})$, $\tau_A$ satisfy the Lie algebra given in Eq. $(\ref{lie})$, 
\begin{equation}
 [\tau_A , \tau_B] = i f^C_{AB}\tau_C. 
\end{equation}

Having  reviewed some properties of $SU(N)$ Lie algebra,  we shall   briefly review $SU(N)$ Yang-Mills theories (see Chapter $9$ of Ref. $\cite{Ircaxial}$).
We start from a matter field $\phi (x)$ which forms a $N$-dimensional vector in the defining representation space of the Lie algebra.
 So it transforms under a spacetime dependent $SU(N)$ transformation $u(x)$ as follows: 
\begin{equation}
 \phi^u (x) = u(x) \phi(x).
\end{equation}
We now also want a derivative of $\phi (x)$ that transforms like $\phi(x)$. 
However $\nabla_a \phi (x)$ does not do so. 
So we define a derivative $D_a \phi (x)$ called the covariant derivative which transforms like $\phi (x)$. It is defined to be  
\begin{equation}
 D_a \phi = [\nabla_a - i g A_a   ]\phi,
\end{equation}
where $A_a$ is defined to be a matrix-valued gauge field. It can be given by
\begin{equation}
A_a = A^A_a T_A.
\end{equation}
Now  we expect the covariant derivative to transform like $\phi (x)$. So if $A^u_a$ is the transformed gauge field, we have
\begin{equation}
 [\nabla_a - ig A_a^u] u \phi = u D_a \phi.
\end{equation}
So we have 
\begin{eqnarray}
  [\nabla_a - ig A_a^u] u \phi - u D_a \phi
&=& [(\nabla_a u) -i g A^u_a u + i g u A_a] \phi= 0.
\end{eqnarray}
 Now we have 
\begin{equation}
 ig A^u_a u  =  (\nabla_a u) + ig  u A_a.
\end{equation}
Multiplying by $- ig^{-1} u^{-1}$ from the right-hand side, we get
\begin{equation}
 A^u_a = u A_a u^{-1} - i g^{-1}(\nabla_a u) u^{-1}.
\end{equation}
Now we expand $u$ and $u^{-1}$ as 
\begin{eqnarray}
 u &=& 1 + i g \Lambda^A T_A + \mathcal{O}  (\Lambda^2),
\\
u^{-1} &=& 1 - i g \Lambda^A T_A + \mathcal{O}  (\Lambda^2).
\end{eqnarray}
The gauge field $A_a^u$ can now be written, to  first order in $\Lambda^A$: as 
\begin{eqnarray}
 A^u_a &=& [1 + i g\Lambda^A T_A] A_a [1 - i  g\Lambda^B T_B] - ig^{-1} (\nabla_a  i g\Lambda^A T_A)  \nonumber  \\ &=& 
 A_a + i g \Lambda^A [T_A, T_B] A^B_a + \nabla_a \Lambda^A T_A \nonumber  \\ &=& A_a  - g f^C_{AB}T_C  \Lambda^A  A^B_a + \nabla_a \Lambda^A T_A \nonumber  \\ &=& 
[A^A_a + \nabla_a \Lambda^A + g f^A_{BC}A_a^B \Lambda^C ] T_A.
\end{eqnarray}
So if the infinitesimal transformation of $A_a$ is given by 
\begin{equation}
 \delta_\Lambda A_a = T_A \delta_\Lambda A^A_a,
\end{equation}
then we have 
\begin{equation}
 \delta_\Lambda A^A_a = \nabla_a \Lambda^A + g f^A_{BC}A_a^B \Lambda^C. \label{gaugena}
\end{equation}
This is the gauge transformation of the gauge fields. 

Now we define $F^{A}_{ab}$ as follows: 
\begin{equation}
 [D_a, D_b ] \phi = i g^{-1}F^A_{ab}\phi,
\end{equation}
where 
\begin{equation}
 F_{ab} = F_{ab}^A T_A. 
\end{equation}
Thus we get 
\begin{eqnarray}
 F_{ab} \phi&=&g^{-1}[ g\nabla_a A_b - g \nabla_b A_a - i g^2[A_a, A_b ]]\phi \nonumber \\ &=& 
T_A[\nabla_a A_b^A - \nabla_b A^A_a + g f^A_{BC}A^B_a A^C_b]\phi.
\end{eqnarray}
So we get 
\begin{equation}
 F_{ab}^A = \nabla_a A_b^A - \nabla_b A^A_a + g f^A_{BC}A^B_a A^C_b.
\end{equation}
Now by construction $[D_a, D_b] \phi$  transform like $\phi$:
\begin{equation}
 [D_a, D_b]\phi \to u [D_a, D_b]\phi. \label{226}
\end{equation}
This implies that $F_{ab}^A $ transforms as follows: 
\begin{equation}
 F_{ab}  = u  F_{ab} u^{-1}.
\end{equation}
This is because it follows from Eq. $(\ref{226})$
\begin{eqnarray}
F_{ab} \phi &\to& u F_{ab}  u^{-1} u \phi =  u F_{ab} \phi. 
\end{eqnarray}
The classical action for the gauge field is given by 
\begin{equation}
 S = \int d^4x \sqrt{-g} \,  \mathcal{L}_c,
\end{equation}
where $  \mathcal{L}_c $ is the classical Lagrangian, which is given by 
\begin{equation}
 \mathcal{L}_c = -\frac{1}{2} Tr (F_{ab} F^{ab}).
\end{equation}
In this thesis we define the Lagrangian to be a scalar rather than a scalar density that would include the measure $\sqrt{-g}.$
It can be seen to be invariant under $u$ transformations, 
\begin{eqnarray}
Tr (F_{ab} F^{ab})   &\to&Tr (uF_{ab}u^{-1}uF^{ab} u^{-1})  \nonumber \\&=&
 Tr (F_{ab}F^{ab}).
\end{eqnarray}
This classical Lagrangian can also be written as
\begin{equation}
  \mathcal{L}_c = -\frac{1}{4}F_{ab}^A F^{ab}_A,\label{lna}
\end{equation}
  because
\begin{eqnarray}
 \mathcal{L}_c &=& - \frac{1}{2} Tr (F_{ab}^A F^{ab B}T_A T_B)  \nonumber  \\&=&  - \frac{1}{4} F_{ab}^A F^{ab B}\delta_{AB}  \nonumber  \\&=& 
-\frac{1}{4}F_{ab}^A F^{ab}_A,
\end{eqnarray}
here we have used Eq. $(\ref{2de})$. 
\section{Gauge Fixing}
In this section we  discuss gauge-fixing   for Yang-Mills theories.
 Fixing a gauge is essential in Yang-Mills theories before they can be quantized. 
To see this we first shall briefly review the quantization of the scalar field theory. 
If we consider a massive free scalar field on a $(3+1)$ dimensional spacetime, then the Lagrangian of this scalar field is given by 
\begin{equation}
 \mathcal{L} = \frac{1}{2}[-\nabla_a \phi \nabla^a\phi - m^2 \phi^2].
\end{equation}
 We can canonically quantize this theory by first defining  conjugate momentum $\pi (x)$  as 
\begin{equation}
 \pi = \sqrt{-g}\frac{\partial \mathcal{L} }{\partial \nabla_0\phi} = -\sqrt{-g} g^{0 a} \nabla_a \phi, 
\end{equation}
then by imposing the following canonical commutation relations: 
\begin{eqnarray}
 [\phi({\bf{x}},t), \pi({\bf{x}}',t)] &=& i \delta({\bf{x}},{\bf{x}}'), \nonumber \\ 
  \left[ \phi({\bf{x}},t), \phi({\bf{x}}',t)\right]  &=& 0, \nonumber \\ 
\left[ \pi({\bf{x}},t), \pi({\bf{x}}',t)\right]  &=&0,
\end{eqnarray}
where $\delta({\bf{x}},{\bf{x}}')$ is defined as 
\begin{equation}
 \int d^3{\bf{x}} \delta({\bf{x}},{\bf{x}}')f({\bf{x}}) = f({\bf{x}}'),
\end{equation}
for any compactly supported smooth function $f({\bf{x}})$. Note here that $d^3{\bf{x}} = dx^1 dx^2 dx^3$. 

To apply the above  procedure to Yang-Mills theories, we would have to define   
\begin{equation}
 \pi^{a}_A = \sqrt{-g} \frac{\partial \mathcal{L} }{\partial \nabla_0 A^A_a}.
\end{equation}
However, the conjugate momentum $\pi^{ 0}_A$ corresponding to $A^A_0$ is constrained to vanish:
\begin{equation}
 \pi^{ 0}_A = \sqrt{-g} \frac{\partial \mathcal{L} }{\partial \nabla_0 A^A_0} =0.
\end{equation}
So we can not quantize Yang-Mills theories simply by imposing the canonical commutation relations, that are the straightforward 
generalization of those used in  the scalar field theory. 
In fact the constraint $\pi^{0}_A =0$ is closely related to the fact that the initial Lagrangian given in Eq.  $(\ref{lna})$ is 
invariant under  gauge transformations $\cite{2ch9}$, 
\begin{equation}
 \delta_\Lambda A_a^A =  \nabla_a \Lambda^A + g f^A_{BC} A_a^B \Lambda^C.
\end{equation}
However as is well known, we can overcome this problem by fixing the gauge and this can be achieved by adding a gauge-fixing   term and a ghost term to 
the original classical Lagrangian $\cite{2ch10}$. 

Let us explain the general procedure for obtaining a Lagrangian which can be quantized following the canonical quantization procedure $\cite{2ch20}$. 
We start by choosing a gauge-fixing   function, $G[A]^A$. In the classical theory one imposes the condition $G[A]^A=0$. 
In the quantum theory one adds a gauge-fixing   term and a ghost term to the classical Lagrangian.
 The gauge-fixing   term is  obtained by first squaring this gauge-fixing   function $G[A]^A$ 
in such a way that it forms a scalar in  the representation space of the Lie algebra and
 then multiplying it by $-1/(2 \alpha)$. 
To find the ghost Lagrangian we first take the gauge transformation of the gauge-fixing   function $G[A]^A$ and then change 
the $\Lambda^A$ to ghost fields $c^A$.
After that we contract any free index left with anti-ghosts $\overline{c}^A$ to form a scalar quantity 
in the representation space of the Lie algebra and then multiply it by $i$.  
The scalar thus obtained is called the Faddeev-Popov ghost Lagrangian   $\cite{2ch10}$. 
Both the ghost fields and anti-ghost fields obey Fermi-Dirac statistics. 
  The Faddeev-Popov ghost Lagrangian is also needed for ensuring the unitarity of the theory (see Chapter $16$ of Ref. $\cite{1chaaa234}$).
In fact Faddeev-Popov ghosts were first conjectured by Feynman using unitarity $\cite{fy}$ and later derived by 
Faddeev and Popov by using path integral $\cite{2ch10}$. 

In  the classical theory, in Lorentz gauge the gauge-fixing   condition is given by
\begin{equation}
 G[A]^A = \nabla_a A^{Aa} = 0.
\end{equation}
In the quantum theory the  gauge-fixing   term $\mathcal{L}_g$  corresponding to this gauge-fixing   condition is 
obtained by first squaring this gauge-fixing   condition 
in such a way that it forms a scalar quantity  and then multiplying it by $-1/(2 \alpha)$.
\begin{equation}
 \mathcal{L}_g = - \frac{1}{2\alpha}[  \nabla_a A^{a}_A \nabla_b A^{Ab}]. 
\end{equation}
To obtain the ghost Lagrangian  we first take the gauge transformation of the gauge-fixing   function: 
\begin{equation}
 \delta_\Lambda G[A]^A = \nabla^a [\nabla_a \Lambda^A + g f^A_{BC} A_a^B \Lambda^C],
\end{equation}
and then change $\Lambda^A$ to $c^A$, 
\begin{equation}
\delta_{(c)}G[A]^A  = \nabla^a [\nabla_a c^A + g f^A_{BC} A_a^B c^C].
\end{equation}
Finally we contract with anti-ghosts $\overline{c}^A$, in such a way that it forms a scalar quantity, 
\begin{equation}
 \overline{c}_A \delta_{(c)} G[A]^A =  \overline{c}_A \nabla^a [\nabla_a c^A + g f^A_{BC} A_a^B c^C].
\end{equation}
After dropping total derivative terms, and multiplying by $i$ we obtain the  ghost Lagrangian,
\begin{equation}
 \mathcal{L}_{gh} =  - i \nabla^a \overline{c}_A D_a c^A, \label{gmy}
\end{equation}
where  $D_a c^A$ is the covariant derivative for the fields in adjoint representation with generators $\tau_A$,
\begin{equation}
 D_a c^A = \nabla_a c^A + g  f^A_{BC} A_a^B c^C.
\end{equation}
The total Lagrangian is given by the sum of the classical Lagrangian, the gauge-fixing   term and the ghost term, 
\begin{equation}
 \mathcal{L} = \mathcal{L}_c +  \mathcal{L}_g + \mathcal{L}_{gh}.
\end{equation}

\section{BRST Symmetry for Yang-Mills Theories}
The original classical Lagrangian is invariant under gauge transformations. After fixing the gauge this invariance is broken. 
However the  total Lagrangian, which is obtained as a sum of the original classical Lagrangian, 
the gauge-fixing   term and the ghost term is now invariant under a transformation called the BRST transformation $\cite{BRST}$. 
To see this we first rewrite the gauge-fixing   term in terms of an auxiliary field $B^A$, as:
\begin{equation}
 \mathcal{L}_g =  \left[ B_A \nabla_a A^{Aa} + \frac{\alpha}{2} B_A B^A \right].
\end{equation}
It is possible to recover the original gauge-fixing   term from this using the field equation for $B^A$.
Now  the BRST transformation  for Yang-Mills theory is given by 
\begin{equation}
\textbf{s} = \epsilon s,
\end{equation}
where $\epsilon $ is an anti-commuting complex number and 
\begin{eqnarray}
  s  A_a^A &=& i  [ \nabla_a c^A + g f^A_{BC} A_a^B c^C], \nonumber \\
s  B^A &=& 0, \nonumber \\
 s  \overline{c}^A &=&  B^A, \nonumber \\
s  c^A &=& -\frac{ig}{2}   f^A_{BC} c^B c^C.
\end{eqnarray}
As the BRST transformation is fermionic in the sense that it depends on a fermionic parameter $\epsilon$,
 it is important to fix a convention as to how the transformation $s$ acts on the fields. We let them  act from the left on the fields, for example
\begin{equation}
 s (\overline{c}^A c_A) = s(\overline{c}^A) c_A - \overline{c}^A s c_A.
\end{equation}
Now these  transformations can be seen to satisfy 
\begin{equation}
 (s )^2=0. \label{nilpotent}
\end{equation}
The action of $(s )^2$ on $B^A$ and $\overline{c}^A$ trivially vanishes. 
\begin{equation}
 s (s  B^A) =0,
\end{equation}
and 
\begin{equation}
 s (s  \overline{c}^A) = s  B^A  =0.
\end{equation}
Its action on $c^A$ is given by 
\begin{eqnarray}
 (s )^2 c^A &=&  - \frac{ig}{2} s  [ f^A_{BC} c^B c^C] \nonumber \\&=& 
\frac{g^2}{4} [- f^A_{BC} f^B_{EF} c^E c^F c^C + f^A_{BC} f^C_{EF} c^B  c^E c^F] \nonumber\\ &=& 0. 
\end{eqnarray}
We have used the Jacobi identity given by Eq. $(\ref{jacobi})$ and the fact that $c^A$ are anti-commutating  in the last equality. 
Similarly we have 
\begin{eqnarray}
(s )^2 A_a^A &=& i s  [ \nabla_a c^A + g f^A_{BC} A_a^B c^C] \nonumber \\
&=&   \frac{g}{2} f^A_{BC}\nabla_a  (c^B c^C) - g f^A_{BC} (\nabla_a c^B) c^C \nonumber \\ && - g^2 f^A_{BC} f^B_{EF} A_a^E c^F c^C + \frac{g^2}{2} f^A_{BC}  
 f^C_{EF} A_a^B c^E c^F \nonumber\\ &=& 0,
\end{eqnarray}
 where we have used the Jacobi identity given by Eq. $(\ref{jacobi})$ and the fact that 
\begin{eqnarray}
 f^A_{BC}\nabla_a (c^B c^C)  &=&   f^A_{BC}(\nabla_a c^B) c^C +  f^A_{BC} c^B \nabla_a c^C \nonumber \\ &=&
f^A_{BC}(\nabla_a c^B) c^C -  f^A_{CB} c^B \nabla_a c^C \nonumber \\&=&f^A_{BC}(\nabla_a c^B) c^C -  f^A_{BC} c^C \nabla_a c^B \nonumber \\ &=&
2 f^A_{BC}(\nabla_a c^B) c^C.
\end{eqnarray}
The original classical Lagrangian is invariant under the BRST transformation, 
as the BRST transformation for the original classical Lagrangian is just the gauge transformations with $\Lambda^A$ replaced by $c^A$:
\begin{equation}
 s  \mathcal{L}_c = 0.
\end{equation}
We also note that 
\begin{eqnarray}
&& s   \left[ \overline{c}_A \left[ \nabla_a A^{Aa} + \frac{\alpha}{2}  B^A \right]  \right]  \nonumber \\
&&= B_A  \left[ \nabla_a A^{Aa} + \frac{\alpha}{2}  B^A \right]  + i  \overline{c}_A \nabla^a D_a c^A   \nonumber \\
&&=\left[ B_A \nabla_a A^{Aa} + \frac{\alpha}{2} B_A B^A\right]  -  i \nabla^a \overline{c}_A D_a c^A   \nonumber \\
&&= \mathcal{L}_g + \mathcal{L}_{gh},
\end{eqnarray}
up to a total divergence term. 
The invariance of the sum of the gauge-fixing   term and the ghost term can now be deduced  from the fact that their sum is written as a total BRST variation. 
Thus the BRST variation of the sum of the gauge-fixing   term and the ghost term  vanishes due to Eq. $(\ref{nilpotent})$, as:
\begin{equation}
  s [\mathcal{L}_g + \mathcal{L}_{gh}] =   (s )^2  \left[   \overline{c}_A \left[ \nabla_a A^{Aa} + \frac{\alpha}{2}  B^A \right] \right]  =0.
\end{equation}
However, as the original classical Lagrangian is also invariant under the BRST transformation, 
so the total Lagrangian will also be invariant under the BRST transformation: 
\begin{equation}
 s  [\mathcal{L}_c + \mathcal{L}_g + \mathcal{L}_{gh} ]= 0.
\end{equation}
The invariance of Yang-Mills theory under BRST symmetry is crucial in selecting the physical states. 
\section{Physical States}
In Abelian gauge theory in flat spacetime, we get negative norm states in the Fock space. 
For example,  for free Abelian gauge theory in the Feynman gauge the creation and annihilation operators, $a_a (k)$ and $a_b^{\dagger}(k')$ satisfy, 
\begin{equation}
 [a_a (k), a_b^{\dagger}(k')] = i g_{ab} \delta (k,k'), 
\end{equation}
 so we have 
\begin{equation}
 [a_0 (k), a_0^{\dagger}(k')] = - i  \delta (k,k'), 
\end{equation}
It is possible to restrict our Fock space to positive norm states in Abelian gauge theory by using Gupta-Bleuler formalism $\cite{2ch19}$.
In this formalism physical states $|\phi_p\rangle$ are defined by requiring that 
\begin{equation}
 \partial_a A^{(+)a} (x)|\phi_p\rangle =0,
\end{equation}
where $\partial_a A^{(+)a} (x)$ is the positive frequency  part of $\partial_a A^{a} (x)$. It can be shown that 
\begin{equation}
 \Box \partial_a A^{a} (x) =0,
\end{equation}
even if the field $A^a$ interacts with charged matter. 
Thus, this method works for Abelian  gauge theory in Minkowski spacetime, even if interactions are included.
 However it does not work for non-abelian gauge theories in Minkowski spacetime  as $\partial_a A^{a} (x)$
does not satisfy the free field equation for non-abelian gauge theories and hence one cannot define its positive frequency part.
 It also fails in case of Abelian gauge theory in general curved spacetime as it is 
not always possible to define the positive frequency part of $\nabla_a A^{a} (x)$ uniquely 
in general curved spacetime.  

However it is possible to remove these negative norm states in case of non-abelian gauge theory 
 in Minkowski spacetime by the Kugo-Ojima criterion $\cite{2ch20}$.  Kugo-Ojima criterion states that  the physical states
 $ |\phi_p \rangle $  must be annihilated by $Q$ $\cite{2ch20}$,
\begin{equation}
 Q |\phi_p \rangle =0,
\end{equation}
where $Q$ is the BRST charge, which is the Noether charge corresponding to 
invariance of the total Lagrangian under the BRST transformation. 
The Kugo-Ojima criterion divides the Fock space into two parts, the physical part which is annihilated by $Q$ and the unphysical part which is not 
annihilated by $Q$. The physical part in turn has a trivial part which is composed of those  states
 which are obtained by the action of $Q$ on  states that are not annihilated by $Q$, $|\phi_t\rangle = Q|\phi_{up}\rangle $, where $Q|\phi_{up}\rangle \neq 0$. 
These trivial physical states are orthogonal to all physical states, 
\begin{equation}
 \langle \phi_p|\phi_t\rangle =  \langle \phi_p | Q |\phi_{up}\rangle =0.
\end{equation}

Thus in Minkowski spacetime the only relevant physical states for 
 non-abelian gauge theory 
 are those that are annihilated by $Q$ and are 
not  obtained by the action of $Q$ on any other state. 

In curved spacetime it is expected that BRST symmetry will play an important role in defining the physical states $\cite{BRSTcurved}$. However, in this 
thesis we shall show that the Feynman propagator for ghost fields in de~Sitter spacetime suffers from IR divergences. Due to this problem the  definition 
of BRST symmetry for Yang-Mills theories in de~Sitter spacetime might become non-trivial. However if BRST symmetry is defined for Yang-Mills theories in 
de~Sitter spacetime then the  Kugo-Ojima criterion can be used to obtain the physical subspace, in analogy to what is done in case of  Minkowski spacetime.    

\chapter{Perturbative Quantum Gravity}
In this chapter we shall  discuss the quantization of perturbative quantum gravity in general curved spacetime.  
We shall  also discuss the BRST symmetry for perturbative quantum gravity in this chapter.
\section{Basic Formalism}
According to General theory of Relativity, gravity is caused by the curvature of spacetime $\cite{5ch1}$.  
The Lagrangian for gravity with a cosmological constant $\lambda$, is given by
\begin{equation}
 \mathcal{L}_c = \frac{\sqrt{-g^{(f)}}}{16\pi G} [R^{(f)} - 2\lambda ].\label{31}
\end{equation}
We have denoted the full metric as $g^{(f)}_{ab} $ to distinguish it from the fixed  background metric $g_{ab}$. The scalar curvature corresponding to this
full metric is  denoted by $R^{(f)}$. 
We adopt units such that
\begin{equation}
 16 \pi G = 1.
\end{equation}
The Lagrangian given by Eq. $(\ref{31})$ is invariant under the following infinitesimal transformations originating from its 
general coordinate invariance: 
\begin{equation}
\delta_\Lambda g^{(f)}_{ab} =  \pounds_\Lambda g^{(f)}_{ab},
\end{equation}
where the Lie derivative $\pounds_\Lambda t_{ab}$ for any tensor $t_{ab}$ is given by
 \begin{equation}
  \pounds_\Lambda t_{ab} = \Lambda^c \nabla_c t_{ab} +t_{ac} \nabla_b \Lambda^c + t_{bc} \nabla_a \Lambda^c.
 \end{equation}

In perturbative gravity one writes the full metric in terms of a fixed background metric and small perturbations around it.
We also denote the  small perturbation around the fixed background metric as $h_{ab}$. So we can now write, 
\begin{equation}
 g^{(f)}_{ab} = g_{ab} + h_{ab}.
\end{equation}
This small perturbation is regarded as a field that is to be quantized. 
 The covariant derivative  along with the raising and lowering of indices will be   with respect to the background metric. 

For perturbative quantum gravity,
 we first write the Lagrangian in terms of the full
 metric $g^{(f)}_{ab}$ and then expand $g^{(f)}_{ab}$ in terms of  $h_{ab}$ and $g_{ab}$. So we get the gravitational  
Lagrangian for perturbative quantum gravity in terms of $h_{ab}$. It will in general contain infinitely many terms.
 Each term  containing two derivatives of $h_{ab}$ comes from the  Ricci scalar $R^{(f)}$, and there are only finitely many terms of this kind.
 However, because  the inverse of
 the metric contains infinitely many terms with  integer powers of $h_{ab}$, we have infinitely many terms in the Lagrangian
 for perturbative quantum gravity. 

Now as $g_{ab}$ is fixed, the transformation of $g^{(f)}_{ab}$ will be attributed to $h_{ab}$. Thus, the transformation of $h_{ab}$ is now given by 

\begin{eqnarray}
\delta_\Lambda h_{ab} &=&\pounds_\Lambda g^{(f)}_{ab}   \nonumber \\  &=& 
  \pounds_\Lambda g_{ab} +   \pounds_\Lambda h_{ab} \nonumber \\  &=&
\nabla_a \Lambda_b + \nabla_b \Lambda_a + \pounds_\Lambda h_{ab}.
\end{eqnarray}
In the last line we have used $\nabla_c g_{ab} =0$. 
 To first order in $\Lambda_a$, the  Lagrangian  for perturbative quantum gravity will be invariant 
 to all orders in $h_{ab}$, under the following transformation, 
\begin{equation}
\delta_\Lambda h_{ab} = \nabla_a \Lambda_b + \nabla_b \Lambda_a + \pounds_\Lambda h_{ab},\label{ggt}
\end{equation}
where the Lie derivative $\pounds_\Lambda h_{ab}$ is given by
 \begin{equation}
  \pounds_\Lambda h_{ab} = \Lambda^c \nabla_c h_{ab} +h_{ac} \nabla_b \Lambda^c + h_{bc} \nabla_a \Lambda^c.
 \end{equation}

Thus gravity is somewhat analogous to Yang-Mills theories. Just as  in the Yang-Mills  theory the Lagrangian was invariant under gauge transformations,
 here the Lagrangian is invariant under general coordinate transformations given by Eq. $(\ref{ggt})$. 
 Due to this analogy  one can  deal with gravity in a similar way to what was done for Yang-Mills theories.

\section{Quantization}
In the previous section we found that the Lagrangian for perturbative quantum gravity is invariant under the following transformations:
\begin{equation}
\delta_\Lambda h_{ab} = \nabla_a \Lambda_b + \nabla_b \Lambda_a + \pounds_\Lambda h_{ab}.
\end{equation}
In the previous chapter we found  that the Lagrangian of the Yang-Mills theory was invariant under a gauge transformation 
 which led to constraints and we were  not able to quantize this theory without fixing a gauge. Now as the Lagrangian for 
perturbative quantum gravity is invariant under the above mentioned transformations, we  expect that constraints will exist for 
perturbative quantum gravity also. In fact it is known that just like in the Yang-Mills case the invariance of the Lagrangian for 
 perturbative quantum gravity leads 
 to the existence of  constraints $\cite{consta}$. We are thus not able to quantize it using canonical commutation relations 
without fixing the gauge. Thus we add a gauge-fixing   term and a ghost term to the classical Lagrangian of perturbative quantum gravity,
 just like what we did in case of Yang-Mills theories 
(see Chapter $2$ of Ref. $\cite{const}$). We shall explain this procedure below.

In classical theory we impose the gauge-fixing   condition,
\begin{equation}
G[h]_a = (\nabla^b h_{ab} - k  \nabla_a h) = 0. \label{gaugef}
\end{equation}
where, 
\begin{equation}
 k \neq 1. 
\end{equation}
For $k=1$, the conjugate momentum for $h_{00}$ still vanishes, so we take $k\neq 1$. For this reason sometimes $k$ is written as $1 + \beta^{-1}$,
 where $\beta$ is an arbitrary finite constant $\cite{15}$. 
In quantum theory we add a gauge-fixing   term and a ghost term to the classical Lagrangian. 
The gauge-fixing   term corresponding to this gauge-fixing   condition is obtained 
by first squaring this gauge-fixing   function $G[h]_a$, in such a way that it form a scalar quantity  and then multiplying it by $-1/(2 \alpha)$:
\begin{equation}
 \mathcal{L}_g  = -\frac{1}{2\alpha}[  \nabla^b h_{ab} - k  \nabla_a h ] \left[  \nabla_c h^{ac} - k  \nabla^a h \right].\label{gaugeft}
\end{equation}
Now to get the ghost action, we follow the same procedure we followed for getting the ghost action in the case of Yang-Mills theory. 
First we take the gauge transformation of the gauge-fixing   function:
\begin{eqnarray}
\delta_\Lambda G[h]_b = \nabla^a [\nabla_a \Lambda_b + \nabla_b \Lambda_a - 2 k g_{ab} \nabla_c \Lambda^c \nonumber \\
 + \pounds_\Lambda h_{ab} - k g_{ab}g^{cd}\pounds_\Lambda h_{cd}]. \label{ghve}
\end{eqnarray}
Next we introduce the  Faddeev-Popov ghost field $c^a$, which is a fermionic vector field, and define $  \pounds_c h_{ab}$ to be 
 \begin{equation}
  \pounds_c h_{ab} = c_c \nabla^c h_{ab} + h_{ac} \nabla_b c^c + h_{bc} \nabla_a c^c.
 \end{equation}
Then we change $\Lambda_a$ to $c_a$ in Eq. $(\ref{ghve})$  
\begin{eqnarray}
\delta_{(c)} G[h]_b =  \nabla^a [\nabla_a c_b + \nabla_b c_a - 2 k g_{ab} \nabla_c c^c \nonumber \\ + \pounds_c h_{ab} - k g_{ab}g^{cd}\pounds_c h_{cd} ].
\end{eqnarray}
Now we contract the free index  with anti-ghost field $\overline{c}^b$, which is also a fermionic vector field:   
\begin{eqnarray}
\overline{c}^b \delta_{(c)} G[h]_b=   \overline{c}^b\nabla^a [\nabla_a c_b + \nabla_b c_a - 2 k g_{ab} \nabla_c c^c \nonumber \\ 
+ \pounds_c h_{ab} - k g_{ab}g^{cd}\pounds_c h_{cd} ].
\end{eqnarray}
Finally we  multiply it by $i$ and  drop total derivatives to get the ghost Lagrangian, 
\begin{eqnarray}
\mathcal{L}_{gh} &=& - i \nabla^a \overline{c}^b[\nabla_a c_b + \nabla_b c_a - 2 k g_{ab} \nabla_c c^c + \nonumber \\  
&&   c_c \nabla^c h_{ab} + h_{ac} \nabla_b c^c + h_{bc} \nabla_a c^c - \nonumber \\ 
 &&  k g_{ab}g^{cd} [c_e \nabla^e h_{cd} + h_{ce} \nabla_d c^e + h_{de} \nabla_c c^e]].\label{gglfp}
\end{eqnarray}
The total Lagrangian is given by the sum of the classical Lagrangian, the gauge-fixing   term and the ghost term: 
\begin{equation}
 \mathcal{L} = \mathcal{L}_c +  \mathcal{L}_g + \mathcal{L}_{gh}.
\end{equation}

\section{BRST Symmetry for Perturbative Quantum Gravity}
The original classical Lagrangian is invariant under gauge transformations.
 However after fixing the gauge this invariance is broken. However the  total Lagrangian for the perturbative quantum gravity, 
which is obtained as a sum of the original classical Lagrangian, the gauge-fixing   term and the ghost term is also invariant under BRST symmetry
 $\cite{BRSTgravity}$. 
To see this we first rewrite the gauge-fixing   term in terms of an auxiliary field $B^a$, as:
\begin{equation}
 \mathcal{L}_g =  B^a \left[  \nabla^b h_{ab} - k  \nabla_a h + \frac{\alpha}{2} B_a  \right].
\end{equation}
The original form of the gauge-fixing   term can be obtained from this one using the field equation for $B^a$.
Now the BRST transformation  for perturbative quantum gravity  is given by 
\begin{equation}
   \textbf{s} = \epsilon s, 
\end{equation}
where again $\epsilon $ is an anti-commuting complex number and 
\begin{eqnarray}
  s  h_{ab} &=&    i[ \nabla_a c_b + \nabla_b c_a +  c_c \nabla^c h_{ab} \nonumber \\ &&+h_{ac} \nabla_b c^c + h_{bc} \nabla_a c^c], \nonumber \\
s  B_{a} &=& 0, \nonumber \\
  s  \overline{c}_{a} &=&  B_{a}, \nonumber \\
s  c^a &=& -  i c^b\nabla_b c^a.
\end{eqnarray}
This transformation also satisfies (see Chapter $5$ of Ref. $\cite{12chcl}$),
\begin{equation}
 (s )^2=0.
\end{equation}
The original classical Lagrangian is again invariant under the BRST transformation, as the BRST transformation for the original
 classical Lagrangian is just the gauge transformation with $\Lambda^a$ replaced by $c^a$:
\begin{equation}
 s  \mathcal{L}_c = 0. 
\end{equation}
We also note that up to a total divergence,
\begin{eqnarray}
&&s \left[  \overline{c}^a\left[ \nabla^b h_{ab} - k  \nabla_a h + \frac{\alpha}{2}  B_a \right] \right]   
 \nonumber \\&=& B^a\left[ \nabla^b h_{ab} - k  \nabla_a h + \frac{\alpha}{2}  B_a \right]  
+ \nonumber \\ && i \overline{c}^a \nabla^b [ \nabla_a c_b + \nabla_b c_a - 2 k g_{ab} \nabla_c c^c 
 \nonumber \\ && +  c_c \nabla^c h_{ab} + h_{ac} \nabla_b c^c + h_{bc} \nabla_a c^c -\nonumber \\ 
 &&  k g_{ab}g^{cd} [c_e \nabla^e h_{cd} + h_{ce} \nabla_d c^e + h_{de} \nabla_c c^e]]  \nonumber \\ &=&\mathcal{L}_g + \mathcal{L}_{gh}.
\end{eqnarray}
The invariance of the gauge-fixing   term and the ghost term can be 
now deduced from the fact that their sum is written as a total BRST variation and that the BRST variation of a total BRST variation vanishes: 
\begin{equation}
 s  [\mathcal{L}_g + \mathcal{L}_{gh}] =   
(s )^2 \left[ \overline{c}^a \left[ \nabla^b h_{ab} - k  \nabla_a h + \frac{\alpha^2}{2}  B_a \right]\right]   = 0. 
\end{equation}
However as the original classical Lagrangian was also 
invariant under the BRST transformation, so the total Lagrangian will also be invariant under the BRST transformation: 
\begin{equation}
 s  [\mathcal{L}_c + \mathcal{L}_g + \mathcal{L}_{gh} ]= 0.
\end{equation}
The invariance of perturbative quantum gravity  under BRST symmetry is also  crucial in selecting the physical states of the theory.
It is possible to define physical states in perturbative quantum gravity, at least formally using Kugo-Ojima criterion
 (see Chapter $5$ of Ref. $\cite{12chcl}$). 
Thus, the state    $ |\phi_p \rangle $ is said to be  a physical states if 
\begin{equation}
 Q |\phi_p \rangle =0.
\end{equation}
The only relevant physical states in perturbative quantum gravity again  are those states which are annihilated by $Q$ and are 
not  obtained by the action of $Q$ on any other state.  

In this 
thesis we will show that the Feynman propagator for ghost fields in perturbative quantum gravity in de~Sitter spacetime also suffers from IR divergences.
Thus just like the Yang-Mills case, due to IR divergences  the  definition 
of BRST symmetry for perturbative quantum gravity in de~Sitter spacetime might become non-trivial.

\chapter{de Sitter Spacetime}
In this chapter, we shall  review some basic properties of de~Sitter spacetime (see Chapter $5$ of Ref. $\cite{5ch1}$). 
We shall  start by reviewing   properties of different tensors in de~Sitter spacetime and then present  some basic coordinate systems used in de~Sitter spacetime.
\section{Properties of de Sitter Spacetime}
The vacuum Einstein equations with a cosmological constant $\Lambda$ are given by (see chapter $3$ of Ref. $\cite{5ch1}$).
\begin{equation}
 G_{ab} = -\Lambda g_{ab},
\end{equation}
where $G_{ab}$ is the Einstein tensor which is given by 
\begin{equation}
 G_{ab} = R_{ab} - \frac{1}{2}R g_{ab}. \label{einstein1}
\end{equation}
 Constant curvature spacetimes are solutions to Eq. $(\ref{einstein1})$, and are characterized by the condition (see Chapter $5$ of Ref. $\cite{5ch1}$)
\begin{equation}
 {R^a}_{bcd} = \frac{1}{12}R[\delta^a_c g_{bd}-\delta^a_d g_{bc}],
\end{equation}
where $R$ is a constant. 
The spacetimes with $R=0$, $R< 0$ and $R>0$ are called  the Minkowski spacetime, the anti-de~Sitter spacetime and  de~Sitter spacetime, respectively. 
Now for the constant curvature spacetimes, we have 
\begin{eqnarray}
 R_{bd} &=& {R^a}_{bad} \nonumber  \\ &=& \frac{1}{12}R [\delta^a_a g_{bd}-\delta^a_{d}g_{ba}] \nonumber \\ &=& \frac{1}{4}R g_{db}.
\end{eqnarray}
So the Einstein tensor $G_{ab}$ for the constant curvature spacetimes is given by 
\begin{eqnarray}
 G_{ab} &=& R_{ab} - \frac{1}{2}R g_{ab} \nonumber \\ 
&=& \frac{1}{4}g_{ab} R - \frac{1}{2}R g_{ab} \nonumber \\
&=& - \frac{1}{4}g_{ab} R. \label{einstein2}
\end{eqnarray}
Now if we compare Eq. $(\ref{einstein1})$ with Eq. $(\ref{einstein2})$, we see that we  can view Eq. $(\ref{einstein2})$
 as a solution of the vacuum Einstein field equations with a  cosmological constant $\Lambda$ given by 
\begin{equation}
 \Lambda = \frac{1}{4}R.
\end{equation}
As de~Sitter  spacetime is defined to be a spacetime of constant positive curvature, 
so for de~Sitter spacetime the cosmological constant is positive.
This cosmological constant is related to a constant called the Hubble constant $H$, as follows:
\begin{equation}
 \Lambda =  3H^2. \label{hubble}
\end{equation}
It is called the Hubble constant as it was  used by Hubble for the measure of the expansion of the universe $\cite{hubble}$. 
So for de~Sitter spacetime, in terms of the Hubble constant $H$, we have 
\begin{eqnarray}
 R &=& 12 H^2,\\  
 R_{ab} &=& 3 H^2 g_{ab},\\
 {R^a}_{bcd}& = &H^2 [\delta^a_{c} g_{bd} - \delta^a_{d} g_{bc} ],\label{jkla} \\
 G_{ab} &=& - 3 H^2 g_{ab}.
\end{eqnarray}
\section{de Sitter Spacetime Metric}
de Sitter spacetime, which is defined to be a spacetime of constant positive curvature,  has the topology $R \times S^3$ 
and  can be viewed as  a hyperboloid  in five dimensional Minkowski spacetime (see Chapter $5$ of Ref. $\cite{5ch1}$).
 If the Cartesian coordinates in the five dimensional Minkowski spacetime are  $X, Y, Z, W, T $  and the metric is given by
\begin{equation}
 ds^2 = -dT^2 + dW^2 +dX^2 + dY^2 + dZ^2,\label{min}
\end{equation}
with $-\infty < W,\, X,\, Y,\, Z,\, T <\infty$ then de~Sitter spacetime  is the hypersurface given by the  following equation,
\begin{equation}
 X^2 + Y^2 +Z^2 + W^2 - T^2 = \alpha^2. \label{desittergroup}
\end{equation}
Here $\alpha$ is called the radius of de~Sitter spacetime and is related to the 
Hubble constant by $\alpha = H^{-1}$, as we shall see. De~Sitter spacetime inherits the five dimensional Lorentz invariance of  five dimensional
 Minkowski spacetime, which is $SO(4,1)$ $\cite{desittergroup}$. 
This is called  de~Sitter group and is defined to be the group of linear transformations in five dimensions
 which preserve Eq. $(\ref{desittergroup})$.

A convenient parametrisation of this hypersurface is given by 
 \begin{eqnarray}
  T & =& \alpha\sinh (\alpha^{-1}t), \nonumber \\
  W & =&\alpha \cosh (\alpha^{-1}t) \cos \psi ,\nonumber \\
  X & =& \alpha \cosh (\alpha^{-1}t) \sin \psi \sin \theta \cos \phi ,\nonumber \\
  Y & =& \alpha \cosh (\alpha^{-1}t) \sin \psi \sin \theta \sin\phi ,\nonumber \\
  Z &=& \alpha\cosh (\alpha^{-1}t) \sin\psi \cos \theta. \label{coordinates}
 \end{eqnarray}
with  $0\leq  \psi, \, \theta \leq \pi, 0\leq  \phi \leq 2 \pi$ and $-\infty < t <\infty $. Then substituting 
 Eq.  $(\ref{coordinates})$ in the Minkowski metric given by Eq. $(\ref{min})$, we get 
 \begin{equation}
  ds^2 =  -dt^2 + \alpha^2 \cosh^2 \alpha^{-1}t [d\psi^2 + \sin^2 \psi (d\theta^2 + \sin^2 \theta d \phi^2 )]. \label{metricabc1}
 \end{equation}
The singularities  at $\psi =0, \pi$ and $\theta = 0, \pi$ are  those singularities that occur in polar
 coordinates. Apart from them this coordinate system is regular in the whole of de~Sitter spacetime and 
covers all of de~Sitter spacetime. If we rescale  by letting $ t \to \alpha t $, 
then the metric given in Eq. $(\ref{metricabc1})$ becomes
\begin{equation}
ds^2 = \alpha^2 [-dt^2 + \cosh^2 t [d\psi^2 + \sin^2 \psi (d\theta^2 + \sin^2 \theta d \phi^2 )]]. \label{metricabc2}
\end{equation}
 If we further perform the  transformation, 
\begin{equation}
 \chi = \frac{\pi}{2} -it,\label{timeimagenary}
\end{equation}
then this rescaled de~Sitter metric given in Eq. $(\ref{metricabc2})$  becomes 
\begin{equation}
 ds^2 = \alpha^2 [d\chi^2 + \sin \chi^2 [d\psi^2 + \sin^2 \psi (d\theta^2 + \sin^2 \theta d \phi^2 )]].\label{timeimagenary1}
\end{equation}
This is in fact the metric on a four dimensional sphere denoted by $S^4$, of radius $\alpha$. It may be noted that for $\chi$ to be real 
in Eq. $(\ref{timeimagenary})$, time $t$ has to be imaginary. So we are using imaginary time in Eq. $(\ref{timeimagenary1})$. 

We can also write the  de~Sitter metric as follows:
\begin{equation}
 ds^2 = -dt^2 + \exp(2\alpha^{-1} t) (dx^2+dy^2+dz^2), \label{desittermetric2}
\end{equation}
where
\begin{eqnarray}
  t & =& \alpha \log\frac{(W+T)}{\alpha},\nonumber \\
  x & =& \alpha \frac{X}{W+T}, \nonumber \\
  y & =&\alpha \frac{Y}{W+T},\nonumber \\
  z & =& \alpha\frac{Z}{W+T}.
 \end{eqnarray}
Here $-\infty < x, \,y,\, z < \infty$ and $ 0 < t < \infty$.
 However these coordinates only cover half of de~Sitter spacetime as $t$ is not defined for $W+T\leq 0$. 

 Now if the FRW metric is given by 
\begin{equation}
 ds^2 = -dt^2 +a^2(t) d\Omega^2, 
\end{equation}
where $d\Omega^2$ is the spatial part of the metric. It can be shown  that  the  Hubble constant given in Eq. $(\ref{hubble})$, is related to $a$ by 
(see Chapter $5$ of Ref.$\cite{frw1}$)
\begin{equation}
 H = \frac{\dot{a}}{a}.
\end{equation}
So for Eq. $(\ref{desittermetric2})$, the Hubble constant is given by 
\begin{equation}
 H = \frac{1}{\alpha} .
\end{equation}
 Unless specified otherwise, from now on  we shall choose units such that $H^2 =1$, to simplify calculations.  
\chapter{Ghosts for Yang-Mills Theories in de~Sitter Spacetime}
In this chapter we shall  examine the ghost propagators  for Yang-Mills theories in de~Sitter spacetime. We shall  
use the fact the Feynman propagators in the Euclidean vacuum $\cite{{7ch31}}$ in de~Sitter spacetime can be obtained from
 the corresponding Green's function  on the $S^4$  by analytical continuation $\cite{aj}$.
\section{Ghosts for Yang-Mills Theories}
We have already presented the ghost Lagrangian for Yang-Mills theories in general spacetime in Eq. $(\ref{gmy})$.
So we can start from the following ghost Lagrangian,
\begin{equation}
 \mathcal{L}_{gh} =  -i \nabla^a \overline{c}_A D_a c^A.
\end{equation}
Now as  the covariant derivative $D_a c^A$ is given by
\begin{equation}
 D_a c^A = \nabla_a c^A + g f^A_{BC} A_a^B c^C,
\end{equation}
 we have 
\begin{equation}
  \mathcal{L}_{gh} = -  i [\nabla^a \overline{c}_A \nabla_a c^A +  g f^A_{BC} \nabla^a\overline{c}_A  A_a^B c^C].
\end{equation}
This can now be written as a free Lagrangian $\mathcal{L}_{gh}^{free}$  and an interaction part  $  \mathcal{L}_{gh}^{int}$, where the free Lagrangian is given by
\begin{equation}
\mathcal{L}_{gh}^{free} =  - i  \nabla_a \overline{c}_A \nabla^a c^A.
\end{equation}
The equation of motion obtained from this free Lagrangian for ghosts and anti-ghosts are 
\begin{eqnarray}
 \Box c^A &=& 0,\\
 \Box \overline{c}^A &=& 0.
\end{eqnarray}
These are the equations of motion for the  minimally-coupled massless scalar fields.  
 
As is well known, there is ambiguity in the selection of positive frequency solutions and thus an ambiguity in selecting
the vacuum state in curved  spacetime $\cite{912}$. 
However, for de~Sitter spacetime there is a de~Sitter invariant vacuum state  called the Euclidean vacuum $\cite{{7ch31}}$ in 
which the Feynman propagator  can be obtained from the Green's function on the four sphere by analytical continuation $\cite{aj}$. 
We shall use the Euclidean  vacuum  state for calculating the Feynman ghost propagator for Yang-Mills theories.
 
Let $|0\rangle$ be the  Euclidean vacuum state. Then the  Feynman propagator for the free ghost fields would be given by
\begin{equation}
 \langle 0| T[c^A(x)\overline{c}^B (x')]|0\rangle = i \delta^{AB} D_0 (x,x'),
\end{equation}
where $D_0 (x,x')$ would satisfy 
\begin{equation}
 \Box  D_0(x,x') = - \delta^4 (x,x').\label{69}
\end{equation}
However, it is  well known that there is no Feynman propagator for minimally-coupled massless scalar fields that respects de~Sitter invariance $\cite{7ch31}$.
We can see that there is no Euclidean vacuum for these fields as follows. To find the Feynman propagator in the Euclidean vacuum, we first find the Green's function
on $S^4$.   
Now a complete set of basis for any scalar function on $S^4$ is the scalar spherical harmonics $Y^{L\sigma}$  which satisfy $\cite{sph}$, 
\begin{equation}
- \Box Y^{L\sigma} = L(L+3)  Y^{L\sigma},\label{701}
\end{equation}
where $L = 0,1,2,3,4 \cdots$ and   $\sigma$ represents all the other labels. The scalar spherical harmonics are normalized as:
\begin{equation}
 \int d^4x \sqrt{g} Y^{L\sigma}Y^{*L'\sigma'} =  \delta^{L L'}\delta^{\sigma \sigma'}.\label{701b}
\end{equation}
We define $\delta(x,x')$ as:
\begin{equation}
 \int \sqrt{g(x')} d^4x' f(x')  \delta^4 (x,x') = f(x),  
\end{equation}
for any function $f(x')$.
We can  mode expand  $\delta(x,x')$ as: 
\begin{equation}
  \delta^4 (x,x')=\sum_{L=0}^{\infty}  \sum_{\sigma}   Y^{L\sigma} (x)Y^{*L\sigma} (x'),\label{68}
\end{equation}
because if we  mode expand any function $f(x')$ on $S^4$ as: 
\begin{equation}
 f(x') = \sum_{L'=0}^{\infty}  \sum_{\sigma'} \lambda_{L'\sigma'}Y^{L'\sigma'} (x'),
\end{equation}
where $ \lambda_{L'\sigma'}$ are constants, then use Eq. $(\ref{701b})$, we  find
\begin{eqnarray}
&& \sum_{L=0}^{\infty}  \sum_{\sigma} \int \sqrt{g} d^4x'  f(x')    Y^{L\sigma} (x)Y^{*L\sigma} (x') 
\nonumber \\ &=& \sum_{L=0}^{\infty}  \sum_{\sigma} \sum_{L'=0}^{\infty}  \sum_{\sigma'} \int \sqrt{g} d^4x' \,
 \lambda_{L'\sigma'}Y^{L'\sigma'} (x') Y^{L\sigma} (x)Y^{*L\sigma} (x') \nonumber \\ &=& \sum_{L=0}^{\infty}  \sum_{\sigma} \lambda_{L\sigma} Y^{L\sigma} (x)
= f(x).
\end{eqnarray}
If Eq. $(\ref{69})$ were to be satisfied, the Green's function $D_0(x,x')$ would have to be decomposed into spherical harmonic modes as follows:  
\begin{equation}
D_0(x,x') = \sum_{L=0}^{\infty}  \sum_{\sigma} k_{L}  Y^{L\sigma} (x)Y^{*L\sigma} (x'). \label{67} \\
\end{equation}
where $k_{L}$ is a constant. Now substituting Eqs. $(\ref{68})$ and  $(\ref{67})$ into Eq. $(\ref{69})$ we get,
\begin{equation}
\Box   \sum_{L=0}^{\infty}  \sum_{\sigma} k_{L} Y^{L\sigma} (x)Y^{*L\sigma} (x') = -\sum_L  \sum_\sigma   Y^{L\sigma} (x)Y^{*L\sigma} (x').\label{701a}
\end{equation}
 From  Eqs. $(\ref{701})$  and $(\ref{701a})$,  we get 
\begin{equation}
 k_{L} = \frac{1}{L(L+3)}.
\end{equation}
So formally we can write 
\begin{equation}
 D_0(x,x')= \sum_{L=0}^{\infty}  \sum_{\sigma} \frac{Y^{L\sigma} (x)Y^{*L\sigma}(x')}{L(L+3)}. \label{propagatormass}
\end{equation}
However, $D_0(x,x')$ is actually not well defined  because $L(L+3) =0$ for $L=0$. If we regulate this propagator by adding a small mass $m^2$ to 
Eq. $(\ref{69})$, then we get  
\begin{equation}
 [ \Box -m^2] D_{m^2}(x,x') = - \delta(x,x'). \label{ns12a}
\end{equation}
Then following what we did for the minimally-coupled  massless scalar field, we get 
 \begin{equation}
  D_{m^2} (x,x') = \sum_{L= 0 }^{\infty}\sum_{\sigma} \frac{Y^{L\sigma} (x)Y^{*L\sigma}(x')}{L(L+3) + m^2}. \label{propagatormass1}
 \end{equation}
This propagator given by Eq. $(\ref{propagatormass1})$ diverges in the zero mass limit. Such divergences are called infrared divergences. 
As the volume of the unit $S^n$ is given by
\begin{equation}
 V = \frac{2\pi^{n+1}}{\Gamma(\frac{n+1}{2}) },
\end{equation}
so the volume of the unit $S^4$ is
\begin{equation}
 V= \frac{8\pi^2}{3}.\label{ns121}
\end{equation}
Now from from Eqs. $(\ref{701b})$ and  $(\ref{ns121})$, the $L= 0$ mode will be given by 
 \begin{equation}
 Y^{0} =  \sqrt{\frac{3}{8\pi^2 }}.\label{zerosac}
 \end{equation}
The $L=0$ mode, which is a constant mode, is the cause of  infrared divergence as its contribution  diverges in the zero mass limit. 
 Thus by substituting Eq $(\ref{zerosac})$ into  Eq. $(\ref{propagatormass1})$, we find 
\begin{equation}
D_{m^2} (x,x') = \frac{3}{8\pi^2 m^2} +  \sum_{L = 1}^{\infty} \sum_\sigma \frac{Y^{L\sigma} (x)Y^{*L\sigma}(x')}{L(L+1) + m^2}.\label{21}
\end{equation}
This propagator obviously diverges in the zero mass limit. 
\section{Effective Propagator}
We saw in the previous section that the propagator for the ghost fields suffers  infrared divergence in
  de~Sitter spacetime as  the free part of the ghost Lagrangian satisfies the minimally-coupled massless scalar field equation, 
which is known to be IR divergent in de~Sitter spacetime. If the ghost fields were physical fields, we would need to break the de Sitter
 invariance of the vacuum for these fields $\cite{6cha}$. However we shall  see that this problem can be circumvented because they are unphysical 
fields and thus do not exist in the final/initial states of any scattering calculations and  they  only appear in internal loops in the Feynman diagrams. 
This can be seen as follows.

First we note that if  we consider the propagator which is regulated by addition of a small mass,  then we observe that 
  the infrared divergence are caused by the $L=0$ mode,
which is a  constant mode, in the zero mass limit. However this constant mode does not contribute to the loops in the Feynman diagrams.  
This is because the interaction part of the ghost Lagrangian is given by 
\begin{equation}
\mathcal{L}_{gh}^{int} =  -  i g f^A_{BC} \nabla^a\overline{c}_A.  A_a^B c^C, 
\end{equation}
so the anti-ghosts  couple to the gauge field  through a derivative coupling and this derivative eliminates 
the constant modes. Now as the ghosts and anti-ghosts  only occur in the internal loops in the Feynman diagrams, they are unphysical fields. 
So the constant modes will always be eliminated in any perturbative calculation. 

We propose therefore to use an effective propagator obtained by first subtracting out 
this constant mode and then taking the zero mass limit of the regulated propagator.
 Note that, we are free to add any finite constant to this effective propagator, as the contribution coming from any constant will not contribute  
in the perturbative calculations, due to the   reasons mentioned above.
It appears likely that the use of this effective propagator will lead to a consistent perturbative theory. 
However the consistency of the theory is not obvious because it is not clear how the removal of the zero modes from the Faddeev-Popov ghosts will 
affect the BRST symmetry of the theory. 
 
Thus, we define $D_{m^2}^{\rm eff} (x,x')$ by
\begin{equation}
D_{m^2}^{\rm eff} (x,x') = D_{m^2}(x,x') - \frac{3}{8\pi^2 m^2}.\label{20}
\end{equation}
Then from Eqs.  $(\ref{21})$ and $(\ref{20})$, we have 
\begin{equation}
D_{m^2}^{\rm eff} (x,x') =  \sum_{L = 1}^{\infty} \sum_\sigma \frac{Y^{L\sigma} (x)Y^{*L\sigma}(x')}{L(L+1) + m^2}.
\end{equation}
Clearly this is convergent  in the zero mass limit, and its zero mass limit plus any arbitrary finite constant $C/16\pi^2$ 
gives us the effective propagator for the ghost fields $D_{0}^{\rm eff}$:
\begin{equation}
\lim_{m^2 \to 0}D_{m^2}^{\rm eff} (x,x')  + \frac{C}{16\pi^2}  = D_{0}^{\rm eff} (x,x').\label{ymp}
\end{equation}
This propagator appears for very different physical reasons in the works of Allen and Turyn on covariant graviton propagator $\cite{a}$. 
\section{Explicit Expression for the Effective Propagator}
In this section we derive an explicit expression for $D_{0}^{\rm eff}$ using the work of  Allen and Jacobson $\cite{aj}$. 
This propagator has been derived and used in a different context, as was mentioned above $\cite{a}$.
Following Allen and Jacobson, we define  $\mu(x, x′)$ to be the geodesic distance between spacelike separated points $x$ and $x'$ in de~Sitter spacetime. 
 Also the variable $z$ is defined as, 
\begin{equation}
 z = \cos^2 \left( \frac{\mu}{2}\right). \label{geodisc}
\end{equation}
 In terms of the variable $z$, the solution to Eq. $(\ref{ns12a})$  is given by (see Sec. $2$ of Ref. $\cite{aj}$) 
 \begin{equation}
  D_{m^2}(z)  = \frac{1}{16\pi^2}\Gamma(a_+)\Gamma(a_-) F[a_+ , a_- ; 2 ; z].\label{coscalr}
 \end{equation}
where  $a_+$ and $a_-$ are given by
\begin{eqnarray}
 a_+ &=& \frac{3}{2} +\sqrt{\frac{9}{4} - m^2},\\
 a_- &=& \frac{3}{2} - \sqrt{\frac{9}{4} - m^2}.
\end{eqnarray}
This propagator clearly diverges in the zero mass limit, because $\Gamma(a_-) \to \infty $ as $m^2 \to 0$.
 However we have seen in the previous section that the effective propagator obtained by subtracting the constant $L=0$ mode from the solution to 
Eq. $(\ref{ns12a})$, has no IR divergences in the zero mass limit.  
So to find the effective propagator explicitly using the work of  Allen and Jacobson   we shall  first verify that the contribution
 of $L=0$ mode subtracted from the constant part of the Eq. $(\ref{coscalr})$ is a finite constant in the zero mass limit. 

Let $D_c$ be the constant $z$-independent part of $D_{m^2} (z)$ given by Eq. $(\ref{coscalr})$,
 \begin{equation}
 D_c =  \frac{1}{16\pi^2} \Gamma(a_+) \Gamma(a_-).\label{zerosca1}
 \end{equation}
We subtract the zero mode contributions given in Eq. $(\ref{zerosac})$ from Eq. $(\ref{zerosca1})$, 
 \begin{equation}
 D_c  -\frac{3}{8\pi^2 m^2} =  \frac{1}{16\pi^2} \Gamma(a_+) \Gamma(a_-) - \frac{3}{8\pi^2 m^2}.
 \end{equation}
We shall  verify that this is finite in the zero mass limit. We define 
\begin{equation}
 p = \sqrt{\frac{9}{4} - m^2}.
\end{equation}
We have 
\begin{equation}
 p - \frac{3}{2}= - \frac{m^2}{3} + \mathcal{O}(m^4),
\end{equation}
We can now write  
\begin{eqnarray}
\Gamma(a_+) \Gamma(a_-) &=&  \left(\frac{1}{2}+ p \right) \left(\frac{1}{2} -p\right) \Gamma\left( \frac{1}{2}+p\right) \Gamma\left( \frac{1}{2}-p\right)
  \nonumber \\
&=&\left(\frac{1}{2}+ p \right) \left(\frac{1}{2} -p\right) \frac{\pi}{\sin \pi \left( \frac{1}{2} -p \right) } 
 \nonumber \\ &=&
 \left(\frac{1}{2}+ p \right) \left(\frac{1}{2} -p\right)\frac{\pi}{-\sin \pi \left( \frac{3}{2} -p \right) } 
\nonumber \\&=&
\left(\frac{1}{2}+ p \right) \left(\frac{1}{2} -p\right)\frac{1}{- \left( \frac{3}{2} -p \right) } + \mathcal{O}(m^2).
\end{eqnarray}
Here we have used 
\begin{equation}
 \Gamma(z)\Gamma(1-z) = \frac{\pi}{\sin (\pi z )}.
\end{equation}
Hence $\Gamma(a_+) \Gamma(a_-)$ can be approximated as,
\begin{equation}
\Gamma(a_+) \Gamma(a_-) = \frac{ 6-3m^2}{m^2} +\mathcal{O}(m^2).
\end{equation}
The zero mass limit of the constant modes contribution subtracted from $D_c$ can be written as follows
\begin{eqnarray}
 && \lim_{m^2 \to 0}\left[ D_c  -\frac{3}{8\pi^2 m^2}\right]  \nonumber \\ 
&=& \lim_{m^2 \to 0} \frac{1}{16\pi^2} \left[ \frac{ 6-3m^2}{m^2} + \mathcal{O}(m^2) - \frac{6}{m^2} \right] \nonumber \\
 &=& - \frac{3}{16\pi^2}.
\end{eqnarray}
Thus this is finite in the zero mass limit and there are no infrared divergences. 
The exact value of this constant is not important as we are free to add any constant we like to this effective propagator, 
because the constant part of the propagator does not contribute in  perturbative calculations. The important point to note here 
is that it is finite and does not diverge.
 
Now we can calculate  the effective propagator as follows. 
We  first define $D'_0 (z)$ as, 
\begin{equation}
 D'_0 (z) = \lim_{m^2 \to 0}\left[  \frac{d}{dz} D_{m^2}(z)\right], \label{hypergeometric}
\end{equation}
and then define the effective propagator by
\begin{equation}
 D^{\rm eff}_0 (z) = \int_0^z dz D'_0(z) + \frac{C}{16\pi^2},\label{hypergeometric1}
\end{equation}
where $C$ is an arbitrary finite constant. 
Now as 
\begin{equation}
 \frac{d}{dz} F[a, b ; c ; z] = \frac{ab}{c}  F[a +1, b +1; c +1; z], 
\end{equation}
so Eq. $(\ref{hypergeometric})$,  becomes  
\begin{equation}
D'_0(z) =   \frac{1}{16\pi^2}\Gamma(4)\Gamma(1) F[4, 1 ; 3 ; z].
\end{equation}
With $C=0$,  Eq. $(\ref{hypergeometric1})$ in terms of elementary functions is given by 
 \begin{equation}
  D_{0}^{\rm eff}(z) =   \frac{1}{16\pi^2} \left[\frac{1}{1-z} -2 \log(1-z)  - \frac{14}{3} \right]. 
 \end{equation}
This is the effective propagator that can be used to do perturbative calculations.  

\chapter{Ghosts for Perturbative Quantum Gravity in de~Sitter Spacetime}
In this chapter we shall  calculate the ghost propagators for perturbative quantum gravity in de~Sitter spacetime.
 We shall  again use the fact the Feynman propagators in the Euclidean vacuum $\cite{{7ch31}}$ in de~Sitter spacetime can be obtained from the
 corresponding Green's function  on the $S^4$  by analytical continuation $\cite{aj}$.  

\section{Ghosts for Perturbative Quantum Gravity}
Since the  ghosts for perturbative quantum gravity are vector fields, we need to review the formalism
 of Allen and Jacobson for the vector propagators (see Sec. $1$ of Ref. $\cite{aj}$). 
Let $x$ and $x'$ be two spacelike separated points and let $\mu(x, x')$ be the geodesic 
distance between them as before. One defines the unit tangent vectors $n_a(x, x')$ at $x$
and $n_{a'}(x, x')$ at $x'$ along the geodesic between these two points as follows,
\begin{eqnarray}
 n_a(x, x') = \nabla_a \mu (x, x'),\\
n_{a'}(x, x') = \nabla_{a'} \mu (x, x').
\end{eqnarray}
In addition one defines a parallel propagator $g_{ac'}$ such that if 
 $A^a$ is a vector at $x$, and if $A^{c'}$  the vector at $x'$ obtained by parallelly transporting $A^a$ along the geodesic, then 
\begin{equation}
 A^{c'} = g^{c'}_a A^a.
\end{equation}
 Now as the unit tangents at $x$ and $x'$ point away from each other, so we have
 \begin{equation}
  g^{a}_{c'} n_a = - n_{c'}.
 \end{equation}
One also writes the metric at $x$ and $x'$ as $g_{ac}$ and $g_{a'c'}$, respectively.
Now any maximally symmetric bi-tensor can be expressed as a linear combination of 
 $g_{ac}$, $g_{a'c'}$, $n_a$, $n_{c'}$ and $g_{ac'}$ with the coefficient of each term depending only on $z = \cos^2 ( \mu / 2)$.
For example a maximally symmetric bi-vector $V_{ac'}(z)$ may be expressed as
\begin{equation}
 V_{ac'} (z) = \alpha(z)g_{ac'} + \beta(z) n_a n_{c'}.
\end{equation}

We have already presented  the ghost Lagrangian for perturbative quantum gravity in general spacetime in Eq. $(\ref{gglfp})$. 
So we can start from the following ghost Lagrangian, 
\begin{eqnarray}
\mathcal{L}_{gh} &=& - i \nabla^a \overline{c}^b[\nabla_a c_b + \nabla_b c_a - 2 k g_{ab} \nabla_c c^c + \nonumber \\  
&&   c_c \nabla^c h_{ab} + h_{ac} \nabla_b c^c + h_{bc} \nabla_a c^c - \nonumber \\  
&&  k g_{ab}g^{cd} [c_e \nabla^e h_{cd} + h_{ce} \nabla_d c^e + h_{de} \nabla_c c^e]].
\end{eqnarray}
This Lagrangian can now be written in terms of a free ghost Lagrangian $\mathcal{L}_{gh}^{free}$ and the Lagrangian for interactions $\mathcal{L}_{gh}^{int}$.
The free part of the ghost Lagrangian is given by 
\begin{equation}
 \mathcal{L}_{gh}^{free} = - i \nabla^a \overline{c}^b[\nabla_a c_b + \nabla_b c_a - 2 k g_{ab} \nabla_c c^c].
\end{equation}
So the free field equations for the ghosts and anti-ghosts are given by
\begin{eqnarray}
 \nabla^d [\nabla_d c_a + \nabla_a c_d - 2 k g_{da} \nabla_c c^c] &=&0, \label{freefieldequations1}\\
 \nabla^d [\nabla_d \overline{c}_a + \nabla_a \overline{c}_d - 2 k g_{da} \nabla_c \overline{c}^c] &=&0.\label{freefieldequations2}
\end{eqnarray}
Now we define $L^b_a(m^2)$ as follows 
\begin{equation}
 L^b_a (m^2) = \nabla^b \nabla_a - \delta^b_a \Box - 2 \beta^{-1}   \nabla_a \nabla^b  - m^2 \delta^b_a,
\end{equation}
where  
\begin{equation}
 \beta = \frac{1}{k-1}.
\end{equation}
Free field equations, Eqs. $(\ref{freefieldequations1})$ and  $(\ref{freefieldequations2})$, in de~Sitter spacetime  can be written as:
\begin{eqnarray}
 L^b_a(-6) c_b &=&0,\\
 L^b_a(-6) \overline{c}_b &=&0,
\end{eqnarray}
where 
\begin{equation}
 L^b_a(-6) = \nabla^b \nabla_a - \delta^b_a \Box - 2 \beta^{-1}   \nabla_a \nabla^b  + 6 \delta^b_a. 
\end{equation}
Here we have used Eq. $(\ref{jkla})$. 
Here again we choose the Euclidean vacuum $\cite{{7ch31}}$ for calculating the time-ordered product of fields. 
Let $|0\rangle$ be the  Euclidean vacuum state, then Feynman propagator for the ghost fields would be given by
\begin{equation}
 \langle 0| T[c_{b} (x) \overline{c}_{c'} (x')]|0\rangle =  i G_{bc'}(x,x'),
\end{equation}
where $G_{bc'}$ would  satisfy 
\begin{equation}
 L^b_a (-6) G_{bc'} (x,x')= - g_{ac'} \delta(x,x').\label{chq}
\end{equation}
Now as in the Euclidean vacuum state the Feynman propagator  can be obtained from  Green's function on the four sphere by analytical
 continuation $\cite{aj}$, so to find the Feynman propagator we shall  first find  Green's function on $S^4$.

On $S^4$ any smooth vector field  can be expressed as a linear combination of  vector spherical harmonics  $A^{L \sigma}_a$ and the gradient of
 the scalar spherical harmonics $\nabla_a Y^{L\sigma}$ (see Sec. $5$ of Ref. $\cite{a}$). 
The vector spherical harmonics satisfy $\cite{sph}$, 
\begin{eqnarray}
 -\Box A^{L \sigma}_a &=& [L(L+3)-1] A^{L \sigma}_a, \nonumber \\
\nabla^a A^{L \sigma}_a &=& 0,  \label{vectormodeabc1a}
\end{eqnarray}
where $L = 1,2,3,4,5 \cdots$ and they  are normalized as
\begin{equation}
 \int d^4x \sqrt{g} A^{L\sigma}A^{L'\sigma'} = \delta^{LL'}\delta^{\sigma \sigma'}. \label{vectorl}
\end{equation}
Here all the quantum numbers on a three-sphere $S^3$ are denoted by $\sigma$.
The degeneracies for these scalar spherical harmonics  $d_s$ and the  vector spherical harmonics  $d_v$ are given by  $\cite{a}$
\begin{eqnarray}
 d_s &=& \frac{1}{6} (L+1)(L+2)(2L+3), \nonumber \\
d_v &=& \frac{1}{2}L(L+3)(2L+3).
\end{eqnarray}
The scalar spherical harmonics satisfy Eq. $(\ref{701})$ and their gradient covectors are normalized as 
$ W^{L\sigma}_a = \nabla_a Y^{L\sigma}/\sqrt{ L(L+3)}$ by using Eq. $(\ref{701})$ and Eq. $(\ref{701b})$ because 
\begin{eqnarray}
\int d^4 x \sqrt{g} g^{ab} W^{L\sigma}_a W^{*L'\sigma'}_{b} &=&\frac{1}{L(L+3)} \int d^4 x \sqrt{g} g^{ab}
 \nabla_a Y^{L\sigma}\nabla_b Y^{*L'\sigma'} \nonumber \\&=&   - \frac{1}{L(L+3)} \int d^4 x \sqrt{g} g^{ab} 
( \Box Y^{L\sigma} ) Y^{*L'\sigma'}  \nonumber \\  &=&   \delta^{LL'}\delta^{\sigma \sigma'}. \label{vectorl1}
\end{eqnarray}
Now on $S^4$, we have (see Sec. $5$ of Ref. $\cite{a}$)
\begin{eqnarray}
  g_{ac'}\delta^4(x,x') &=& \delta_{ac'}^{(V)4}(x,x') + \delta_{ac'}^{(S)4}(x,x'),
\end{eqnarray}
where 
\begin{eqnarray}
  \delta_{ac'}^{(V)4}(x,x')&=& \sum_{L = 1}^{\infty} \sum_{\sigma} A^{L \sigma}_{a}(x) A^{*L \sigma}_{c'}(x'),
\\
 \delta_{ac'}^{(S) 4}(x,x') &=& \sum_{L = 1}^{\infty}\sum_{\sigma}  W^{L\sigma}_a(x) W^{*L'\sigma'}_{c'}(x'). \label{s1}
\end{eqnarray}
We can show that this is a valid expression for delta-function by first  expanding any vector function $f(x')_{c'}$ on $S^4$ in 
terms  of $A^{L \sigma }_{c'}(x')$ and $ W^{L\sigma}_{c'}(x')$,  then using Eq. $(\ref{vectorl})$ and Eq. $(\ref{vectorl1})$. That is,  following 
 a similar line of argument as was used in the scalar delta-function case, we can show that
\begin{equation}
 \int d^4x'\sqrt{g(x')}f^{c'} (x')g_{ac'}\delta^4(x,x') = f_a(x).
\end{equation}
We also have  (see Sec. $5$ of Ref. $\cite{a}$)
\begin{equation}
  G_{ac'}(x,x') = G_{ac'}^V(x,x') + G_{ac'}^S(x,x'),
\end{equation}
where 
\begin{eqnarray}
  G_{ac'}^V(x,x')&=&  \sum_{L = 1}^{\infty} \sum_{\sigma} k_1A^{L \sigma}_{a}(x) A^{*L \sigma}_{c'}(x'),
\\
 G_{ac'}^S(x,x') &=&  \sum_{L = 1}^{\infty} \sum_{\sigma} k_2  W^{L\sigma}_a (x)W^{*L'\sigma'}_{c'}(x'). \label{s2}
\end{eqnarray}
Here $k_1$ and $k_2$ are $L$-dependent constants which can be determined from the following equations: 
\begin{eqnarray} 
 L^a_b(-6) G^V_{ac'} (x,x')&=& - \delta^{(V)4}_{bc'} (x,x'),\label{zero123} \\ 
 L^a_b(-6) G^S_{ac'} (x,x')&=& - \delta^{(S)4}_{bc'} (x,x'). \label{zero}
\end{eqnarray}
Thus we get,  
\begin{eqnarray}
k_1 &=& \frac{1}{(L+1)(L+2)-6}, \\
  k_2 &=& -\frac{1}{6}\left[  1 -  \frac{L(L+3)}{L(L+3) + 3 \beta}\right].
\end{eqnarray}
So we have,  
\begin{eqnarray}
 G_{ac'}^V(x,x') &=&  \sum_{L = 1}^{\infty}\sum_{\sigma} \frac{A^{L \sigma}_{a}(x) A^{*L \sigma}_{c'}(x')}{(L+1)(L+2)-6}, \label{irv}\\
  G_{ac'}^S (x,x') &=& - \frac{1}{6} \sum_{L = 1}^{\infty} \sum_{\sigma} -\nabla_a\nabla_{c'}\left[  \frac{ Y^{L\sigma}(x) Y^{*L\sigma}(x')}{L(L+3)} \right.
  \nonumber \\ && \left.-  \frac{Y^{L\sigma}(x) Y^{*L\sigma}(x')}{L(L+3) + 3 \beta}\right] \label{irs}.
\end{eqnarray}

  The right-hand side of Eq. $(\ref{irv})$ is infrared divergent and this infrared divergence is 
caused by the contribution coming from the $L=1$ modes because  $(L+1)(L+2)-6 =0$ for $L=1$. 
  But if we regulate  Eq. $(\ref{zero123})$ by changing $m^2 =-6$ to $m^2 = -6 + \rho^2$ thus adding a small mass $\rho^2$ to $m^2$, then we get
\begin{equation}
 L^a_b ( -6 + \rho^2 ) G^{V(\rho^2)}_{ac'} (x,x')=- \delta^{(V)4}_{bc'} (x,x'). \label{killingmodeir}
\end{equation}
By repeating the above procedure  we find that the regulated propagator  $G^{V(\rho^2)}_{ac'}$ is given by
\begin{equation}
 G_{ac'}^{V(\rho^2)}(x,x') =  \sum_{L = 1}^{\infty} \sum_{\sigma} \frac{A^{L \sigma}_{a}(x) A^{*L \sigma}_{c'}(x')}{(L+1)(L+2)-6 + \rho^2}.
\label{zeromodedivergence}
\end{equation}
In the zero $\rho^2$ limit, $G^{V(\rho^2)}_{ac'}(x,x') \to G^{V}_{ac'}(x,x')$ and  $G^{V(\rho^2)}_{ac'}(x,x')$ thus
 diverges in the zero $\rho^2$ limit. Furthermore, 
we have seen that this divergence is caused by the $L=1$ modes. So we can write Eq. $(\ref{zeromodedivergence})$ as, 
\begin{equation}
  G_{ac'}^{V(\rho^2)}(x,x') = \sum_{\sigma} \frac{A^{1 \sigma}_{a}(x) A^{*1 \sigma}_{c'}(x')}{  \rho^2} +  
\sum_{L = 2}^{\infty} \sum_{\sigma} \frac{A^{L \sigma}_{a}(x) A^{*L \sigma}_{c'}(x')}{(L+1)(L+2)-6 + \rho^2 }. \label{divegrende}
\end{equation}
Let $Q^{\rho^2}_{ac'}$ be the solution to (see Sec. $3$ of Ref. $\cite{aj}$), 
\begin{equation}
 L^a_b(-6 + \rho^2) Q^{\rho^2}_{ac'} (x,x')=- g_{bc'} \delta^4 (x,x'). \label{vectorproppa}
\end{equation}
Then by comparing Eq. $(\ref{killingmodeir})$ with Eq.  $(\ref{vectorproppa})$, we can write the 
solution to Eq. $(\ref{killingmodeir}) $ as  (see Sec. $5$ of Ref. $\cite{a}$), 
\begin{equation}
 G^{V(\rho^2)}_{ac'} (x, x')= Q^{\rho^2}_{ac'} (x,x') + \frac{1}{6 + \rho^2} \nabla_a \nabla_{c'} D_0^{\rm eff} (x,x'). \label{gv}
\end{equation}
Here we have used Eq. $(\ref{s1})$ and the fact that  $L^a_b(-6 + \rho^2)$ is invertible. Here $D_0^{\rm eff} (x,x')$ is the 
effective propagator given by Eq. $(\ref{ymp})$, as there is no contribution coming from the 
zero modes for $D_0 (x,x')$ due to the action of derivatives on it. 

In   Eq. $(\ref{irs})$ there are no zero modes because of  the action of derivative on the scalar spherical harmonics. 
 So the only IR divergence in Eq. $(\ref{irs})$ are caused if we choose   $\beta = -n(n+3)/3$, where $(n =1,2,3, \cdots)$. 
The covariant graviton propagator is IR divergent for the same values of $\beta$ $\cite{7cha4}$. The graviton propagator used by Antoniadis and Mottola
 $\cite{7cha5}$ corresponds to the $n=1$ case and is IR divergent as a result. If we avoid these value of $\beta$ there will be no IR divergences left in 
Eq. $(\ref{irs})$.  

Now because of Eq. $(\ref{irs})$  we can write the scalar  part of the ghost propagator as
\begin{equation}
 G^{S}_{ac'} (x, x')= -\frac{1}{6} \nabla_a \nabla_{c'}[ D^{\rm eff}_0 (x,x') - D_{3\beta} (x,x')]. \label{efpr}
\end{equation}
There is no contribution from the $L=0$ mode in Eq. $(\ref{irs})$ because of the action of the derivative. So we have used in Eq. $(\ref{efpr})$,
 the same effective propagator that was obtained in  Eq. $(\ref{ymp})$ for the Yang-Mills case. 
 
\section{Effective Propagator}
We have seen in the previous section that the vector part of the ghost propagator is
 infrared divergent due to the $L=1$ modes. We shall  now argue that the $L=1$ modes do not contribute to the calculations of 
 the time-ordered product of the ghost fields. To see that we note that  the $L =1$ modes are the Killing vectors  on $S^4$ $\cite{7ch6aa}$.
 This can be seen as follows. 
Let a vector $f_a$ satisfy the Killing equation on $S^4$, 
\begin{equation}
 \nabla_a f_b + \nabla_b f_a =0. \label{kiabd}
\end{equation}
 Taking the trace of Eq. $(\ref{kiabd})$, we get
\begin{equation}
 \nabla^a f_a =0.
\end{equation}
So $f_a$ is a divergencesless vector. Now taking the divergence of Eq. $(\ref{kiabd})$, we find 
\begin{equation}
 \left(-\Box -3\right)f_a =0. \label{kiabdpq}
\end{equation}
Which is the equation for $L=1$ mode. Conversely, suppose $f_a$ satisfies Eq.~$(\ref{kiabdpq})$, then 
\begin{eqnarray}
&& \int d^4x \sqrt{g}  (\nabla_a f_b + \nabla_b f_a )( \nabla^a f^b + \nabla^b f^a )\nonumber \\ &=&
-2 \int d^4x \sqrt{g}   f_a \nabla_b( \nabla^a f^b + \nabla^b f^a ) \nonumber \\ 
&=&2 \int d^4x \sqrt{g}   f^a (-\Box -3)f_a
\nonumber \\ &=&0. \label{kiabd1}
\end{eqnarray}
We can conclude from this equation that 
$\nabla_a f_b + \nabla_b f_a =0$. Thus,  $(-\Box -3) f_a =0$ implies $\nabla_a f_b + \nabla_b f_a =0$. In other words, each $L=1$ mode is a 
Killing vector and vice versa. 

The interaction part of the ghost Lagrangian can  be written as:
\begin{equation}
 \mathcal{L} = -i \nabla^a \overline{c}^b [ \pounds_c h_{ab} - k g_{ab}g^{cd}\pounds_c h_{cd}].\label{km}
\end{equation}
Since the coupling term of the  anti-ghosts  to the metric perturbation $h_{ab}$ in Eq. $(\ref{km})$ is proportional 
to $\nabla^a \overline{c}^b +\nabla^b \overline{c}^a $, and since the $L=1$ modes are the Killing vectors, the $L=1$ modes
 do not contribute to the loop diagrams. Now as ghosts and anti-ghosts are unphysical fields they only occur in loop diagrams and thus there is no 
contribution from the $L=1$ modes in any perturbative calculation.

For this reason, as in the Yang-Mills case, we  propose to define  the  vector part of the effective ghost 
propagator by first subtracting out the contributions coming from $L=1$ modes from the vector part of the 
ghost propagator and then taking the $\rho^2 \to 0$ limit. 

  We denote the contribution coming from the $L=1$ modes by $Q^{k\rho^2}_{ac'}(x,x')$: 
\begin{equation}
 Q^{k\rho^2}_{ac'}(x,x')= \sum_{\sigma} \frac{A^{1 \sigma}_{a}(x) A^{*1 \sigma}_{c'}(x')}{  \rho^2}. \label{vecpart}
\end{equation}
Now from  Eqs. $(\ref{divegrende})$ and $(\ref{vecpart})$, we have 
\begin{equation}
 \lim_{\rho^2 \to 0}\left[ G^{V(\rho^2)}_{ac'} (x, x') -  Q^{k\rho^2}_{ac'}(x,x')\right] = 
 \sum_{L = 2}^{\infty} \sum_{\sigma} \frac{A^{L \sigma}_{a}(x) A^{*L \sigma}_{c'}(x')}{(L+1)(L+2)-6 }.
\end{equation}
Clearly this is convergent and has no infrared divergence. As the contribution  $  \nabla_a \nabla_{c'} {D_0^{\rm eff}(x,x')/(6 + \rho^2)}$ had no 
infrared divergence in the $\rho^2 \to 0$ limit, the infrared divergences in $G^{V(\rho^2)}_{ac'}(x,x')$ come from $Q^{\rho^2}_{ac'}(x,x')$. 
So we define the effective propagator $ Q^{\rm eff}_{ac'}(x,x')$ as follows:
\begin{equation}
 Q^{\rm eff}_{ac'}(x,x') =\lim_{\rho^2 \to 0} \left[Q^{\rho^2}_{ac'}(x,x') - Q^{k\rho^2}_{ac'}(x,x')\right]. \label{effectiveprop}
\end{equation}
Now the total effective ghost propagator will be given by 
\begin{eqnarray}
G^{\rm eff}_{ac'}(x,x') &=& \lim_{\rho^2 \to 0}[ G^{V(\rho^2)}_{ac'} (x, x') + G^{S}_{ac'} (x, x') -  Q^{k\rho^2}_{ac'}(x,x')] \nonumber \\ 
&=& \lim_{\rho^2 \to 0} \left[Q^{\rho^2}_{ac'}(x,x') - Q^{k\rho^2}_{ac'}(x,x')\right] \nonumber\\ &&
+ \lim_{\rho^2 \to 0} \frac{1}{6 + \rho^2 }\nabla_a \nabla_{c'} D^{\rm eff}_0(x,x')\nonumber\\ && 
- \frac{1}{6}\nabla_a \nabla_{c'}\left[ D^{\rm eff}_0 (x,x') - D_{3\beta}(x,x')\right]
 \nonumber\\
 &=&Q^{\rm eff}_{ac'}(x,x') +  \frac{1}{6} \nabla_a \nabla_{c'} D_{3\beta}(x,x').
\end{eqnarray}
This propagator can be used to do  calculations in perturbative quantum gravity. 
However just like in the Yang-Mills case the consistency of the perturbation theory is not obvious as it is not clear how the subtraction of the Killing 
modes will affect the BRST symmetry of the theory. 
 \section{Explicit Expression for the Vector Part of the Effective Propagator}
To find the effective ghost  propagator in terms of the variable $z$ we can proceed as follows:
The solution to Eq. $(\ref{vectorproppa})$,  in terms of $z$, is given by (see Sec. $3$ of Ref. $\cite{aj}$)
\begin{equation}
 Q^{\rho^2}_{ac'}(z) = \alpha^{V(\rho^2)} (z) g_{ac'} + \beta^{V(\rho^2)} (z)n_a n_{c'}, \label{651}
\end{equation}
where
\begin{eqnarray}
  \alpha^{V(\rho^2)} &=& \left[ \frac{-2z(1-z)}{3}\frac{d}{dz} + 2z -1\right ] \gamma^{\rho^2}(z),\label{vec1}
\\
 \beta^{V(\rho^2)} &=& \alpha^{V(\rho^2)} - \gamma^{\rho^2}(z).\label{vec2}
\end{eqnarray}
Here $\gamma^{\rho^2}(z)$ is given by
\begin{equation}
 \gamma^{\rho^2}(z) = -\frac{3 \Gamma(b_+)\Gamma(b_-)}{64\pi^2 m^2}F[b_+, b_- ; 3 ; z],\label{gamm}
\end{equation}
where
\begin{eqnarray}
b_+ &=& \frac{5}{2} +\sqrt{\frac{1}{4} +6 - \rho^2 },
\\
b_- &=& \frac{5}{2} -\sqrt{\frac{1}{4} +6 - \rho^2 }.
\end{eqnarray}
It may be noted that we have changed the mass from $m^2 =-6$ to $m^2 =-6 + \rho^2$. 
The propagator given by Eq. $(\ref{651})$ clearly diverges as $\rho^2 \to 0$ because then $b_- \to 0$ and so $\Gamma (b_-) \to \infty$. 

The contribution from the  Killing vectors  in the Allen-Jacobson formalism is  given by  (see Sec. $5$ of Ref. $\cite{a}$)
\begin{equation}
 Q^{k}_{ac'} (z) = \frac{15}{16\pi^2\rho^2}[(2z-1)g_{aa'} + 2(z-1)n_a n_{a'}].
\end{equation}
This can be obtained by   substituting  
\begin{equation}
 \gamma^k =  \frac{15}{16\pi^2 \rho^2}, 
\end{equation}
into Eqs. $(\ref{vec1})$ and  $(\ref{vec2})$.
Just like the Yang-Mills case, we first verify  that $\gamma^k$ subtracted from the constant part of $\gamma^{\rho^2}(z)$ given in
 Eq. $(\ref{gamm})$ is a finite constant in the zero $\rho^2$ limit. 
Let $\gamma_c$ be the constant part of $\gamma^{\rho^2}(z)$ given in Eq. $(\ref{gamm})$, then we have 
\begin{equation}
 \gamma_c = \frac{3\Gamma\left(\frac{5}{2} + q \right)\Gamma\left(\frac{5}{2} - q \right)}{64\pi^2(6 - \rho^2) }, 
\end{equation}
where $q$ is given by 
\begin{equation}
q = \frac{\sqrt{25 - 4 \rho^2 }}{2}.
\end{equation}
We subtract the $\gamma^k$ from $\gamma_c$:
\begin{equation}
 \gamma_c -\gamma^k = \frac{3\Gamma\left(\frac{5}{2} + q \right)\Gamma\left(\frac{5}{2} - q \right)}{64\pi^2(6 - \rho^2) }- \frac{15}{16\pi^2 \rho^2}. 
\end{equation}
 We shall  verify that $\gamma_c -\gamma^k$  has a finite zero $\rho^2$ limit. 
Now we have 
\begin{equation}
 q -\frac{5}{2}  = - \frac{\rho^2}{5} + \mathcal{O} (\rho^4).
\end{equation}
We also have 
\begin{eqnarray}
  \Gamma\left(\frac{5}{2} + q \right)\Gamma\left(\frac{5}{2} - q\right) &=&
 \left[ \left(\frac{5}{2} + q \right)\left(\frac{5}{2} - q \right)\left(\frac{3}{2} + q \right)\left(\frac{3}{2} - q \right)\right] 
\nonumber \\ &&\times \left[ \Gamma\left(\frac{1}{2} + q \right)\Gamma\left(\frac{1}{2} - q \right)\right].
\end{eqnarray}
We can use the following approximation 
\begin{eqnarray}
  \Gamma\left(\frac{1}{2} + q \right)\Gamma\left(\frac{1}{2} -q \right) =  \frac{\pi}{\sin\pi \left( \frac{1}{2} - q\right)} &=&
\frac{\pi}{\sin\pi \left( \frac{5}{2} - q \right)}\nonumber\\ &=&
\frac{1}{ \left( \frac{5}{2} - q \right)} +  \mathcal{O} (\rho^2).
\end{eqnarray}
Here we have used 
\begin{equation}
 \Gamma(z)\Gamma(1-z) = \frac{\pi}{\sin (\pi z )}.
\end{equation}
Thus we find
\begin{equation}
 \gamma_c =  \frac{1}{64\pi^2 \rho^2} \left[ 60 -\frac{87}{5} \rho^2 + \mathcal{O}(\rho^4)\right].
\end{equation}
Hence we obtain  
\begin{eqnarray}
   \lim_{\rho^2 \to 0}[\gamma_c - \gamma^k]   
 &=&\lim_{\rho^2 \to 0} \frac{1}{64\pi^2}\left[ \frac{1}{\rho^2} \left[60 -\frac{87}{5} \rho^2  + \mathcal{O}(\rho^4) \right] - \frac{60}{\rho^2}\right]\nonumber \\
&=&- \frac{87}{320 \pi^2 }.
\end{eqnarray}
The exact value of this constant  is not important because we are free to add any finite arbitrary 
  constant to $\gamma^{\rho^2}(z)$, as the addition of any finite constant will only generate Killing contributions 
which will not contribute to the perturbative calculations. The important point to note here is that $\gamma^{\rho^2}(z) - \gamma^k$ 
is finite and so there are no infrared divergences.  

Now we can calculate the vector part of the effective ghost propagator  by first defining $\gamma'(z)$ as, 
\begin{equation}
 \gamma'(z) = \lim_{\rho^2 \to 0} \frac{d}{dz}\left[ \gamma^{\rho^2}(z)\right], \label{abcde1234}
\end{equation}
and then defining the effective  $\gamma^{\rm eff}$ by
\begin{equation}
 \gamma^{\rm eff} (z) = \int_0^z dz \gamma'(z) + \frac{C}{64\pi^2}, 
\end{equation}
where $C$ is a  arbitrary finite constant. 
Now as 
\begin{equation}
 \frac{d}{dz} F[a, b ; c ; z] = \frac{ab}{c}  F[a +1, b +1; c +1; z], 
\end{equation}
so Eq. $(\ref{abcde1234})$ now becomes 
\begin{equation}
 \gamma'(z) =\frac{1}{64\pi^2} \frac{ \Gamma(6)\Gamma(1)}{6}F[6, 1 ; 4 ; z].
\end{equation}
With $C=11$,   $ \gamma^{\rm eff}(z)$ in terms of elementary functions is given by 
\begin{equation}
 \gamma^{\rm eff}(z) = \frac{1}{64\pi^2}\left[\frac{1}{(1-z)^2}+ \frac{6}{(1-z)}-12 \log(1-z) + 4 \right].
\end{equation}
The advantage of this choice of $C$ is that it eliminates  terms proportional to $z$
 in $ \alpha^{V\rm(eff)} (z)$ and $ \beta^{V\rm(eff)} (z)$, 
thus simplifying the calculations. 
After substituting this value of $\gamma^{\rm eff}$ into Eqs. $(\ref{vec1})$ and $(\ref{vec2})$, we get 
\begin{equation}
 Q^{\rm eff}_{ac'} (z) = \alpha^{V\rm (eff)} (z) g_{ac'} + \beta^{V\rm (eff)} (z)n_a n_{c'},\label{7}
\end{equation}
 where 
\begin{eqnarray}
 \alpha^{V\rm(eff)} (z) &=& \frac{1}{64\pi^2}\left[-\frac{1}{3(1-z)^2}-12(2z-1) \log(1-z)\right. \nonumber \\ 
 && \left.  -12 +\frac{4}{3(1-z)}  \right],
\\
 \beta^{V\rm(eff)} (z) &=& \frac{1}{64\pi^2}\left[-\frac{4}{3(1-z)^2}  -24(z-1) \log(1-z) \right. \nonumber \\ 
&& \left.  -16 -\frac{14}{3(1-z)} \right].
\end{eqnarray}
\section{Explicit Expression for  the Total Effective Propagator}
To calculate the total effective propagator we have to add the contribution coming from the scalar part to the effective vector part of the ghost propagator. 
The scalar part of the ghost propagator cannot be written in terms of elementary functions for a general value of $\beta$.
 However the corresponding scalar field equation for $\beta = 2/3$ is the conformally-coupled massless scalar field equation. 
(An equation of the form $[\Box - R/6] \phi =0$ is called the conformally-coupled massless scalar field equation as it is 
invariant under conformal transformations).  
For this value of $\beta$ the mass takes the value $m^2 =2$ and the scalar propagator takes a simple form $\cite{ch7ab}$:
\begin{equation}
 D_2 (z) = \frac{1}{16\pi^2}\frac{1}{1-z}.\label{9}
\end{equation}

Now  (see Sec. $1$ of Ref. $\cite{aj}$), we have 
\begin{equation}
 \nabla_{c'} f(z)  = - \sqrt{z(1-z)} \frac{d f(z) }{dz}  n_{c'},
\end{equation}
and
\begin{equation}
 \nabla_a n_{c'} = - \frac{1}{2 \sqrt{z(1-z)}}(g_{ac'} +n_a n_{c'}).
\end{equation}
So we obtain the following result,
\begin{equation}
 \nabla_a \nabla_{c'} f(z) = \frac{1}{2} \frac{df(z) }{dz} g_{ac'} + (1-z)\frac{d}{dz}\left(z\frac{d f(z)}{dz}\right) n_a n_{c'}.\label{8}
\end{equation}
Now from  Eqs. $(\ref{9})$ and $(\ref{8})$ we get
\begin{equation}
 \frac{1}{6} \nabla_a \nabla_{c'} D_{2}(z) = \alpha^{S} (z) g_{ac'} + \beta^{S} (z) n_a n_{c'},\label{6}
\end{equation}
where
\begin{eqnarray}
 \alpha^{S} (z) &=& \frac{1}{64\pi^2}\left[\frac{1}{3(1-z)^2} \right],
\\
 \beta^{S} (z) &=& \frac{1}{64\pi^2}\left[-\frac{2}{3(1-z)}+\frac{4}{3(1-z)^2}  \right].
\end{eqnarray}

So the full effective ghost propagator will be given by the sum of the effective vector contributions given by Eq. $(\ref{6})$ and the scalar contributions
 given by Eq. $(\ref{7})$ 
\begin{eqnarray}
 G^{\rm eff}_{ac'} (z)&=& (\alpha^{V\rm(eff)} +\alpha^{S}) g_{ac'} + (\beta^{V \rm(eff)} +\beta^{S}) n_a n_{c'}\nonumber 
\\ &=& \alpha^{\rm eff}(z) g_{ac'} + \beta^{\rm eff}(z) n_a n_{c'},
\end{eqnarray}
 where
\begin{eqnarray}
  \alpha^{\rm eff}(z) &=&\frac{1}{16\pi^2}\left[ \frac{1}{3(1-z)} -3-3(2z-1)\log(1-z)\right],
\\
 \beta^{\rm eff}(z) &=& \frac{1}{16\pi^2}\left[ -\frac{4}{3(1-z)}-4+6(1-z)\log(1-z)\right].
\end{eqnarray}
This is the  effective ghost propagator for perturbative quantum gravity with $\beta = 2/3$ in de~Sitter spacetime, which can be used to do perturbative calculations .

\chapter{Symplectic Product and Wightman Two-Point Function}
In this chapter we shall  first discuss canonical quantization of free scalar field theory and then generalize these 
results to higher spin fields. We  shall first construct the Wightman two-function for free scalar field theory in de~Sitter spacetime
 by canonical quantization. Then we shall  formally see how we can use this method to construct the Wightman two-function for higher spin fields. 
From now on two-point function will mean the Wightman two-function.
\section{Symplectic Product for Scalar Field Theory }
We shall  review the scalar field theory in  de~Sitter spacetime in this section $\cite{912}$. 
However we shall proceed in a  way slightly different from  $\cite{912}$, so that our procedure is easily generalized to higher  spin fields.
  The Lagrangian for a  minimally-coupled real massive scalar field theory is given by  
\begin{equation}
 \mathcal{L} =\frac{1}{2}\left[ - \nabla_a\phi \nabla^a \phi - m^2 \phi^2\right],
\end{equation}
 and the classical equation of motion is given by
\begin{equation}
 (\Box- m^2) \phi (x) = 0.\label{sysysy}
\end{equation}
Now we can define  a quantity we call the momentum current $\pi^c$ as follows:
 \begin{equation}
  \pi^c = \sqrt{-g}\frac{\partial \mathcal{L}}{\partial \nabla_c \phi} = -\sqrt{-g} \nabla^c \phi.
 \end{equation}
Using Eqs. $(\ref{sysysy})$, we get 
\begin{equation}
 \nabla_c \pi^c  = -\sqrt{-g} m^2 \phi (x), \label{mn1}
\end{equation}
where 
\begin{equation}
 \nabla_c \pi^c  = \sqrt{-g} \nabla_c [ (-g)^{-1/2} \pi^c].
\end{equation}

If $\phi , \phi'$ are two solutions of the field equations,  and $\pi^c , \pi'^c$ the  momentum currents conjugate to them, 
then we define a current $J^c$ as follows:
 \begin{equation}
  J^c = - \frac{i}{\sqrt{-g}} [\phi^* \pi'^c - \phi' \pi^{*c}].\label{b123}
 \end{equation}
If $d\Sigma_a$ is a surface element of the   spacelike hypersurface $\Sigma$, then the  symplectic product on this hypersurface can  be defined as follows:
 \begin{equation}
  (\phi, \phi') = \int_\Sigma d\Sigma_c J^c. \label{sympcprod}
 \end{equation}
Using Eq. $(\ref{mn1})$, we can show that the current $J^c$ is conserved,
 \begin{eqnarray}
\nabla_c J^c   &=& - \frac{i}{\sqrt{-g}} \nabla_c [\phi^* \pi'^c - \phi' \pi^{*c}] 
= i\nabla_c [\phi^* \nabla^c \phi' - \phi' \nabla^c \phi^*  ] \nonumber\\ &=&
 i[ (\nabla_c\phi^{*}.\nabla^c \phi' - \nabla_c\phi^{*} .\nabla^c\phi') + m^2 (\phi^* \phi' - \phi^* \phi' )] =  0.\label{eqm1}
 \end{eqnarray}
Now if we consider a spacetime region with volume $V$, bounded by a future spacelike hypersurface $\Sigma^{+}$ and a past spacelike hypersurface $\Sigma^{-}$, 
then by Gauss theorem we have
\begin{eqnarray}
 && \int_V d^4x \sqrt{-g} \nabla_c J^c  \nonumber \\  &=&\int_{\Sigma^{+}} d\Sigma_c J^c  - \int_{\Sigma^{-}} d\Sigma_c J^c \nonumber \\ &=& 0,
\end{eqnarray}
and so we have
\begin{equation}
 \int_{\Sigma^{+}} d\Sigma_c J^c  = \int_{\Sigma^{-}} d\Sigma_c J^c.
\end{equation}

Let us now consider  de~Sitter metric which was given by Eq. $(\ref{metricabc2})$:
\begin{equation}
ds^2 = -dt^2 + \cosh^2 t [d\psi^2 + \sin^2 \psi (d\theta^2 + \sin^2 \theta d \phi^2 )].
\end{equation}
If we define $n_c = (1, 0)$ to be the past pointing unit normal to the constant-time spacelike hypersurface $\Sigma$ and $\gamma_{ij}$ to be the metric on it,
then we have
\begin{equation}
 n_c n^c = g^{ab} n_a n_b  = -1,
\end{equation}
and
\begin{eqnarray}
 d\Sigma_a J^a &=& d^3 {\bf{x}}\sqrt{\gamma} n_a J^a \nonumber \\ &=& d^3 {\bf{x}}\sqrt{\gamma} \alpha J^0 \nonumber \\ &=& d^3 {\bf{x}} \sqrt{-g} J^0.\label{d3x}
\end{eqnarray}
So the symplectic product given in Eq. $(\ref{sympcprod})$ can now be written as: 
\begin{equation}
 (\phi, \phi') = -i \int d^3 {\bf{x}} [\phi^* ({\bf{x}}, t)\pi'^0 ({\bf{x}}, t)- \phi'({\bf{x}}, t) \pi^{*0}({\bf{x}}, t)]. \label{hermihermi}
\end{equation}

\section{Fock Space}
Let $\{\phi_n\}$ and $\{\phi^*_n\}$ form a complete set of solutions to the field equation $(\ref{sysysy})$, and suppose 
\begin{eqnarray}
 (\phi_n,\phi_m) &=& M_{nm}, \label{conditions1}\\ 
(\phi_n,\phi^*_m) &=& 0, \label{conditions2}\\ 
(\phi^*_n,\phi^*_m) &=& - M_{nm}.\label{conditions3}
\end{eqnarray}
The condition given in Eq. $(\ref{conditions2})$ does not hold in general and so this is a requirement on the 
complete set of solutions to the field equation, Eq. $(\ref{sysysy})$. 
We also chose  $M_{nm}$ to have positive eigenvalues only. This again is not always
 true and so this is again a requirement on the complete set of solutions to the field equation, Eq. $(\ref{sysysy})$.
Also in general  $n, m$ can be continuous labels. However in de~Sitter spacetime these are actually discrete labels.

Now using Eq. $(\ref{hermihermi})$ the infinite dimensional matrix $M_{nm}$ can be shown to be  Hermitian: 
\begin{eqnarray} 
 M_{nm} &=& -i \int d^3 {\bf{x}} [ \phi_n^*({\bf{x}}, t)\pi^0_m({\bf{x}}, t) - \phi_m({\bf{x}}, t) \pi_n^{*0}({\bf{x}}, t)] \nonumber \\ 
&=&\left[ -i  \int d^3 {\bf{x}}  [ \phi_m^* ({\bf{x}}, t)\pi^0_n ({\bf{x}}, t)- \phi_n ({\bf{x}}, t)\pi_m^{*0}({\bf{x}}, t)] \right]^* \nonumber \\ 
&=& M^*_{mn }.\label{bov21}
\end{eqnarray}

As $\phi$ is a real field, we can now expand it as follows:
\begin{equation}
\phi = \sum_n [{\rm{a}}_n \phi_n + {\rm{a}}^*_n \phi^*_n ].
\end{equation}

In quantum field theory when $\phi$ and $\pi^c$ are promoted to  Hermitian operators $\hat{\phi}$ and $\hat{\pi}^c$, respectively, 
and  the following equal-time commutation relations are imposed, 
\begin{eqnarray}
 \left[ \hat{\phi}({\bf{x}},t ), \hat{\pi}^0 ({\bf{x}}', t)\right]  &=& i \delta ({\bf{x}},{\bf{x}}'), \nonumber \\ 
  \left[  \hat{\phi}({\bf{x}}, t), \hat{\phi} ({\bf{x}}', t)\right]  & =& 0,\nonumber \\  
\left[ \hat{\pi}^0({\bf{x}}, t), \hat{\pi}^0 ({\bf{x}}', t)\right]  &=& 0,\label{ccr3}
\end{eqnarray}
then ${\rm{a}}^*_n$ and ${\rm{a}}_n $ become operators $a^{\dagger}_n$ and $a_n$ respectively.
Here $\delta({\bf{x}},{\bf{x}}')$ is defined by 
\begin{equation}
 \int d^3 {\bf{x}} \delta({\bf{x}},{\bf{x}}') f({\bf{x}}) = f({\bf{x}}'),
\end{equation}
for any compactly supported smooth function $f({\bf{x}})$. 
So we  can now express $\hat{\phi}$ as follows: 
 \begin{equation}
  \hat{\phi} = \sum_n [a_n \phi_n + a^{\dagger}_n \phi^*_n ]. \label{bov}
 \end{equation}
For this choice of complete set of solutions to the field equation, Eq. $(\ref{sysysy})$,
 we define a state called the vacuum state $|0\rangle$, as the state  that  is annihilated by $a_n$:
 \begin{equation}
  a_n |0\rangle = 0.
 \end{equation}
It may be noted that here $a_n^{\dagger}$ and $a_n$ will not satisfy the standard commutation relations, however we still 
call them 
the creation and annihilation operators respectively, in analogy  with those for the simple quantum harmonic oscillator. 
Many particle states can be built  by repeated action of $a^{\dagger}_n$ on the vacuum state.

It may be noted that  the division between $\{\phi_n\}$ and $\{\phi^*_n\}$ is not unique even after imposing conditions given by
 Eqs. $(\ref{conditions1})$-$(\ref{conditions3})$  $\cite{912}$. Due to this non-uniqueness in division between
 $\{\phi_n\}$ and $\{\phi^*_n\}$, there is   non-uniqueness in the definition of the vacuum state also. 
This can be seen by considering  $\{\phi'_n\}$ and $\{\phi'_m\}$ as another  
 complete set of solutions to the field equation, Eq. $(\ref{sysysy})$, satisfying
 conditions given by Eqs. $(\ref{conditions1})$-$(\ref{conditions3})$. Now we have 
\begin{equation}
   \hat{\phi} = \sum_n [a'_n \phi'_n + a'^{\dagger}_n \phi'^*_n]. \label{bov1}
\end{equation}
Here the  vacuum state $|0'\rangle$ is   the state annihilated by $a'_n$,
\begin{equation}
 a'_n|0'\rangle =0.
\end{equation}
Now many particle states can be built  by repeated action of $a'^{\dagger}_n$ on $|0'\rangle$.
As $\phi_n$ and $\phi^*_n$ form a complete set of  solutions to the field equation, Eq.~$(\ref{sysysy})$, we can  express $\phi'_n$  
as a linear combination of $\phi_n$ and $\phi^*_n$, 
\begin{equation}
 \phi'_n = \sum_m [\alpha_{nm}\phi_m + \beta_{nm}\phi^*_m].\label{bovv2} 
\end{equation}
By substituting Eq. $(\ref{bovv2})$ in Eq. $(\ref{bov1})$ and comparing the resulting expression with Eq. $(\ref{bov})$, we find 
\begin{eqnarray}
 a_n &=& \sum_m [\alpha_{nm}a'_m + \beta^*_{nm}a'^{\dagger}_m],\\
a^{\dagger}_n &=& \sum_m [\alpha^*_{nm} a'^{\dagger}_m + \beta_{nm}a'_m].
\end{eqnarray}
The two Fock spaces based on these choices of complete set of  solutions to the field equation, 
Eq. $(\ref{sysysy})$, are different as long as $\beta_{nm} \neq 0$. In particular $a_n|0'\rangle$ does not vanish because  
\begin{eqnarray}
  a_n|0'\rangle &=& \sum_m [\alpha_{nm}a'_m + \beta^*_{nm}a'^{\dagger}_m]|0'\rangle  \nonumber \\
&=&\sum_m \beta^*_{nm}a'^{\dagger}_m |0'\rangle \neq 0,
\end{eqnarray}
but, 
\begin{equation}
 a_n |0\rangle =0.
\end{equation}
Thus $a_n|0'\rangle$ is  a one-particle state. In fact we have 
\begin{equation}
\langle 0'| a_n^{\dagger} a_n|0'\rangle = \sum_m \sum_k \beta_{nm}\beta^*_{nk}M_{mk}.
\end{equation}

We  shall use a de~Sitter invariant vacuum state called the Euclidean vacuum 
state and its higher-spin analogues in our analysis $\cite{{7ch31}}$.  The Euclidean vacuum is often referred to
 as the Bunch-Davies vacuum $\cite{dei2}$. An advantage of using this vacuum state is that the two-point functions  reduce to the standard 
Minkowski two-point functions  when  de~Sitter  radius is taken to infinity, after fixing the geodesic distance between two points 
$\cite{dei3}$.
 \section{Wightman Two-Point Function for Scalar Field Theory}
The Wightman two-point function  for scalar field theory is given by
\begin{equation}
 G(x, x') = \langle 0| \hat{\phi}(x)  \hat{\phi}(x')|0\rangle.
\end{equation}
This can be written as:
\begin{eqnarray}
 G(x, x') &=& \sum_{n , m}\langle 0|(a_n \phi_n (x) +  a_n^{\dagger}\phi^*_n(x)) (a_m\phi_m(x') + a_m^{\dagger}  
\phi_m(x')|0\rangle  \nonumber \\ &=&\sum_{n , m}\phi_n(x) \phi^*_m (x')C_{nm},
\end{eqnarray}
where $C_{nm}$ is the commutator,
\begin{equation}
 C_{nm} = \langle 0|[a_n, a_m^{\dagger}]|0\rangle. \label{tion}
\end{equation}
Now as $\hat{\phi}$ and $\hat{\pi}^{0}$ are Hermitian, by using the canonical commutation relations given in Eq. $(\ref{ccr3})$, we get,
\begin{eqnarray}
&&[(\phi_n, \hat{\phi}),(\hat{\phi}, \phi_m)]   \nonumber \\ &= &  -\int d^3 {\bf{x}} d^3 {\bf{x'}}   [\phi_n^* ({\bf{x}}, t)\hat{\pi}^0({\bf{x}}, t) 
 - \hat{\phi} ({\bf{x}}, t)\pi_n^{*0}({\bf{x}}, t) , 
\nonumber \\ &&  \hat{\phi}({\bf{x}}', t) \pi_m^0({\bf{x}}', t) -\phi_m({\bf{x}}', t) \hat{\pi}^{0}({\bf{x}}', t)]\nonumber  \\  
&=& - i \int d^3 {\bf{x}} d^3 {\bf{x'}} [\phi_n^* ({\bf{x}}, t)\pi_m^0 ({\bf{x}}', t)- \phi_m ({\bf{x}}', t)\pi_n^{*0}({\bf{x}}, t)] \delta({\bf{x}},{\bf{x}}')
 \nonumber \\ &=& - i \int d^3 {\bf{x}}  [\phi_n^*({\bf{x}}, t)\pi_m^0({\bf{x}}, t) - \phi_m ({\bf{x}}, t)\pi_n^{*0}({\bf{x}}, t)] 
\nonumber \\&=&  M_{nm}.\label{bov7}
\end{eqnarray} 
Now   we have 
\begin{equation}
 (\phi_n, \hat{\phi}) =  \sum_k a_k M_{nk}. \label{bov2}
\end{equation}
Using Eqs. $(\ref{bov21})$ and $(\ref{bov2})$, we also get
 \begin{equation}
 ( \hat{\phi}, \phi_m) = [(\phi_m, \hat{\phi})]^{\dagger}  =   \sum_l a^{\dagger}_l M^*_{ml}  = \sum_l a^{\dagger}_l M_{lm}. \label{bov6}
\end{equation}
Now from Eqs. $(\ref{bov7})$-$(\ref{bov6})$, we get 
\begin{equation}
 \sum_{kl} M_{nk}[a_k, a_l^{\dagger}]M_{lm} = M_{nm}.\label{tion1}
\end{equation}
Using Eqs. $(\ref{tion})$ and $(\ref{tion1})$, we get
\begin{equation}
  \sum_{kl} M_{nk}C_{kl}M_{lm} = M_{nm}.
\end{equation}
 This equation  in matrix notation is written as, 
\begin{equation}
 MCM = M.
\end{equation}
Assuming that $M_{nm}$ has only positive eigenvalues, i.e., that it is  invertible, we get
\begin{equation}
 C= M^{-1}.
\end{equation}
 Therefore the two-point function is given by
 \begin{equation}
   G(x,x') = \sum_{nm} \phi_n (x) \phi^*_m(x') M^{-1}_{nm}. \label{twopointfunctionscalar}
 \end{equation}
 \section{Wightman Two-Point Function for Tensor Fields}
In this section we shall  formally generalize what we did for scalar fields to general non-interacting higher-spin fields, described by tensor fields.
  Let us denote a tensor field by a shorthand notation, $A_{a_1 a_2 a_3 a_4...a_n} = A_I$. 
The Lagrangian $\mathcal{L}$ for this field $A_I$ will be assumed to be a scalar function of only  $A_I$ and $\nabla_c A_I$, 
as the inclusion of higher derivative terms will lead to problems like non-unitarity of the theory $\cite{unitarity}$. 
As we are considering only free field theories, we can write the most general Lagrangian for higher-spin free field theories as follows:
\begin{equation}
 \mathcal{L} = -\frac{1}{2}T^{IcJd}\nabla_c A_I \nabla_d A_J - \frac{1}{2}S^{IJ}A_I A_J,\label{classiclag}
\end{equation}
where $T^{IcJd}$ and $S^{IJ}$ do not depend on $A_I$ and satisfy 
\begin{eqnarray}
T^{Ic Jd} &=& T^{Jd Ic},\\
S^{IJ}&=& S^{JI}. 
\end{eqnarray}
The equation of motion is now given by 
\begin{equation}
 \nabla_c[ T^{IcJd} \nabla_d A_J] - S^{IJ}A_J =0. \label{higherspineq}
\end{equation}
We define a momentum current  $\pi^{cI}$ conjugate  to $A_I$ as follows:
 \begin{equation}
  \pi^{cI} = \sqrt{-g}\frac{\partial \mathcal{L}}{\partial \nabla_c A_I } = -\sqrt{-g} T^{IcJd}\nabla_d A_J.
 \end{equation}
From Eq. $(\ref{higherspineq})$, we have 
\begin{equation}
 \nabla_c \pi^{cI} =  -\sqrt{-g} S^{IJ}A_J.
\end{equation}
Now if $A_{I}$ and $A'_{I}$ are two solutions to the field equations, Eq. $(\ref{higherspineq})$,  then we define the current $J^c$ as follows:
 \begin{equation}
  J^c = - \frac{i}{\sqrt{-g}} [A_{I}^* \pi'^{cI} - A'_{I} \pi^{*cI}].\label{jijiji1}
 \end{equation}
 Now we  define a symplectic product on a spacelike hypersurface $\Sigma_c$ with surface element $d\Sigma_c $, as follows:
 \begin{equation}
  (A , A') = \int d\Sigma_c J^c.
 \end{equation}
By using  Eq. $(\ref{higherspineq})$, we can show that $J^c$ is conserved,  
\begin{eqnarray}
 \nabla_c J^c &=& - \frac{i}{\sqrt{-g}} \nabla_c[A_{I }^* \pi'^{cI} - A'_{I } \pi^{*cI}] \nonumber \\ &=&
 i \nabla_c [ A_{I}^* T^{IcJd} \nabla_d A'_{J } - A'_{I } T^{IcJd}\nabla_d A^*_{J }  ] \nonumber \\  &=& 
i[\nabla_c A^*_{I }. T^{IcJd} \nabla_d A'_{I} - \nabla_d A'_{I }. T^{IcJd} \nabla_c A^*  \nonumber \\ 
&&+ S^{IJ} (A^*_{I } A'_{J } - A'_{J } A^*_{I})] \nonumber \\ &=&  0. \label{conhspinab}
 \end{eqnarray}
Now if we again consider a spacetime region bounded by a future spacelike hypersurface $\Sigma^{+}$ and a
 past spacelike hypersurface $\Sigma^{-}$, then using Gauss theorem we can show that 
\begin{equation}
 \int_{\Sigma^{+}} d\Sigma_c J^c  = \int_{\Sigma^{-}} d\Sigma_c J^c. \label{sp12abab}
\end{equation}

Let $\{A_{In}\}$ and $\{A^*_{Im}\}$ be a  complete set of solutions to the field equation $(\ref{higherspineq})$,  and suppose 
\begin{eqnarray}
 (A_n,A_m) &=& M_{nm}, \label{condhighspin1}\\ 
(A_n,A^*_m) &=& 0,\label{condhighspin2} \\ 
(A^*_n,A^*_m) &=&- M_{nm}.\label{condhighspin3}
\end{eqnarray}
Here again  Eq. $(\ref{condhighspin2})$ is a requirement on the complete set of solutions to the field equation $(\ref{higherspineq})$. 
If the theory has gauge symmetry, $M_{nm}$ will contain zero eigenvalues and thus will not be invertible $\cite{2ch9}$. However after  a suitable 
gauge-fixing   term  is added to the classical Lagrangian given in Eq. $(\ref{classiclag})$, $M_{nm}$ will contain no zero eigenvalues $\cite{2ch10}$ and thus be 
invertible. In this thesis, we will explicitly invert $M_{nm}$ for linearized quantum gravity in de~Sitter spacetime, in the covariant gauge.  

 It can also  be shown that $M_{nm}$ is a  Hermitian matrix  by repeating the argument used in the scalar field theory case.
 Now after adding a suitable gauge-fixing   term and imposing canonical commutation  relations, we can also expand $A_{I}$ in terms of 
creation and annihilation operators  as follows: 
 \begin{equation}
  \hat{A}_I = \sum_n [a_n A_{In} + a^{\dagger}_n A^*_{In} ].
 \end{equation}
Here the vacuum state is defined by requiring it to satisfy
\begin{equation}
 a_n |0 \rangle =0.
\end{equation}
It may be noted that, just like in the scalar field theory case, there is still an ambiguity in the division
 between $\{A_{In}\}$ and $\{A^*_{I n}\}$, and thus there is also an ambiguity in the definition of the vacuum state $\cite{912}$.
The Wightman two-point function is now given by 
\begin{eqnarray}
 G_{II'} (x,x') &=& \langle 0|  \hat{A}_I(x)  \hat{A}_{I'}(x')|0\rangle \nonumber \\&=& \sum_{mn}A_{In}(x)A_{I' m}(x')\langle 0| [a_n, a_m^{\dagger}]|0\rangle.
\end{eqnarray}
As after choosing a suitable gauge-fixing  term $M_{nm}$ becomes invertible, thus following  a similar line of argument as was used in 
the scalar field theory case, we can show that   
\begin{equation}
 M^{-1}_{nm} = \langle 0| [a_n, a_m^{\dagger}]|0\rangle. 
\end{equation}
Therefore we can now write the two-point function as follows: 
 \begin{equation}
   G(x,x')_{II'} = \sum_{nm} A_{In}(x)A_{I'm}(x') M^{-1}_{nm}. \label{wighthig}
 \end{equation}
 
It will also be useful to define a Klein-Gordon type product  as follows:
\begin{equation}
  \langle A_{1}, A_{2} \\  \rangle =  -i \int  d\Sigma_a [A_{I1}^* \nabla^a A_2^I - A_{I2} \nabla^a A ^{I*}_1]. \label{kl}
\end{equation}
In the above mentioned scalar field theory case  this Klein-Gordon type  product is equal to the symplectic product given in  
 Eq. $(\ref{sympcprod})$, but this is not always the case. For example, for a scalar field theory given by the following Lagrangian, 
\begin{equation}
 \mathcal{L} = \frac{1}{2\mu} [- \nabla_a\phi \nabla^a \phi - m^2 \phi^2], 
\end{equation}
we have 
\begin{equation}
 (\phi_1 , \phi_2 ) = \mu \langle \phi_1 , \phi_2  \rangle. 
\end{equation}
\chapter{Spin-2 Field in de~Sitter Spacetime}
In the previous chapter we derived a formal expression for the Wightman two-point function of tensor fields. In this chapter we
shall review certain properties  of linearized quantum gravity in de~Sitter spacetime, which will be used to explicitly  calculate the graviton  
 Wightman two-point in covariant gauge in the next chapter. 
 \section{Linearized Quantum Gravity in de~Sitter Spacetime}
In this section we review the linearized quantum gravity in de~Sitter spacetime.
After dropping total divergences,  the Lagrangian for linearized quantum gravity in de~Sitter spacetime can be written as  $\cite{15}$,
\begin{eqnarray}
\mathcal{L}_{gr} & =&  -\frac{1}{4} \nabla_a h^{bc}\nabla^a h_{bc}+ \frac{1}{2} \nabla_a h^{ac}\nabla^b h_{bc} +  \frac{1}{4} \nabla^c h \nabla_c h - 
  \nonumber \\&& \frac{1}{2} \nabla^a h \nabla^b h_{ab}   -\frac{1}{2} \left[h_{ab}h^{ab}+\frac{1}{2}h^2\right].
\end{eqnarray}
This Lagrangian is invariant under the following gauge transformation,
\begin{equation}
\delta_\Lambda h_{bc} = \nabla_c\Lambda_b + \nabla_b \Lambda_c,
\end{equation} 
where $\Lambda_c$ is any vector field on de~Sitter spacetime. Now we need to add a  gauge-fixing   term  to this Lagrangian for linearized quantum gravity
 in de~Sitter spacetime and we choose to add the gauge-fixing   term   given in Eq. $(\ref{gaugeft})$, 
\begin{equation}
 \mathcal{L}_g  = -\frac{1}{2\alpha}\left[  \nabla^b h_{ab} -  k \nabla_a h \right] \left[  \nabla_c h^{ac} -  k \nabla^a h \right].
\end{equation}
So after neglecting total derivatives, the sum of this gauge-fixing   term and classical  Lagrangian for linearized quantum gravity in de~Sitter 
spacetime can be written as 
\begin{eqnarray}
 \mathcal{L}_{gr} + \mathcal{L}_g  &=&  -\frac{1}{4} \nabla_a h^{bc}\nabla^a h_{bc}+ \left(\frac{1}{2}-\frac{1}{2\alpha} \right) \nabla_a h^{ac}\nabla^b h_{bc}
 \nonumber \\ &&- \left( \frac{(\beta+ 1)^2}{2\alpha \beta^2} - \frac{1}{4}\right) \nabla^c h   \nabla_c h  - \left( \frac{1}{2}-\frac{\beta +1}{\beta \alpha}\right)
\nonumber \\ &&
\times  \nabla^a h \nabla^b h_{ab}   -\frac{1}{2} \left[h_{ab}h^{ab}+\frac{1}{2}h^2\right].
\end{eqnarray}
This is the Lagrangian of massless spin-$2$ field in de~Sitter spacetime.  It will be seen that it is of 
calculational  advantage to first add a mass term and then 
take the zero mass limit of this massive theory at the end of our calculations. So during most calculations, we shall  start from massive spin-$2$ 
field and then set the mass to zero at the end of the calculation.  The Lagrangian $\mathcal{L}$ of massive spin-$2$ field in de~Sitter spacetime is 
obtained by adding a mass term called the Fierz-Pauli term, to the massless spin-$2$ field Lagrangian:
\begin{eqnarray}
 \mathcal{L} &=&  \mathcal{L}_{gr} +\mathcal{L}_g  +\frac{1}{4} M^2 \left[h_{ab}h^{ab}- h^2\right]   \nonumber \\ 
&=& -\frac{1}{4} \nabla_a h^{bc}\nabla^a h_{bc}+ \left(\frac{1}{2}-\frac{1}{2\alpha} \right) \nabla_a h^{ac}\nabla^b h_{bc}
 \nonumber \\ &&-\left( \frac{(\beta+ 1)^2}{2\alpha \beta^2} - \frac{1}{4}\right) \nabla^c h   \nabla_c h  - \left( \frac{1}{2}-\frac{\beta +1}{\beta \alpha}\right)
\nonumber \\ &&
\times  \nabla^a h \nabla^b h_{ab}    -\frac{1}{2} \left[h_{ab}h^{ab}+\frac{1}{2}h^2\right] \nonumber \\ &&   + \frac{1}{4} M^2 \left[h_{ab}h^{ab}- h^2\right].
 \label{lagrangianmass}
\end{eqnarray}
The equation of motion obtained from this Lagrangian is
\begin{eqnarray}
\frac{1}{2} \Box h_{bc} - \left(\frac{1}{2}-\frac{1}{2\alpha} \right) [\nabla_b\nabla_a h^a_c + \nabla_c\nabla_a h^a_b] && 
  \nonumber \\ + \left( \frac{(\beta+ 1)^2}{\alpha \beta^2} - \frac{1}{2}\right) g_{bc}\Box  h 
 +  \left( \frac{1}{2}-\frac{\beta +1}{\beta \alpha}\right)&&   \nonumber \\ \times  [\nabla_c \nabla_b h + g_{bc}\nabla_a\nabla_d  h^{ad} ] 
 -  \left[h_{bc}+ \frac{1}{2}g_{bc}h\right]&&   \nonumber \\
- \frac{1}{2}  M^2 [h_{bc} - g_{bc}h] &=& 0.\label{eqmm}
\end{eqnarray}
\section{Symplectic Product}
Now to calculate the symplectic product, we first calculate the  momentum current $\pi^{abc}$, which is given by
 \begin{equation}
 \pi^{abc} =\sqrt{-g} \frac{\partial  \mathcal{L}}{\partial\nabla_a h_{bc}}.
\end{equation}
It can be  explicitly written as
\begin{eqnarray}
\pi^{abc}&=&\sqrt{-g}\left[- \frac{1 }{2}\nabla^a h^{bc} + \left(\frac{1}{2}-\frac{1}{2\alpha} \right)[g^{ac}\nabla_d h^{db} + g^{ab}\nabla_d h^{dc}]
\right. \nonumber \\&& \left.- \left( \frac{(\beta+ 1)^2}{\alpha \beta^2} - \frac{1}{2}\right) g^{bc} \nabla^a h 
 - \left( \frac{1}{4}-\frac{\beta +1}{2\beta \alpha}\right)\right. \nonumber \\ && 
\left. \times  [2 g^{bc}\nabla_d h^{ad} + g^{ac}\nabla^b h + g^{ab}\nabla^c h ]\right]. \label{p2ia}
\end{eqnarray}
Now if $h_{bc}$ and $h'_{bc}$ are two solutions to the field equation, Eq. $(\ref{eqmm})$, and if $\pi^{abc}$ and $\pi'^{abc}$ are the momentum current 
conjugate to them,
 then we find using Eq. $(\ref{jijiji1})$ that the  conserved current $J^a$ is given by
\begin{equation}
    J^a = - \frac{i}{\sqrt{-g} }[h^*_{cb} \pi'^{abc} - h'_{cb}\pi^{abc*}]. \label{jj}
\end{equation}
The symplectic product defined on a spacelike hypersurface $\Sigma$  is given by
\begin{equation}
  (h, h') = \int d \Sigma_a J^a.
 \end{equation}
Now because of Eq. $(\ref{conhspinab})$, we have 
\begin{equation}
 \nabla_c J^c =0.
\end{equation}
So for a  spacetime region bounded by a future spacelike hypersurface $\Sigma^{+}$ and a past spacelike hypersurface $\Sigma^{-}$,
 because of Eq. $(\ref{sp12abab})$, we have  
\begin{equation}
 \int_{\Sigma^{+}} d\Sigma_c J^c  = \int_{\Sigma^{-}} d\Sigma_c J^c.
\end{equation}

\section{Mode Decomposition}
In this section we shall  decompose $h_{bc}$ into modes. In the next chapter we shall calculate the  graviton   Wightman two-point in covariant 
gauge by calculating the contribution to it coming from each of these modes. 

We can  decompose the field  $h_{bc}$ into scalar, vector and tensor parts $\cite{15}$,  
\begin{equation}
 h_{bc} =  C_{bc} + A_{bc} + E_{bc}  .
\end{equation}
Here  the scalar part $C_{bc}$  is given by, 
\begin{equation}
C_{bc} = (\nabla_b \nabla_c - \frac{1}{4}g_{bc}\Box )B + \frac{1}{4}g_{bc}h. \label{cab}
\end{equation}
The vector part, $A_{bc}$ is given by  
\begin{equation}
 A_{bc} = \nabla_c A_b + \nabla_b A_c, \label{cab1} 
\end{equation}
and satisfies, 
\begin{equation}
 A^b_b = 2 \nabla^b A_b  = 0.\label{cab1a} 
\end{equation}
The tensor part $E_{bc}$ satisfies 
\begin{eqnarray}
  E_b^b = 0, \label{cab2}\\
 \nabla^c E_{bc} = 0.\label{cab2a}
\end{eqnarray}

Now we want to find the equations of motion for each of these parts. We start from the scalar part $C_{ab}$.
First by substituting  $h_{ab}= C_{ab}$, where $C_{ab}$ is defined in  Eq. $(\ref{cab})$,  into the equation of motion  $(\ref{eqmm})$
and then using the formula for $R^d_{bac}$ given in Eq. $(\ref{jkla})$,
we find the following equation, 
\begin{equation}
 g_{ab}[X_1 h +  X_2 B] + \nabla_a \nabla_b [Y_1 B + Y_2 h] = 0, \label{app11}
\end{equation}
where
\begin{eqnarray}
X_1 &=& \left[\frac{(\beta +1)^2}{\alpha \beta^2} -\frac{\beta +1}{4\alpha \beta} - \frac{1}{4}\right]\Box - \frac{3}{4} +\frac{3M^2}{8}  , \\
X_2 &=& \left[\frac{1}{4}-\frac{3 (1 + \beta)}{4\alpha \beta}\right]\Box^2 + 3  \left[\frac{1}{4}- \frac{1 +\beta}{\alpha\beta}\right]\Box 
 +\frac{M^2}{8}\Box, \\
Y_1& =& \frac{3- \alpha}{4 \alpha} \Box + \frac{3}{\alpha}-\frac{M^2}{2},\\
 Y_2 &=&   \frac{\alpha\beta - 3 \beta - 4}{4 \alpha\beta}.
\end{eqnarray}
Now if we define 
\begin{eqnarray}
  X_1 h +  X_2 B &=&\mu,  \label{824eq} \\ 
Y_1 B + Y_2 h &=&\nu, \label{825eq}
\end{eqnarray}
then we can write Eq. $(\ref{app11})$ as
\begin{equation}
 g_{ab}\mu + \nabla_a \nabla_b \nu = 0. \label{a1pp11}
\end{equation}
Taking the trace of Eq. $(\ref{a1pp11})$, we obtain  
\begin{equation}
\mu = -\frac{1}{4} \Box \nu. \label{numunu1}
\end{equation}
 Now  its divergence, we obtain 
\begin{equation}
 \nabla_a [\Box  + 4 ] \nu =0.
\end{equation}
So  we have
\begin{equation}
 [\Box + 4] \nu =4k, \label{bree}
\end{equation}
where $k$ is a constant. Thus we get 
\begin{equation}
 \nu = \mu  + k.
\end{equation}
Now a particular solution to Eqs. $(\ref{824eq})$ and $(\ref{825eq})$ can be written as 
\begin{eqnarray}
 h &=& h^{(p)} = 0, \\ 
 B &=& B^{(p)} = \frac{2\alpha}{6 -M^2 \alpha} k + \frac{2 \mu}{2-M^2}.
\end{eqnarray}
Any solution Eqs. $(\ref{824eq})$ and $(\ref{825eq})$ can be written as a sum of this particular solution and the general solution. The 
general solution is the 
solution to the homogeneous equations obtained from Eqs. $(\ref{824eq})$ and $(\ref{825eq})$ by setting $\mu = \nu =0$. 
However, $C_{ab} =0$ for the particular  solution $h=h^{(p)} $ and $ B= B^{(p)}$. Since we take the $M^2 \to 0$ 
limit in the end, the special cases $M^2 =2$  and $M^2 = 6/\alpha$ are not relevant here. 
Hence we only need to consider the solution to the homogeneous equations
given by 
\begin{eqnarray}
 X_1 h + X_2 B &=&0,  \\ 
 Y_1 B + Y_2 h &=& 0.
\end{eqnarray}
Thus we have 
\begin{eqnarray}
 \left[\left(\frac{1}{4}-\frac{3 (1 + \beta)}{4\alpha \beta}\right)\Box^2 + 3  \left(\frac{1}{4}- \frac{1 +\beta}{\alpha\beta}\right)\Box 
 +\frac{M^2}{8}\Box\right] B&& \nonumber \\+ 
\left[\left(\frac{(\beta +1)^2}{\alpha \beta^2} -\frac{\beta +1}{4\alpha \beta} - \frac{1}{4}\right)\Box - \frac{3}{4} +\frac{3M^2}{8} \right] h 
   &=&0,\\
\left[ \frac{3- \alpha}{4 \alpha} \Box + \frac{3}{\alpha}-\frac{M^2}{2}\right] B + \left[ \frac{\alpha\beta - 3 \beta - 4}{4 \alpha\beta}\right] h &=& 0.
\end{eqnarray}
After eliminating the fourth-order derivative of $B$, we can rearrange these equations as: 
\begin{eqnarray}
 \Box B & =& a_1 B + a_2 h, \label{B12} \label{xcv}\\
 \Box h &= & a_3 h + a_4 B,\label{h12} \label{xcv1}
\end{eqnarray}
where
\begin{eqnarray}
 a_1 &=& \frac{12 - 2 \alpha M^2}{\alpha - 3}, \nonumber \\
 a_2 &=& \frac{\alpha\beta - 3 \beta - 4}{\beta(\alpha - 3)},\nonumber \\
a_3 &=& \frac{6(\alpha\beta - 3 \beta - 2)}{\alpha - 3}\left(1- \frac{M^2}{2}\right),\nonumber \\
 a_4 &=& \frac{-9\beta (\alpha\beta - 3 \beta - 4)}{\alpha - 3}\left(1 - \frac{M^2}{2}\right)\left(1 - \frac{\alpha}{6}M^2\right).
\end{eqnarray}

The equations of motion for the vector part are obtained by   substituting   $h_{ab}= A_{ab}$,  where $A_{ab}$ 
is defined by  Eqs. $(\ref{cab1})$ and $(\ref{cab1a})$,  into the equations of motion
 $(\ref{eqmm})$. Thus we get,
\begin{equation}
 \nabla_b V_a + \nabla_a V_b  = 0,
\end{equation}
where, 
\begin{equation}
 V^a = \nabla_b [\nabla^b A^a + \nabla^a A^b] - \alpha M^2 A^a.
\end{equation}
The general solution to this equation is 
\begin{equation}
 \nabla_b [\nabla^b A^a + \nabla^a A^b] - \alpha M^2 A^a = f^a
\end{equation}
where $f_a$ is a Killing vector and thus satisfies 
\begin{equation}
 \nabla_a f_b + \nabla_b f_a =0.
\end{equation}
So a particular solution is given by 
\begin{equation}
 A^{(p)}_{a} = - \frac{f_a}{M^2 \alpha}.
\end{equation}
However $A_{ab}=0$ for this solution, $A_a = A_a^{(p)}$. So following  a similar line of argument as was used for 
the scalar part, we can set $f^a=0$.
Thus we get 
\begin{equation}
  \nabla_b [\nabla^b A^a + \nabla^a A^b] - \alpha M^2 A^a =0.
\end{equation}
Now by using Eq. $(\ref{jkla})$, we can write this equation as   
\begin{equation}
 [\Box +3  -\alpha M^2 ] A_a =0. \label{aeq1a}
\end{equation}
The equation for the tensor part is obtained by   substituting  $ h_{ab} = E_{ab}$, where $E_{ab}$ is defined by Eqs. $(\ref{cab2})$ and $(\ref{cab2a})$,
into the equation of motion $(\ref{eqmm})$. Thus we get
 \begin{equation}
 [\Box - 2 -M^2] E_{ab} =0.\label{eeq1a}
\end{equation}

\section{Harmonic Modes on de~Sitter Spacetime}
To find the contribution of the vector and tensor parts to the two-point function we expand the vector and tensor parts into
 harmonic modes on de~Sitter spacetime. The spatial part of these modes is given by the spherical harmonics on $S^3$ and its temporal part 
is given in terms of  the associated Legendre function. We denote the coordinates on $S^3$ by $\bf{x}$ and define $\chi$ using  Eq. $(\ref{timeimagenary})$  
as $\chi = \pi/2 -it$.

Now we review some properties of scalar, vector  and tensor spherical harmonics on $S^3$ $\cite{sph}$. Let the indices on de~Sitter 
spacetime be denoted by  $(a,b, c \cdots)$ and on $S^3$ be denoted by $(i, j, k \cdots)$.
 We denote the scalar, vector and tensor spherical harmonics  
 on $S^3$ by $Y^{\ell \sigma} ({\bf{x}}), Y_i^{\ell \sigma}(\bf{x})$ and $Y_{ij}^{\ell \sigma}(\bf{x})$, respectively. 
Here $\sigma$ denotes all the indices other than $\ell$. Thus $\sigma$ denotes all the quantum numbers 
on a two-sphere $S^2$ from now on. It may be noted that $\sigma$ previously denoted all the quantum numbers on a three-sphere 
(in the chapters on ghost propagators).  We also denote the covariant derivative on $S^3$ by  $\tilde{\nabla}_i$, 
the metric on $S^3$ by $\eta_{ij}$ and $\tilde{\nabla}_i \tilde{\nabla}^i$ by $\tilde{\nabla}^2$. 
For scalar spherical harmonics on $S^3$,  we have  $\ell =0,1,2,3 \cdots$ and  
\begin{equation}
 -\tilde{\nabla}^2 Y^{\ell\sigma} = \ell (\ell +2)Y^{\ell\sigma}. 
\end{equation}
For vector spherical harmonics on $S^3$, we have $\ell =1,2,3,4 \cdots$ and
\begin{eqnarray}
 -\tilde{\nabla}^2 Y^{\ell\sigma}_i &=& [\ell (\ell +2)-1]Y^{\ell\sigma}_i, \nonumber \\
 \tilde{\nabla}^i Y^{\ell\sigma}_i &=& 0. 
\end{eqnarray}
For tensor spherical harmonics on $S^3$, we have $\ell = 2,3,4,5 \cdots$ and 
\begin{eqnarray}
 -\tilde{\nabla}^2 Y^{\ell\sigma}_{ij} &=& [\ell (\ell +2)-2]Y^{\ell\sigma}_{ij}, \nonumber \\ 
 \tilde{\nabla}^i Y^{\ell\sigma}_{ij} &=& 0, \nonumber \\ 
   \eta^{ij} Y^{\ell\sigma }_{ij} &=& 0. 
\end{eqnarray}
Next we review some properties of the  associated Legendre functions $P^{-\mu}_\nu (x )$ $\cite{6ch1}$. 
The degree $\nu$ can be lowered or raised as 
\begin{eqnarray}
  \left(  (1-x^2)\frac{d}{dx} +\nu x\right)P^{-\mu}_{\nu}(x) &=& (\nu-\mu)P_{\nu-1}^{-\mu}(x),\nonumber \\ 
 \left(  (1-x^2)\frac{d}{dx} -(\nu+1) x\right)P^{-\mu}_{\nu}(x) &=& -(\nu+\mu+1)P_{\nu+1}^{-\mu}(x). \label{5212a}
\end{eqnarray}
Now we define $D_n$ as follows:
\begin{equation}
 D_n = \frac{d}{d\chi} + n\cot\chi. \label{D}
\end{equation}
Then we have 
\begin{equation}
 -\sin \chi D_n P^{\mu}_{\nu} (\cos \chi)= \left[ (1- \cos^2\chi)\frac{d}{d\cos\chi} - n \cos\chi \right]P^{-\mu}_{\nu} (\cos \chi). \label{52121a}
\end{equation}
 So now because of Eqs. $(\ref{5212a})$ and  $(\ref{52121a})$, we have 
\begin{eqnarray}
 -\sin\chi D_{-\nu}P^{-\mu}_{\nu} &=& (\nu-\mu)P_{\nu-1}^{-\mu}(x),\nonumber \\ 
-\sin\chi D_{\nu + 1}P^{-\mu}_{\nu} &=& -(\nu+\mu+1)P_{\nu+1}^{-\mu}(x). \label{D1}
\end{eqnarray}
We also have
\begin{equation}
 D_m \sin^n \chi P^{-\mu}_{\nu} (\cos \chi) = \sin^n \chi D_{m+n} P^{-\mu}_{\nu} (\cos \chi),
\end{equation}
because 
\begin{eqnarray}
 && \left[ \frac{d}{d\chi} + m\cot\chi \right] \sin^n\chi P^{-\mu}_{\nu} (\cos \chi)  \nonumber \\  
 &=& \sin^n\chi \left[ \frac{d}{d\chi} + (m+n)\cot\chi \right] P^{-\mu}_{\nu} (\cos \chi).
\end{eqnarray}

There are two types of solutions to Eq. $(\ref{aeq1a})$ denoted by $A_b^{n \ell \sigma} $  with $n =0, 1$. These modes have $\ell = 1, 2, 3 \cdots $.  
There are three types of solutions to Eq. $(\ref{eeq1a})$ denoted by $E_{bc}^{n \ell \sigma} $  
with $n =0, 1, 2 $. These modes have  $\ell = 2, 3, 4 \cdots $ 
(see Sec. $2, 4, 5, 6$ of Ref $\cite{6ch3}$).  Here $\sigma$ denotes all the quantum numbers on a two-sphere. 
We let the solutions to the minimally-coupled massive scalar field equation in de~Sitter spacetime be denoted by  $S^{ \ell \sigma}$
 with $\ell =0, 1, 2, 3, \cdots $, where 
\begin{equation}
 [\Box - m^2 ] S^{\ell \sigma} =0.\label{scans}
\end{equation}

We shall write the  solutions on de~Sitter spacetime in analogy with the solutions on $S^4$, exploiting 
 the similarity of  de~Sitter metric given in Eq. $(\ref{metricabc2})$ to the metric on $S^4$, after defining 
 $\chi$ by  Eq. $(\ref{timeimagenary})$  as $\chi = \pi/2 -it$.

So first we review the defining properties of scalar,  vector and tensor spherical harmonics on $S^4$  $\cite{sph}$.
For scalar spherical harmonics on $S^4$, we have 
\begin{equation}
 -\Box S^{L\ell\sigma} = [ \tilde{L} ( \tilde{L} +3)]S^{L\ell\sigma},\label{scansa}
\end{equation}
with $\tilde{L} = 0,1,2, \cdots$ and $\ell \leq \tilde{L}$.
For vector spherical harmonics on $S^4$, we have 
\begin{eqnarray}
-\Box A^{L\ell\sigma}_b &=& [L (L +3)-1]A^{L\ell\sigma}_b, \nonumber \\
\nabla^b A^{L\ell\sigma}_b &=& 0, \label{vectortensorharmonicmodes1a}
\end{eqnarray}
with $L = 1,2,3, \cdots$ and $\ell \leq L$.
For tensor spherical harmonics on $S^4$, we have 
\begin{eqnarray}
 -\Box E^{\overline{L}\ell\sigma}_{bc} &=& [\overline{L} (\overline{L} +3 )-2]E^{\overline{L}\ell\sigma}_{bc}, \nonumber \\ 
 \nabla^b E^{\overline{L}\ell\sigma}_{bc} &=& 0, \nonumber \\ 
 g^{bc}E^{\overline{L}\ell\sigma }_{bc} &=& 0, \label{vectortensorharmonicmodes2a}
\end{eqnarray}
with $\overline{L} = 1,2,3 \cdots$ and $\ell \leq \overline{L}$.

What we use is the fact that the positive-frequency mode function for the Euclidean vacuum 
in de~Sitter spacetime are obtained by solving the same equations as these 
spherical harmonics under the transformation $\chi = \pi/2 - it$, if we allow $\tilde{L}, L $ and $\overline{L}$ to be non-integers and unconstrained 
by the condition $\tilde{L}, L, \overline{L} \geq \ell$ (see Ref. $\cite{6ch3}$). 
Comparing  Eq. $(\ref{aeq1a})$ with Eq. $(\ref{vectortensorharmonicmodes1a})$, Eq. $(\ref{eeq1a})$ with Eq. $(\ref{vectortensorharmonicmodes2a})$,
 Eq. $(\ref{scans})$ with Eq. $(\ref{scansa})$ and  
 using the similarity of de~Sitter spacetime with $S^4$, we can write $ \tilde{L}$, $\overline{L}$ and $L$ as,
\begin{eqnarray}
\overline{L}(\overline{L}+3)&=& -M^2, \nonumber \\ 
L(L+3) &=& -\alpha M^2 + 4, \nonumber \\
\tilde{L} ( \tilde{L} +3)& =& -m^2. \label{inpd10}
\end{eqnarray}
Now $\tilde{L}, L, \overline{L}$ to the first  order in $M^2$ or $m^2$, are given by 
\begin{eqnarray}
\tilde{L}&\approx& -\frac{M^2}{3},\\
 L&\approx& 1 -\frac{\alpha M^2}{5},\\
\overline{L}&\approx& -\frac{m^2}{3}.
\end{eqnarray}
In the limit $M^2 \to 0$, we have $ \overline{L} =0$ and $L=1$, for de~Sitter spacetime. 
Now the scalar, vector and tensor harmonic modes in de~Sitter spacetime are given by (see Sec. $2, 4, 5, 6$ of Ref. $\cite{6ch3}$),

\begin{eqnarray}
 S^{\ell \sigma} ( \chi, {\bf{x}}; m^2) &=& (\sin \chi)^{-1}  P^{-(\ell +1)}_{\tilde{L}+1}(\cos \chi) Y^{\ell \sigma}( {\bf{x}}),\label{sp0}\\
 A^{1\ell \sigma}_\chi ( \chi, {\bf{x}}; M^2) &=& 0 , \nonumber \\ 
A^{1 \ell \sigma}_i( \chi, {\bf{x}}; M^2) & =&  P_{L+1}^{-(\ell +1)} (\cos \chi) Y_i^{\ell \sigma}( {\bf{x}}), 
\label{sp1}\\
A^{0\ell \sigma}_\chi ( \chi, {\bf{x}}; M^2)&=&  (\sin\chi)^{-2} P_{L+1}^{-(\ell +1)} (\cos \chi)Y^{\ell \sigma}( {\bf{x}}), \nonumber \\ A^{0\ell \sigma}_i ( \chi, {\bf{x}}; M^2)&=&   \frac{1}{\ell(\ell+2)} D_1 P_{L+1}^{-(\ell +1)} (\cos \chi)\tilde{\nabla}_i Y^{\ell \sigma}( {\bf{x}}),
  \label{sp2}\\
 E^{2\ell \sigma}_{\chi\chi}( \chi, {\bf{x}}; M^2)& =& 0,
  \nonumber \\ E^{2\ell \sigma}_{\chi i} (\chi, {\bf{x}}; M^2)&=&0 , 
\nonumber \\ E^{2\ell \sigma}_{ij}( \chi, {\bf{x}}; M^2) &=&  (\sin \chi)^{-1} P_{\overline{L}+1}^{-(\ell +1)}  (\cos \chi) Y_{ij}^{\ell \sigma}( {\bf{x}}),
 \label{sp3}\\
 E^{1\ell\sigma}_{\chi\chi}( \chi, {\bf{x}}; M^2) &=& 0,\nonumber \\
 E^{1\ell\sigma}_{\chi i}( \chi, {\bf{x}}; M^2)&=&  (\sin \chi)^{-1} P_{\overline{L}+1}^{-(\ell +1)} (\cos \chi)Y_i^{\ell \sigma}( {\bf{x}}),\nonumber \\
 E^{1\ell\sigma}_{ij}( \chi, {\bf{x}}; M^2)& = & \frac{1}{(\ell-1)(\ell+3)}\sin \chi D_2 P_{\overline{L}+1}^{-(\ell +1)} (\cos \chi)\nonumber\\ 
&& \times   (\tilde{\nabla}_i Y_j^{\ell \sigma}( {\bf{x}})  +\tilde{\nabla}_j Y_i^{\ell \sigma}( {\bf{x}})), 
\label{sp4}\\
 E^{0\ell\sigma}_{\chi\chi} ( \chi, {\bf{x}}; M^2)&=&  (\sin\chi)^{-3} P_{\overline{L}+1}^{-(\ell +1)} (\cos \chi)Y^{\ell \sigma}( {\bf{x}}),
 \nonumber\\ E^{0\ell\sigma}_{\chi i} ( \chi, {\bf{x}}; M^2)&=&  \frac{1}{\ell(\ell+2)} (\sin\chi)^{-1} 
D_1 P_{\overline{L}+1}^{-(\ell +1)} (\cos \chi)\tilde{\nabla}_i Y^{\ell \sigma}( {\bf{x}}) ,
\nonumber \\ E^{0\ell\sigma}_{ij}( \chi, {\bf{x}}; M^2)& =&  ( b_1^{\ell} \tilde{T}^\ell_{ij} +   b_2^{\ell}\eta_{ij} )Y^{\ell \sigma}( {\bf{x}}),\nonumber \\
 b_1^{ \ell}(\chi, M^2) &=& \frac{3}{2(\ell+3)(\ell-1)}  \left( -\frac{1}{3}(\sin\chi)^{-1} \right. \nonumber \\ && \left. + \frac{1}{\ell(\ell+2)}\sin\chi D_2  D_1  \right)P_{\overline{L}+1}^{-(\ell +1)} (\cos \chi),\nonumber  \\
 b_2^{\ell}(\chi, M^2)&=& -\frac{1}{3\sin\chi} P_{\overline{L}+1}^{-(\ell +1)} (\cos \chi),\nonumber \\
  \tilde{T}^\ell_{ij} &=& \tilde{\nabla}_i \tilde{\nabla}_j + \frac{\ell( \ell+ 2)}{3}\eta_{ij}, \label{sp6}
\end{eqnarray}

We emphasize again that we have merely used the analogy between de~Sitter spacetime and $S^4$ and 
not really analytically continued from $S^4$ to de~Sitter spacetime.
If for the modes given by Eqs. $(\ref{sp0})$-$(\ref{sp6})$, we let   $\tilde{L} = 0,1,2,3,4\cdots, L = 1,2,3,4,5\cdots, \overline{L} = 2,3,4,5,6\cdots$ and
$\tilde{L}, L, \overline{L} \geq \ell$, where $\ell$ is a positive integer, then they   would be regular at $\chi =\pi$ 
and thus Eqs. $(\ref{sp0})$-$(\ref{sp6})$ would 
define valid spherical harmonics. However, in  case of de~Sitter spacetime these modes do not satisfy all of these 
conditions, and as a result they are singular at $\chi =\pi$. 
But, this point corresponds to imaginary time $t = i \pi/2$, and hence it is not physically relevant in de~Sitter spacetime and so we
 can safely use Eqs. $(\ref{sp0})$-$ (\ref{sp6})$ for calculating graviton  Wightman two-point function in de~Sitter spacetime. 

These modes given by  Eqs. $(\ref{sp0})$-$(\ref{sp6})$, regarded as functions on  de~Sitter spacetime, 
can be used to decompose the scalar $C_{ab}$, vector $A_{ab}$ and tensor $E_{ab}$ parts of
the field $h_{ab}$ in terms of mode functions. The scalar, vector and tensor mode functions that constitute the fields $C_{ab}$, $A_{ab}$ and $E_{ab}$, 
respectively, are orthogonal to one another with respect to the symplectic product  as long as $M\neq 0$ $\cite{14}$.
 This implies that  $C_{ab}$, $A_{ab}$ and $E_{ab}$
commute with one another upon quantization. Thus the quantization of each of these fields can be considered separately. However in the limit 
$M \to 0$, the vector modes constituting $A_{ab}$, and tensor modes constituting $E_{ab}$, with $n =0,1$ coincide $\cite{phy}$.  For this reason, we shall 
consider the quantization of the vector and tensor parts, $A_{ab}$ and $E_{ab}$ together, but consider the quantization of the scalar part $C_{ab}$, separately    
from the vector and tensor parts.

\chapter{Wightman Graviton Two-Point Function}
In this chapter we shall  explicitly calculate the contributions to the covariant  Wightman graviton 
two-point function coming from the scalar and vector-tensor sectors. 

\section{Physical Equivalence}
In linearized quantum gravity the  Wightman two-point function of the graviton field $h_{ab}$ has no physical 
meaning by itself because this theory has gauge invariance under the gauge transformation,
\begin{equation}
\delta_\Lambda h_{bc} = \nabla_c\Lambda_b + \nabla_b \Lambda_c,
\end{equation}
where $\Lambda_c$ is any vector field on de~Sitter spacetime.
One can find tensor fields at $x$ that are linear in $h_{ab}$ and are invariant under this gauge
transformations $\cite{gi}$. An example of such a tensor field is the linearized Weyl tensor $W_{abcd}(x)$ given by 
\begin{eqnarray}
W_{abcd}(x) & = & \tilde{W}_{[ab][cd]}(x),\\
\tilde{W}_{abcd}(x) & = & \nabla_{c} \nabla_{b}h_{ad}(x) + h_{cb}(x)g_{ad}(x).
\end{eqnarray}

 Now, if a   Wightman graviton two-point function $  G_{bcb'c'}(x,x')$ can be written as
\begin{equation}
  G_{bcb'c'}(x,x') = P_{bcb'c'}(x,x') + Q_{bcb'c'}(x,x'),
 \end{equation}
 where
\begin{eqnarray}
Q_{aba'b'}  &=& \nabla_{a} K^1_{ba'b'}(x,x') + \nabla_{b} K^1_{aa'b'}(x,x')\nonumber \\ && 
 +\nabla_{a'} K^2_{abb'}(x,x') +\nabla_{b'} K^2_{aba'}(x,x'), \label{Q}
\end{eqnarray}
for  some $ K^1_{ba'b'}(x,x')$  and $  K^2_{abb'}(x,x')$,
then the two-point function of a local gauge-invariant tensor field linear in $h_{bc}$ will be the same whether
 one uses $G_{bcb'c'}(x,x')$ or $P_{bcb'c'}(x,x')$ as the graviton two-point function. 

This motivates the following definition: 
We say that the two graviton two-point functions,
$G_{bcb'c'}(x,x')$ and $P_{bcb'c'}(x,x')$ are physically equivalent in linearized gravity if they differ
 from each other by a two-point function of pure gauge form $ Q_{bcb'c'} (x,x')$.
\section{Symplectic Product for  Vector and Tensor Modes}
The Klein-Gordon type product given in Eq. $(\ref{kl})$,  for vector and   tensor harmonic modes  on de~Sitter spacetime is given by
 (see Sec. $8, 9$ of Ref. $\cite{6ch3}$),
\begin{eqnarray}
\langle E^{1 \ell \sigma}, E^{1 \ell' \sigma'} \rangle 
& = & - \frac{4 \overline{L}(\overline{L}+3)\delta^{ll'} \delta^{\sigma \sigma'}}{(\ell-1)(\ell+3)\Gamma(\ell-\overline{L})\Gamma(\ell+\overline{L} + 3)},
\nonumber \\
 \langle E^{0 \ell \sigma}, E^{0 \ell' \sigma'} \rangle 
& =& \frac{3\left[\overline{L}(\overline{L}+3)+2\right]\overline{L}(\overline{L}+3)\delta^{ll'} \delta^{\sigma \sigma'}}{\ell(\ell+2)(\ell-1)(\ell+3)
\Gamma(\ell-\overline{L})\Gamma(\ell+\overline{L} +3)}
,\nonumber \\
\langle E^{2 \ell \sigma}, E^{2 \ell' \sigma'} \rangle 
& = &  \frac{\delta^{ll'} \delta^{\sigma \sigma'}}{\Gamma(\ell-\overline{L})\Gamma(\ell+\overline{L} + 3)},\nonumber \\
\langle A^{0 \ell \sigma}, A^{0 \ell' \sigma'} \rangle 
& = & - \frac{2(\overline{L}(\overline{L}+3)+2)\delta^{ll'} \delta^{\sigma \sigma'}}
{\ell(\ell+2)\Gamma(\ell- L)\Gamma(\ell+3+L)},\nonumber \\
\langle A^{1 \ell \sigma}, A^{1 \ell' \sigma'} \rangle  &= &  \frac{2\delta^{ll'} \delta^{\sigma \sigma'}}{\Gamma(\ell - L )\Gamma(\ell+3+L)}. \label{inpd1}
\end{eqnarray}
The Klein-Gordon type product $\langle E^{n \ell \sigma}, E^{n'\ell' \sigma'} \rangle $ and $\langle A^{n \ell \sigma}, A^{n'\ell' \sigma'} \rangle $ 
vanish for $n \neq n'$. 
The vector mode functions corresponding to $A^{n \ell \sigma}_a$ are given by 
\begin{equation}
 A^{n \ell \sigma}_{ab} = \nabla_aA^{n \ell \sigma}_b + \nabla_b A^{n \ell \sigma}_a.
\end{equation}
As shown in  Appendix A, the symplectic product for the vector  and tensor modes is related to their Klein-Gordon type product  as follows: 
\begin{eqnarray}
(A^{n \ell \sigma}, A^{n' \ell' \sigma'})&=& - M^2 \langle A^{n \ell \sigma}, A^{n' \ell' \sigma'} \rangle,\\
(E^{n \ell \sigma},E^{n' \ell' \sigma'})& =&  \frac{1}{2}\langle E^{n \ell \sigma}, E^{n' \ell' \sigma'} \rangle. 
 \end{eqnarray}

Now recall 
\begin{eqnarray}
 L&\approx& 1 -\frac{\alpha M^2}{5}, \label{xabcd1} \\
\overline{L}&\approx& -\frac{M^2}{3}.  
\end{eqnarray}
It will turn out that we need $L$ and $\overline{L}$ only to the first order in $M^2$. Using Eqs. $(\ref{inpd10})$ and $(\ref{inpd1})$, we get
\begin{eqnarray}
(E^{1 \ell \sigma}, E^{1 \ell' \sigma'}) &\approx& 
\frac{2M^2\delta^{ll'} \delta^{\sigma \sigma'} }{(\ell-1)(\ell+3)\Gamma(\ell+M^2/3)\Gamma(\ell+ 3 - M^2/3)},\nonumber \\
(E^{0 \ell \sigma}, E^{0 \ell' \sigma'})
&\approx&  \frac{3M^2(M^2-2)\delta^{ll'} \delta^{\sigma \sigma'}}{2\ell(\ell+2)(\ell-1)(\ell+3)\Gamma(\ell+M^2/3)\Gamma(\ell+ 3-M^2/3)},\nonumber\\
(E^{2 \ell \sigma}, E^{2 \ell' \sigma'}) 
&\approx& \frac{\delta^{ll'} \delta^{\sigma \sigma'}}{\Gamma(\ell+M^2/3)\Gamma(\ell+ 3 - M^2/3)},\nonumber \\
(A^{0 \ell \sigma}, A^{0 \ell' \sigma'} )
&\approx& -\frac{2M^2(\alpha M^2-6)\delta^{ll'} \delta^{\sigma \sigma'}}{\ell(\ell+2)\Gamma(\ell-1+\alpha M^2/5)\Gamma(\ell+4-\alpha M^2/5)},\nonumber \\ 
(A^{1 \ell \sigma}, A^{1 \ell' \sigma'} )
 &\approx& - \frac{2M^2\delta^{ll'} \delta^{\sigma \sigma'}}{\Gamma(\ell-1+\alpha M^2/5)\Gamma(\ell+4-\alpha M^2/5)} . \label{inner}
\end{eqnarray}
To simplify these expression further, we note 
\begin{eqnarray}
&&\Gamma(\ell-M^2/3)\Gamma(\ell+3+M^2/3)\nonumber \\  &=& (\ell+2-M^2/3)(\ell+1-M^2/3)(\ell-M^2/3)\nonumber  \\ &&\times  \Gamma(\ell-M^2/3)\Gamma(\ell+M^2/3),
\end{eqnarray}
and since $\Gamma(\ell-M^2/3)\Gamma(\ell+M^2/3)$ is analytic and even in $M^2$, we have
\begin{equation}
\Gamma(\ell-M^2/3)\Gamma(\ell+M^2/3) \approx \left[\Gamma(\ell)\right]^2 + \mathcal{O}(M^4).
\end{equation}
Hence, to the first order in $M^2$,  we find
\begin{eqnarray}
&& \Gamma(\ell+M^2/3)\Gamma(\ell+3-M^2/3)  \nonumber \\  &\approx& (\ell+2)!(\ell-1)!
\left[ 1 - \left( \sum_{k=0}^2 \frac{1}{\ell+k} \right)\frac{M^2}{3}\right].\label{msallmass2}
\end{eqnarray}
Similarly,  we have 
\begin{eqnarray}
&&\Gamma(\ell -1 +\alpha M^2/5)\Gamma(\ell+4 -\alpha M^2/5)  \nonumber \\ 
&\approx& (\ell+3)!(\ell-2)! \left[ 1 - \left( \sum_{k=-1}^3\frac{1}{\ell+k}\right) \frac{\alpha M^2}{5} \right]. \label{msallmass1}
\end{eqnarray}

We have found the symplectic product for  vector-tensor harmonic modes in de~Sitter spacetime, in this section. We have also seen
  that the Wightman two-point function for any free tensor field   can be obtained from this symplectic product by using  Eq. $(\ref{wighthig})$. 
So in the next section we shall use the symplectic product calculated here to find the vector-tensor contribution to the graviton  Wightman two-point function.  
\section{Vector-Tensor Contributions}
Now we can write  the vector-tensor contribution to the Wightman two-point function by using Eq. $(\ref{wighthig})$ as follows:
\begin{eqnarray}
 G^{(TV)}_{aba'b'}(x,x') &=& \sum_{n =0}^1 \sum _{\ell}\sum_{\sigma} 
\frac{A^{n \ell \sigma}_{ab} (x)A^{*n \ell \sigma}_{a'b'}(x')}{(A^{n \ell \sigma}, A^{n \ell \sigma} )} \nonumber \\ 
&& + \sum_{n =0}^2 \sum _{\ell}\sum_{\sigma} \frac{E^{n \ell \sigma}_{ab} (x)E^{*n \ell \sigma}(x')_{a'b'}}{(E^{n \ell \sigma}, E^{n \ell \sigma} )},
\end{eqnarray}
where for $(n =0,1)$, we have defined $A_{ab}^{n\ell \sigma}$ as, 
\begin{equation}
 A_{ab}^{n\ell \sigma} = \nabla_a A_b^{n\ell \sigma} +  \nabla_b A_a^{n\ell \sigma}.
\end{equation}
In Ref. $\cite{phy}$ a physical  Wightman graviton two-point function $P_{aba'b'}(x,x') $ was obtained. To do so first  de Donder gauge
 $\nabla_b h^{ab} - \nabla^a h/2 =0$ was imposed. Then it was shown that the trace could be gauged away and thus the conditions 
 $h^a_a=0$ and $\nabla^a h_{ab} =0$ were also imposed. Finally it was shown that all the modes apart from $E^{2\ell \sigma}_{ab}(x)$ could be gauged away. Thus
 it was shown that  only $E^{2\ell \sigma}_{ab}(x)$ modes
 contribute to the physical part of the   Wightman graviton two-point function.  So  $P_{aba'b'}(x,x') $ thus obtained was given by 
\begin{equation}
 P_{aba'b'}(x,x') = \sum_{\ell} \sum_{\sigma}\frac{ 2 E^{2\ell \sigma}_{ab}(x)
E^{*2\ell \sigma}_{a'b'}(x')}{\langle E^{2 \ell \sigma}, E^{2 \ell \sigma}\rangle}.
\end{equation}
We will show here that this physical  Wightman graviton two-point function given in Ref. $\cite{phy}$ is physically equivalent to the   Wightman graviton two-point
 function in covariant gauge.  
To do so, we  write down the  tensor-vector part of the  Wightman graviton two-point function in covariant gauge,
\begin{equation}
 G[h]_a = \nabla^b h_{ab} - (1 + \beta^{-1}) \nabla_a h, 
\end{equation}
 as
\begin{equation}
G^{(TV)}_{aba'b'}(x,x')=  P_{aba'b'}(x,x') + M_{aba'b'}(x,x')+ L_{aba'b'}(x,x'),
\end{equation}
where,
\begin{eqnarray}
M_{aba'b'}(x,x') &=&  \sum_{n =0}^1 \sum_{\sigma} \frac{A^{n 1 \sigma}_{ab} (x)A^{*n 1 \sigma}_{a'b'}(x')}{(A^{n 1 \sigma},
 A^{n \ell \sigma} )}, \\ 
L_{aba'b'}(x,x') &=& \sum_{n =0}^1 \sum _{\ell=2}^{\infty}\sum_{\sigma} \frac{A^{n \ell \sigma}_{ab} (x)A^{*n \ell \sigma}_{a'b'}(x')}{(A^{n \ell \sigma},
 A^{n \ell \sigma} )} \nonumber \\ && 
+ \sum_{n =0}^1 \sum _{\ell=2}^{\infty}\sum_{\sigma} \frac{E^{n \ell \sigma}_{ab} (x)E^{*n \ell \sigma}(x')_{a'b'}}{(E^{n \ell \sigma},
 E^{n \ell \sigma} )}.\label{outer}
\end{eqnarray}
Now   substituting  Eq. $(\ref{inner})$ into Eq. $(\ref{outer})$, we get 
\begin{eqnarray}
 L_{aba'b'}(x,x') &\approx&   \sum _{\ell=2}^{\infty}\sum_{\sigma}
\left[ \frac{(\ell-1)(\ell+3)}{2M^2}\Gamma\left(\ell-\frac{M^2}{3}\right)\Gamma\left(\ell+ 3+\frac{M^2}{3}\right)\right. \nonumber \\&&\left. \times
E^{1\ell \sigma}_{ab}(x)
E^{*1\ell \sigma}_{a'b'}(x') \right.\nonumber \\
&& \left. - \frac{1}{2M^2}\Gamma\left(\ell-1+\frac{\alpha M^2}{5}\right)\Gamma\left(\ell+4-\frac{\alpha M^2}{5}\right)\right. \nonumber \\&&\left.\times
A^{1\ell \sigma}_{ab}(x)
A^{*1\ell \sigma}_{a'b'}(x')\right] \nonumber \\
&& + \sum _{\ell=2}^{\infty} \sum_{\sigma}
 \left[ - \frac{2\ell(\ell+2)(\ell-1)(\ell+3)}{3(2-M^2)M^2} \right. \nonumber \\&&\left. \times
\Gamma\left(\ell+\frac{M^2}{3}\right)
\Gamma\left(\ell+ 3-\frac{M^2}{3}\right) E^{0\ell \sigma}_{ab}(x)
E^{*0\ell \sigma}_{a'b'}(x')  \right. \nonumber \\
&& \left. +\,  \frac{\ell(\ell+2)}{2M^2(6-\alpha M^2)}\Gamma\left(\ell-1+\frac{\alpha M^2}{5}\right)
\Gamma\left(\ell+4-\frac{\alpha M^2}{5}\right)\right. \nonumber \\&&\left. \times
A^{0\ell \sigma}_{ab}(x)
A^{*0\ell \sigma}_{a'b'}(x')\right], \label{msallmass3}
\end{eqnarray}
The contribution from each mode  to $L_{aba'b'}$ is divergent in the zero mass limit. However we shall see that the terms of order $1/M^2$  
cancel out. Now for small $M^2$,   substituting  Eq. $(\ref{msallmass1})$ and Eq. $(\ref{msallmass2})$ into Eq. $(\ref{msallmass3})$, we get 
\begin{eqnarray}
 L_{aba'b'}(x,x')
&\approx&   \frac{1}{2}\sum _{\ell=2}^{\infty}\sum_{\sigma}
(\ell+3)!(\ell-2)!\left[(\ell-1)^2\left( \frac{1}{M^2} -
\frac{1}{3}\sum_{k=0}^2 \frac{1}{\ell+k}\right) \right.\nonumber \\ && \left. \times E^{1\ell \sigma}_{ab}(x)
E^{*1\ell \sigma}_{a'b'}(x')\right.\nonumber \\
&& \left.
- \left( \frac{1}{M^2} - \frac{\alpha}{5}\sum_{k=-1}^3\frac{1}{\ell+k}\right)
A^{1\ell \sigma}_{ab}(x) A^{*1\ell \sigma}_{a'b'}(x')\right] \nonumber \\
&& + \frac{1}{12}\sum _{\ell=2}^{\infty} \sum_{\sigma}
 \ell(\ell+2)(\ell+3)!(\ell-2)!\nonumber \\
 && \times
 \left[ - 4(\ell-1)^2\left(\frac{1}{M^2} + \frac{1}{2}-\frac{1}{3}\sum_{k=0}^2 \frac{1}{\ell+k}\right) \right.\nonumber \\ && \left. \times E^{0\ell \sigma}_{ab}(x)
E^{*0\ell \sigma}_{a'b'}(x')\right. \nonumber \\
&& \left. + \left(\frac{1}{M^2} + \frac{\alpha}{6} - \frac{\alpha}{5}\sum_{k=-1}^3\frac{1}{\ell+k}\right)
A^{0\ell \sigma}_{ab}(x)
A^{*0\ell \sigma}_{a'b'}(x')\right]. \label{k3}
\end{eqnarray}
It may be noted that if $E^{0\ell \sigma}_{ab} = A^{0\ell \sigma}_{ab}/ 2(\ell-1)$ and $E^{1\ell \sigma}_{ab} = A^{1\ell \sigma}_{ab}/ (\ell-1)$, 
in the zero mass limit then the $1/M^2$ singularity will be absent in this limit. 
In fact we show in Appendix B that to the first order in $M^2$,   we have 
\begin{eqnarray}
E^{0\ell \sigma}_{ab} &=& \frac{1}{2(\ell-1)}\left[A^{0\ell \sigma}_{ab} + M^2 k_{ab}^{0\ell \sigma}\right], \nonumber \\
E^{1\ell \sigma}_{ab}&=&\frac{1}{(\ell-1)}\left[A^{1\ell \sigma} _{ab} + M^2 k_{ab}^{1\ell \sigma}  \right].\label{k2}
\end{eqnarray}
The exact form of $k^{n\ell \sigma}_{ab}$ which is $M^2$-independent, for $n =0, 1 $ is also given in Appendix B.
Now if we substitute  Eq. $(\ref{k2})$ into Eq. $(\ref{k3})$, and define $N^{n\ell \sigma}_{aba'b'}$ for $n =0, 1$ as   
\begin{equation}
 N^{n\ell \sigma}_{aba'b'}(x,x') =
 A^{n\ell \sigma}_{ab}(x)A^{*n\ell \sigma}_{a'b'}(x') + A^{n\ell \sigma}_{ab}(x)k^{*n\ell \sigma}_{a'b'}(x') + k^{n\ell \sigma}_{ab}(x)A^{*n\ell \sigma}_{a'b'}(x'),
\end{equation}
then in the $M^2\to 0$ limit,  we have 
\begin{equation}
L_{aba'b'}(x,x') = \sum _{\ell=2}^{\infty}\sum_{\sigma}\left[w_{0\ell} N^{0\ell \sigma}_{aba'b'}(x,x') + w_{1\ell} N^{1\ell \sigma}_{aba'b'}(x,x')\right],
\end{equation}
where 
\begin{eqnarray}
w_{0\ell} &=&  \frac{1}{12}
 \ell(\ell+2)(\ell+3)!(\ell-2)!\times \nonumber \\
 && 
\left[-\frac{1}{2}+\frac{1}{3}\sum_{k=0}^2 \frac{1}{\ell+k}
 +\frac{\alpha}{6} - \frac{\alpha}{5}\sum_{k=-1}^3\frac{1}{\ell+k}\right],\nonumber \\
 w_{1\ell} &=&  \frac{1}{2}
(\ell+3)!(\ell-2)!
\left[\frac{\alpha}{5}\sum_{k=-1}^3\frac{1}{\ell+k}
- \frac{1}{3}\sum_{k=0}^2 \frac{1}{\ell+k}\right].
\end{eqnarray}
We can now write $L_{aba'b'}$ as, 
\begin{eqnarray}
 L_{aba'b'}(x,x') &=& \nabla_{a} v^1_{ba'b'}(x,x') + \nabla_{b} v^1_{aa'b'}(x,x')\nonumber \\ &&  +\nabla_{a'} v^2_{abb'}(x,x') +\nabla_{b'} v^2_{aba'}(x,x'), 
\end{eqnarray}
where for $n =1,2 $, 
\begin{eqnarray}
 v^1_{ba'b'}(x,x') &=& 
\sum_{n =0}^1\sum _{\ell=2}^{\infty}\sum_{\sigma} w_{n\ell }A^{n\ell \sigma}_{b} (x)\left[\frac{1}{2} A^{n\ell \sigma}_{a'b'}(x')  +  k^{n\ell \sigma}_{a'b'}(x')
 \right],\nonumber \\
 v^2_{abb'}(x,x') &=& 
\sum_{n =0}^1\sum _{\ell=2}^{\infty}\sum_{\sigma} w_{n\ell }A^{n\ell \sigma}_{b'}(x')\left[\frac{1}{2} A^{n\ell \sigma}_{ab}(x)   +  k^{n\ell \sigma}_{ab}(x)   \right].
\end{eqnarray}

There  is a  finite contribution  to the 
Wightman graviton two-point function 
coming from the $\ell= 1$ vector modes. In the small mass approximation
this finite contribution  can be  written as 
\begin{eqnarray}
  M_{aba'b'}(x,x') &\approx&   \sum_{\sigma}
- \Gamma\left(\frac{\alpha M^2}{5}\right)\Gamma\left(5 -\frac{\alpha M^2}{5}\right)
 \nonumber \\&&\times
\frac{A^{11 \sigma}_{ab}(x)
A^{*11 \sigma}_{a'b'}(x')}{2M^2} \nonumber \\
&&  + \sum_{\sigma} \,  \Gamma\left(\frac{\alpha M^2}{5}\right)
\Gamma\left(5 -\frac{\alpha M^2}{5}\right)\nonumber \\&& \times
\frac{3A^{01 \sigma}_{ab}(x)
A^{*01 \sigma}_{a'b'}(x')}{2M^2(6-\alpha M^2)} \nonumber \\ 
&\approx& \frac{1}{\alpha M^4} \sum_{n=0}^{1} \sum_{\sigma} b_n A^{n 1 \sigma}_{ab}(x)A^{n 1 \sigma * }_{a'b'}(x'),
\end{eqnarray}
where $b_0 =30$ and $b_1 =-60$.  
This  has a finite zero mass limit because
 $A^{n 1 \sigma}_{ab} = \nabla_a A^{n 1 \sigma}_b + \nabla_b A^{n 1 \sigma}_a =0$ in the zero mass  limit for $ \ell =1$,
since for $L=1$ these are Killing vectors. 
Expanding $A^{n 1 \sigma}_{ab}$,  we get 
\begin{equation}
 A^{n 1 \sigma}_{a} (x)= A^{n 1 \sigma}_{a}(x)\left.\right|_{L=1} + (L-1)\frac{\partial A^{n 1 \sigma}_a(x)}{\partial L}\left.\right|_{L=1}. 
\label{xabcd}
\end{equation}
Now from Eqs. $(\ref{xabcd1})$ and $(\ref{xabcd})$, we get 
\begin{equation}
 A^{n 1 \sigma}_{a}(x) = A^{n 1 \sigma}_{a}(x)\left.\right|_{L=1} + \alpha M^2 X^{n 1 \sigma}_a(x),
\end{equation}
where
\begin{equation}
 X^{n 1 \sigma}_a(x) = -\frac{1}{5} \frac{\partial A^{n 1 \sigma}_a(x)}{\partial L}\left.\right|_{L=1}.
\end{equation}
Thus  $A^{n 1 \sigma}_{ab}$ is of order $\alpha M^2$. 
So in the zero mass limit, we obtain  
\begin{eqnarray}
M_{aba'b'}(x,x') &=& \sum_{n =0}^1 \sum_\sigma \alpha b_n  (\nabla_a X^{n 1 \sigma}_b + \nabla_b X^{n 1 \sigma}_a)
(\nabla_{a'}X^{n 1 \sigma *}_{b'} + \nabla_{b'}X_{a'}^{n 1 \sigma *}) \nonumber \\
&=&    \nabla_{a} x^1_{ba'b'}(x,x') + \nabla_{b} x^1_{aa'b'}(x,x')\nonumber \\ && 
 +\nabla_{a'} x^2_{abb'}(x,x') +\nabla_{b'} x^2_{aba'}(x,x'), 
\end{eqnarray}
where
\begin{eqnarray}
   x^1_{ba'b'}(x,x') &=&\sum_{n =0}^1 \sum_\sigma  \frac{b_n \alpha}{2} X^{n 1 \sigma}_b(\nabla_{a'}X^{n 1 \sigma *}_{b'} + 
\nabla_{b'}X^{n 1 \sigma *}_{a'}), \nonumber \\
x^2_{abb'}(x,x')&=&\sum_{n =0}^1 \sum_\sigma \frac{b_n \alpha}{2}  X^{n 1 \sigma *}_{b'} 
(\nabla_{a}X^{n 1 \sigma}_{b} + \nabla_{b}X^{n 1 \sigma}_{a}).
\end{eqnarray}

Thus, we have calculated the vector-tensor contribution to the
  Wightman graviton two-point function in this section.  In particular, we have shown that it is physically equivalent 
in linearized gravity to the physical Wightman two-point function of Ref. $\cite{phy}$.  
In the next section we shall calculate the scalar contribution to the  Wightman   graviton two-point function and find that it is pure gauge.

\section{Equations of Motion for Scalar Modes}
The scalar contribution to the Wightman two-point function has been 
calculated for $\alpha\beta - 3 \beta - 4 =0$ in Ref. $\cite{15}$. 
The scalar contribution to the  Wightman graviton two-point function  was found 
to be a pure gauge contribution  for these value of $\alpha$ and $\beta$. We shall calculate the scalar  contribution to the graviton 
Wightman two-point function for  arbitrary  values of $\alpha$ and $\beta$.
 
 Let us recall that the field equations for scalar modes,  given by  Eqs. $(\ref{xcv})$ and $(\ref{xcv1})$, are
\begin{eqnarray}
 \Box B & =& a_1 B + a_2 h, \label{B1} \\
 \Box h &= & a_3 h + a_4 B,\label{h1}
\end{eqnarray}
where
\begin{eqnarray}
 a_1 &=& \frac{12 - 2 \alpha M^2}{\alpha - 3}, \label{1}\\
 a_2 &=& \frac{\alpha\beta - 3 \beta - 4}{\beta(\alpha - 3)}, \label{12} \\
 a_3 &=& \frac{6(\alpha\beta - 3 \beta - 2)}{\alpha - 3}\left(1- \frac{M^2}{2}\right), \\
 a_4 &=& \frac{-9\beta (\alpha\beta - 3 \beta - 4)}{\alpha - 3}\left(1 - \frac{M^2}{2}\right)\left(1 - \frac{\alpha}{6}M^2\right).\label{2}
\end{eqnarray}
In deriving these equations we have assumed that $\alpha \neq 3, 0$ and $\beta \neq 0$.
 However, the final result for the scalar contribution to the  Wightman graviton two-point function is well
 defined even for these values of $\alpha$ and $\beta$. 

  Eqs. $(\ref{B1})$ and $(\ref{h1})$ suggest  that $B$ and $h$ are non-trivial linear combinations of two scalar 
fields with different masses unless $\alpha\beta - 3 \beta - 4 =0$, which leads to $a_2 =a_4 =0$. Let 
 a linear combination of $B$ and $h$, $h + \lambda B$, satisfy the free scalar field equation with a definite mass.  
Now we have
\begin{eqnarray}
 \Box (h +\lambda B ) &= & a_3 h + a_4 B + \lambda( a_1 B + a_2 h) \nonumber \\&=& (a_3 + \lambda a_2 )h + (a_4 + \lambda a_1) B. \label{xyzt}
\end{eqnarray}
As we want Eq. $(\ref{xyzt})$ to satisfy a free scalar field equation, with a definite mass,  the right hand side
 of Eq. $(\ref{xyzt})$ must be proportional to $h + \lambda B $. That is, 
\begin{equation}
 \frac{(a_4 + \lambda a_1) }{(a_3 + \lambda a_2 )} = \lambda.
\end{equation}
Thus we get the following quadratic equation in $\lambda$,
\begin{equation}
 \lambda^2 -\frac{a_1 - a_3}{a_2}\lambda -\frac{a_4}{a_2} = 0.
\end{equation}
Let $\lambda_1$ and $\lambda_2$  be the two roots of this equation.
Then the product of the roots is given by
\begin{equation}
 \lambda_1\lambda_2 = -\frac{a_4}{a_2}, \label{a44}
\end{equation}
and the sum of the roots is given by
\begin{equation}
 \lambda_1 + \lambda_2 = \frac{a_1 - a_3}{a_2}.
\end{equation}
Now the scalar fields  $\phi_1$ and $\phi_2$ defined by 
\begin{eqnarray}
 \phi_1 &=& h + \lambda_1 B, \label{sfe1}\\
 \phi_2 & =& h +\lambda_2 B,\label{sfe2}
\end{eqnarray}
satisfy 
\begin{eqnarray}
 \Box \phi_1 &=& (a_3 + \lambda_1 a_2 ) \phi_1, \label{mases}  \\
 \Box \phi_2 & =&(a_3 + \lambda_2 a_2 ) \phi_2. \label{mases1}
\end{eqnarray}
Solving Eqs. $(\ref{sfe1})$ and $(\ref{sfe2})$ for  $B$ and $h$, we get  
\begin{eqnarray}
  B &=& \frac{\phi_2 -\phi_1}{\lambda_2- \lambda_1},\label{sfe1a}
\\
 h &=& \frac{\lambda_2 \phi_1 - \lambda_1\phi_2}{\lambda_2 - \lambda_1 }.\label{sfe2a}
\end{eqnarray}
\section{Scalar Contributions }
Since the scalar part $C_{ab}(x)$ depends on two scalar fields $B(x)$ and $h(x)$, we need to find  $\langle 0|h(x)h(x')|0\rangle$, 
$\langle 0|B(x)B(x')|0\rangle$ and $\langle 0|B(x)h(x')|0\rangle$ and take the limit $M\to0$ in order to find $\langle 0|C_{ab}(x)C_{a'b'}(x')|0\rangle$.

Let two solutions to the equations of motion Eq. $(\ref{eqmm})$ be  $C^{\ell \sigma}_{bc}$ and $C^{\ell'\sigma'}_{b'c'}$,  given by 
\begin{eqnarray}
 C^{\ell\sigma}_{bc} & =& (\nabla_b \nabla_c - \frac{1}{4}g_{bc}\Box )B^{\ell\sigma} + \frac{1}{4}g_{bc}h^{\ell\sigma}, \nonumber \\
C^{\ell'\sigma'}_{b'c'} &= &(\nabla_{b'} \nabla_{c'} - \frac{1}{4}g_{b'c'}\Box' )B^{\ell'\sigma'} + \frac{1}{4}g_{b'c'}h^{\ell'\sigma'}. 
\end{eqnarray}
Here the covariant derivatives  $\nabla_{a}$ and $\nabla_{a'}$ only act on quantities at $x$ and $x'$, respectively. 
 We also define $\Box' = \nabla_{a'}\nabla^{a'}$ and $\Box = \nabla_{a}\nabla^{a}$. 
Also let  $\pi^{(\ell\sigma)abc}_S (x)$ and $\pi^{(\ell'\sigma')a'b'c'}_S(x')$ be momentum current conjugate  to
 $C^{\ell\sigma}_{bc}$  and $C^{\ell'\sigma'}_{b'c'}$, respectively. We can now calculate the scalar contribution to the symplectic product 
by substituting these in Eq. $(\ref{jj})$. This  symplectic product  for the scalar part is calculated in  Appendix A, and is given by 
\begin{equation}
 (C^{\ell\sigma}, C^{\ell'\sigma'}) = - K_1 \langle B^{\ell\sigma}, B^{\ell'\sigma'} \rangle - K_2\langle h^{\ell\sigma}, h^{\ell'\sigma'}\rangle, \label{spsp}
\end{equation}
where
\begin{equation}
 K_1 = \frac{3}{4(\alpha - 3)}(2 - M^2)(6 - \alpha M^2 ), \label{k1a}
\end{equation}
\begin{equation}
 K_2 = -\frac{1}{\beta^2 (\alpha - 3)},\label{k1ab}
\end{equation}
and where $\langle B^{\ell\sigma}, B^{\ell'\sigma'} \rangle$ and $\langle h^{\ell\sigma}, h^{\ell'\sigma'}\rangle$ are defined by Eq. $(\ref{kl})$. 
We note that $B^{\ell\sigma}$ and $h^{\ell\sigma}$ are linear combinations of modes with different masses, 
so the Klein-Gordon-type product $\langle B^{\ell\sigma}, B^{\ell'\sigma'} \rangle$ and $\langle h^{\ell\sigma}, h^{\ell'\sigma'}\rangle$
are  not conserved separately. 

Now let $\phi_1^{\ell\sigma}$ and $\phi_2^{\ell\sigma}$ be positive frequency modes satisfying Eqs. $(\ref{mases})$ and $(\ref{mases1})$,
 respectively. Let us also require that 
\begin{equation}
 \langle \phi_1^{\ell\sigma}, \phi_1^{\ell'\sigma'} \rangle = \langle\phi_2^{\ell\sigma},\phi_2^{\ell'\sigma'} \rangle
 = \delta^{\ell\ell'}\delta^{\sigma \sigma'},\label{aaaaaaa1}
\end{equation}
then by using Eq. $(\ref{twopointfunctionscalar})$, we have 
\begin{eqnarray}
 \Delta_{m_1^2}(x,x') &=&\sum_{\ell}\sum_{\sigma} \phi_1^{\ell\sigma }(x) \phi^{*\ell\sigma }_1(x'),\nonumber \\
\Delta_{m_2^2}(x,x') &=&\sum_{\ell}\sum_{\sigma} \phi_2^{\ell\sigma }(x) \phi^{*\ell\sigma }_2(x').\label{aaaaaaa2}
\end{eqnarray}
Here  $\Delta_{m^2}(x,x')$ is the standard two-point function of the scalar field  with squared mass $m^2$,  (see Sec. $2$ of Ref. $\cite{aj}$) and 
\begin{eqnarray}
 m_1^2 &=& a_3 + \lambda_1 a_2, \nonumber \\
 m_2^2 &=& a_3 + \lambda_2 a_2. \label{12we}
\end{eqnarray}
Now we can expand $\phi_1 (x)$ and $\phi_2 (x)$ as 
\begin{eqnarray}
 \phi_1 (x) &=& \sum_{\ell}\sum_{\sigma} \phi_1^{\ell\sigma}(x)a_{1\ell \sigma} +  \phi_1^{*\ell\sigma} (x)a^{\dagger}_{1\ell \sigma}, \nonumber \\ 
\phi_2 (x) &=&\sum_{\ell}\sum_{\sigma} \phi_2^{\ell\sigma}(x)a_{2\ell \sigma} +  \phi_2^{*\ell\sigma} (x)a^{\dagger}_{2\ell \sigma}. \label{pqr1}
\end{eqnarray}
We now  write the field $C_{ab}(x)$ as follows:
\begin{equation}
 C_{ab}(x)= C^{1}_{ab} (x)+ C^{2}_{ab}(x),
\end{equation}
where 
\begin{eqnarray}
 C^{1}_{ab} (x)&=& \hat{x}^1_{ab} \phi_1(x), \nonumber \\
C^{2}_{ab} (x) &=& \hat{x}^2_{ab} \phi_2(x), \label{aaaaaaa2a}
\end{eqnarray}
with the differential operators $\hat{x}^1_{ab}$ and $\hat{x}^2_{ab}$ defined by 
\begin{eqnarray}
\hat{x}^1_{ab} &=& \frac{-1}{\lambda_2 -\lambda_1} \left[ \left( \nabla_a \nabla_b -\frac{1}{4}g_{ab} \Box\right) - \frac{1}{4}g_{ab} \lambda_2  \right], 
\nonumber  \\ 
 \hat{x}^2_{ab} &=& \frac{1}{\lambda_2 -\lambda_1} \left[ \left( \nabla_a \nabla_b -\frac{1}{4}g_{ab} \Box\right) - \frac{1}{4}g_{ab} \lambda_1  \right].
\end{eqnarray}
We now expand the fields $C^1_{ab}(x)$ and $C^2_{ab}(x)$, as
\begin{eqnarray}
C^1_{ab}(x) &=&\sum_{\ell}\sum_{\sigma}C^{1\ell\sigma}_{ab}(x)a_{1\ell \sigma} + C^{*1\ell\sigma}_{ab} (x)a^{\dagger}_{1\ell \sigma}, \nonumber  \\
C^2_{ab} (x)&=& \sum_{\ell}\sum_{\sigma}C^{2\ell\sigma}_{ab}(x) a_{2\ell \sigma} + C^{*2\ell\sigma}_{ab}(x) a^{\dagger}_{2\ell \sigma}, \label{c1}
\end{eqnarray}
where 
\begin{eqnarray}
 C^{1\ell\sigma}_{ab} (x)&=& \hat{x}^1_{ab} \phi_1^{\ell\sigma}(x), \nonumber \\
C^{2\ell\sigma}_{ab} (x) &=& \hat{x}^2_{ab} \phi_2^{\ell\sigma}(x). \label{aaaaaaa2aa}
\end{eqnarray}
Now from Eq. $(\ref{spsp})$, we get 
\begin{eqnarray}
 (C^{1\ell\sigma} ,C^{1\ell'\sigma'})&=& - \frac{K_1+\lambda_2^2 K_2}{(\lambda_2-\lambda_1)^2}\langle \phi_1^{\ell\sigma}, \phi_1^{\ell'\sigma'} \rangle, \label{2121h} \\
 (C^{2\ell\sigma} ,C^{2\ell'\sigma'})&=&- \frac{K_1 + \lambda_1^2 K_2}{(\lambda_2-\lambda_1)^2}\langle \phi_2^{\ell\sigma}, \phi_2^{\ell'\sigma'}\rangle,\\
(C^{1\ell\sigma} ,C^{2\ell'\sigma'})&=&0. \label{2121h1}
\end{eqnarray}
Now we note from Eqs. $(\ref{12})$, $(\ref{2})$ and $(\ref{a44})$, 
\begin{eqnarray}
- \lambda_1 \lambda_2 K_2 &=& \frac{a_4}{a_2}K_2  \nonumber \\ &=& \frac{3}{4(\alpha - 3)}(2 - M^2)(6 - \alpha M^2 ) \nonumber \\ &=&  K_1. \label{944a}
\end{eqnarray}
 Substituting Eq. $(\ref{944a})$ into Eqs. $(\ref{2121h})$-$(\ref{2121h})$,and using Eq. $(\ref{aaaaaaa1})$,
we get 
\begin{eqnarray}
(C^{1\ell\sigma},C^{1\ell'\sigma'})&= & \frac{\lambda_2 K_2}{\lambda_1-\lambda_2}\langle\phi_1^{\ell\sigma},\phi_1^{\ell'\sigma'}\rangle \nonumber \\ 
&=&  \frac{\lambda_2 K_2}{\lambda_1-\lambda_2}\delta^{\ell\ell'} \delta^{\sigma \sigma' }
,\nonumber \\
(C^{2\ell\sigma},C^{2\ell'\sigma'})&=& \frac{\lambda_1 K_2}{\lambda_2-\lambda_1}\langle\phi_2^{\ell\sigma}, \phi_2^{\ell' \sigma'}\rangle\nonumber \\ 
&=&  \frac{\lambda_1 K_2}{\lambda_2-\lambda_1} \delta^{\ell\ell'}\delta^{\sigma \sigma' }.
\end{eqnarray}
Thus  the only non-zero commutators are given by 
\begin{eqnarray}
 \left[a_{1\ell \sigma}, a_{1\ell'\sigma'}^{\dagger }\right] &=& \frac{\lambda_1-\lambda_2}{\lambda_2 K_2}\delta_{\ell\ell'}\delta_{\sigma \sigma'}, \nonumber \\ 
 \left[a_{2\ell\sigma}, a_{2\ell'\sigma'}^{\dagger }\right]&=& \frac{\lambda_2 - \lambda_1}{\lambda_1 K_2}\delta_{\ell\ell'}\delta_{\sigma \sigma'}. \label{c2}
\end{eqnarray}
Now using Eqs. $(\ref{wighthig})$ and $(\ref{c2})$,   we get
\begin{eqnarray}
 \langle 0|C_{ab}(x)C_{a'b'}(x')|0\rangle 
& = & \sum_{\ell}\sum_{\sigma}\left[ \frac{\lambda_1-\lambda_2}{\lambda_2 K_2}C^{1\ell\sigma}_{ab}(x)C^{*1\ell\sigma}_{a'b'}(x') 
\right.\nonumber \\ && \left. + \frac{\lambda_2 - \lambda_1}{\lambda_1 K_2}C^{2\ell\sigma}_{ab}(x)C^{*2\ell\sigma}_{a'b'}(x') \right]. \label{956}
\end{eqnarray}
Now from Eqs. $(\ref{aaaaaaa2})$, $(\ref{aaaaaaa2a})$ and $(\ref{956})$, we get 
\begin{eqnarray}
&&\langle 0|C_{ab}(x)C_{a'b'}(x')|0\rangle \nonumber \\
 &=& \sum_{\ell}\sum_{\sigma}\left[ \frac{\lambda_1-\lambda_2}{\lambda_2 K_2}x^1_{ab}\phi^{\ell\sigma}_{1 }(x)x^1_{a'b'}\phi^{*\ell\sigma}_{1 }(x')  
+ \frac{\lambda_2 - \lambda_1}{\lambda_1 K_2}x^2_{ab}\phi^{\ell\sigma}_{2 }(x)x^2_{a'b'}\phi^{*\ell\sigma}_{2 }(x')  \right]\nonumber \\ 
&=& \left[ \frac{\lambda_1-\lambda_2}{\lambda_2 K_2}x^1_{ab}x^1_{a'b'}\Delta_{m_1^2}(x,x')  
+ \frac{\lambda_2 - \lambda_1}{\lambda_1 K_2}x^2_{ab}x^2_{a'b'}\Delta_{m_2^2}(x,x')   \right]. 
\end{eqnarray}
Thus  we have 
\begin{eqnarray}
\langle 0|\phi_1(x)\phi_1(x')|0\rangle
& = & \frac{\lambda_1-\lambda_2}{\lambda_2 K_2}\Delta_{m_1^2}(x,x'), \nonumber \\
\langle 0|\phi_2(x)\phi_2(x')|0\rangle
& = &  \frac{\lambda_2 - \lambda_1}{\lambda_1 K_2}\Delta_{m_2^2}(x,x'). \label{z01}
\end{eqnarray}

Next we use these to find the Wightman functions for $B(x)$ and $h(x)$.  
We define $\Delta^{\rm av}(x,x') $ and $\Delta^{\rm diff}(x,x')$ as: 
\begin{eqnarray}
\Delta^{\rm av}(x,x') & = & \frac{\Delta_{m_1^2}(x,x')+\Delta_{m_2^2}(x,x')}{2},\\
\Delta^{\rm diff}(x,x') & = & \frac{\Delta_{m_2^2}(x,x')-\Delta_{m_1^2}(x,x')}{m_2^2-m_1^2} \nonumber \\ &=&
\frac{\Delta_{m_2^2}(x,x')-\Delta_{m_1^2}(x,x')}{a_2 (\lambda_2 - \lambda_1)}.
\end{eqnarray}
Then using Eqs. $(\ref{sfe1a})$, $(\ref{sfe2a})$ and $(\ref{z01})$, we obtain
\begin{eqnarray}
\langle 0|h(x)h(x')|0\rangle & = & -\frac{1}{K_2}\left[
 \Delta^{\rm av}(x,x')  \right.  \nonumber \\&& \left. -  a_2\frac{\lambda_1+\lambda_2}{2}\Delta^{\rm diff}(x,x')\right],\\
\langle 0|B(x)h(x')|0\rangle & = & \frac{-a_2}{K_2}\Delta^{\rm diff}(x,x'),\\
\langle 0|B(x)B(x')|0\rangle & = &
\frac{1}{K_2}\left[  \frac{1}{\lambda_1\lambda_2}\Delta^{\rm av}(x,x')
\right. \nonumber \\ &&\left. + \frac{a_2}{2}\left(\frac{1}{\lambda_1} + \frac{1}{\lambda_2}\right)
\Delta^{\rm diff}(x,x')\right].
\end{eqnarray}
We have, from Eqs. $(\ref{12})$, $(\ref{k1a})$ and  $(\ref{k1ab})$
\begin{eqnarray}
\frac{1}{K_2} & = & - \beta^2(\alpha-3),\\
\frac{a_2}{K_2} & = & - \beta(\alpha\beta-3\beta-4).
\end{eqnarray}
In the zero mass  limit, we obtain from Eqs $(\ref{1})$-$(\ref{2})$
\begin{eqnarray}
\lim_{M^2 \to 0}\lambda_1\lambda_2 &= & 9\beta^2,\\
\lim_{M^2 \to 0}\frac{\lambda_1+\lambda_2}{2} & =& - 3\beta,
\end{eqnarray}
This implies 
\begin{equation}
 \lim_{M^2 \to 0}\lambda_1 = \lim_{M^2 \to 0}\lambda_2 = -3 \beta,
\end{equation}
and hence by Eq. $(\ref{12we})$, we get 
\begin{equation}
 \lim_{M^2 \to 0}m_1^2  = \lim_{M^2 \to 0}m_2^2 = 3\beta .
\end{equation}
Now we define $\Delta^{(1)}_{m^2}(x,x')$ as
\begin{equation}
\Delta^{(1)}_{m^2}(x,x') = -\frac{\partial\ }{\partial m^2}\Delta_{m^2}(x,x').
\end{equation}
Note that 
\begin{eqnarray}
 \lim_{M^2 \to 0}\Delta^{\rm av}(x,x') & = &\Delta_{3\beta} (x,x'), \nonumber \\ 
 \lim_{M^2 \to 0}\Delta^{\rm diff}(x,x') & = &\Delta^{(1)}_{3\beta} (x,x').
\end{eqnarray}
 So   in the zero mass limit, we have
\begin{eqnarray}
\langle 0|h(x)h(x')|0\rangle & = & \beta^2(\alpha-3)\Delta_{3\beta}(x,x')\nonumber \\ &&  -3\beta^2(\alpha\beta-3\beta-4)\Delta^{(1)}_{3\beta }(x,x'),\\
\langle 0|B(x)h(x')|0\rangle  & = &
-\beta(\alpha\beta-3\beta-4)\Delta^{(1)}_{3\beta }(x,x'),\\
\langle 0|B(x)B(x')|0\rangle & = & -\frac{\alpha-3}{9}\Delta_{3\beta }(x,x')\nonumber \\ &&- \frac{\alpha\beta-3\beta-4}{3}\Delta^{(1)}_{3\beta }(x,x').
\end{eqnarray}
By differentiating the formula $(\Box - m^2)\Delta_{m^2}(x,x')=0$ with respect to $m^2 $, we obtain
\begin{equation}
\Box \Delta^{(1)}_{m^2}(x,x') = m^2\Delta^{(1)}_{m^2}(x,x')-\Delta_{m^2}(x,x').
\end{equation}
So the   scalar part of the graviton Wightman two-point function, which is
\begin{equation}
  G^{(S)}_{aba'b'}(x,x') = \langle 0| C_{ab}(x) C_{a'b'} (x') |0 \rangle,
\end{equation}
can be written as 
\begin{equation}
   G^{(S)}_{aba'b'}(x,x') = G^{(S0)}_{aba'b'}(x,x') + G^{(S1)}_{aba'b'}(x,x') + G^{(S2)}_{aba'b'}(x,x'),
\end{equation}
where $G^{(S0)}_{aba'b'}(x,x'), G^{(S1)}_{aba'b'}(x,x')$ and $G^{(S2)}_{aba'b'}(x,x')$ are given by
\begin{eqnarray}
G^{(S0)}_{aba'b'}(x,x') &=& \left(\nabla_{a'}\nabla_{b'}-\frac{1}{4}g_{a'b'}(x')\Box' \right)\nonumber \\
&&\times\left(\nabla_a \nabla_b - \frac{1}{4}g_{ab}(x)\Box\right) \langle 0|B(x)B(x')|0\rangle 
\\ \nonumber & =& [\nabla_a\nabla_b\nabla_{a'}\nabla_{b'}\langle 0|B(x)B(x')|0\rangle \nonumber \\
&& \,\,\, + \frac{1}{4}(g_{ab}\nabla_{a'}\nabla_{b'}+g_{a'b'}\nabla_{a}\nabla_b)]
\left[ \frac{4}{3}\Delta_{3\beta }(x,x') \right. \nonumber \\
&& \left.+ \beta(\alpha\beta-3\beta-4)\Delta_{3\beta }^{(1)}(x,x')\right]
\nonumber \\
&& \,\,\,-\frac{1}{16}g_{ab}g_{a'b'}
[ -\beta(\alpha\beta-3\beta-8)\Delta_{3\beta }(x,x')\nonumber \\
&& + 3\beta^2(\alpha\beta-3\beta-4)
\Delta_{3\beta }^{(1)}(x,x')], \\ 
G^{(S1)}_{aba'b'}(x,x')&=& \frac{1}{4}g_{a'b'}\left(\nabla_a\nabla_b - \frac{1}{4}g_{ab}\Box\right)\langle 0|B(x)h(x')|0\rangle \nonumber \\
&=&  - \frac{1}{4}\beta(\alpha\beta-3\beta-4)g_{a'b'}\nabla_a\nabla_b \Delta_{3\beta }^{(1)}(x,x')\nonumber \\
&&\,\,\, + \frac{1}{16}\beta(\alpha\beta-3\beta-4)g_{ab}g_{a'b'}
\left[ - \Delta_{3\beta }(x,x') \right. \nonumber \\  
&&\left.+ 3\beta \Delta_{3\beta }^{(1)}(x,x')\right], \\  
G^{(S2)}_{aba'b'}(x,x')&=& \frac{1}{16}g_{ab}(x)g_{a'b'}(x')\langle 0|h(x)h(x')|0\rangle \nonumber\\
 & = & \frac{1}{16}g_{ab}(x)g_{a'b'}(x') \left[ \beta^2(\alpha-3)\Delta_{3\beta }(x,x')\right.\nonumber \\ && \left.
  -3\beta^2(\alpha\beta-3\beta-4)\Delta^{(1)}_{3\beta }(x,x')\right].  
\end{eqnarray}
The contributions proportional to $g_{ab}g_{a'b'}$ cancel out  and the scalar part of the  Wightman graviton two-point function can now be written as 
\begin{eqnarray}
  G^{(S)}_{aba'b'} (x,x') &=&  \nabla_a \nabla_b \nabla_{a'}\nabla_{b'}\langle 0|B(x)B(x')|0\rangle  \nonumber \\&& +
\frac{1}{3}[ g_{ab}\nabla_{a'}\nabla_{b'} + g_{a'b'}\nabla_a \nabla_b]  \Delta_{3\beta }(x,x') \nonumber \\ &=& 
\nabla_{a} s^1_{ba'b'}(x,x') + \nabla_{b} s^1_{aa'b'}(x,x')\nonumber \\ &&  +\nabla_{a'} s^2_{abb'}(x,x') +\nabla_{b'} s^2_{aba'}(x,x'),
\end{eqnarray}
where 
\begin{eqnarray}
 s^1_{ba'b'}(x,x')& =& \frac{1}{4} \nabla_b \nabla_{a'}\nabla_{b'}\langle 0|B(x)B(x')|0\rangle  \nonumber \\&& + \frac{1}{6}g_{a'b'}\nabla_b  \Delta_{3\beta }(x,x'),
\nonumber \\
s^2_{abb'}(x,x')& =& \frac{1}{4} \nabla_a \nabla_{b}\nabla_{a'}\langle 0|B(x)B(x')|0\rangle  \nonumber \\&& + \frac{1}{6}g_{ab}\nabla_{b'}  \Delta_{3\beta }(x,x').
\end{eqnarray}

Thus, we have calculated the scalar contribution to the  Wightman graviton two-point function and found it to be of pure gauge form.  

\section{Full Graviton Two-Point Function}
Now the full  Wightman graviton two-point function can be written as:
\begin{eqnarray}
 G_{aba'b'} (x,x')&=&G^{(S)}_{aba'b'}(x,x')  +  G^{(TV)}_{aba'b'}(x,x')  \nonumber \\ &=& 
 P_{aba'b'}(x,x') +  \nabla_{a} s^1_{ba'b'}(x,x') + \nabla_{b} s^1_{aa'b'}(x,x') \nonumber \\ &&  +\nabla_{a'} s^2_{abb'}(x,x') 
+\nabla_{b'} s^2_{aba'}(x,x')
  + \nabla_{a} v^1_{ba'b'}(x,x')\nonumber \\ && + \nabla_{b} v^1_{aa'b'}(x,x') +\nabla_{a'} v^2_{abb'}(x,x') 
 +\nabla_{b'} v^2_{aba'}(x,x')\nonumber \\ 
&& + \nabla_{a} x^1_{ba'b'}(x,x') + \nabla_{b} x^1_{aa'b'}(x,x') +\nabla_{a'} x^2_{abb'}(x,x') \nonumber \\ &&
 +\nabla_{b'} x^2_{aba'}(x,x')  \nonumber \\ &= &  
                                              P_{aba'b'} (x,x')+ Q_{aba'b'}(x,x'),
\end{eqnarray}
 where $Q_{aba'b'}(x,x')$ is a pure gauge contribution  given by  Eq. $(\ref{Q})$  with $K^1_{ba'b'} = s^1_{ba'b'} + v^1_{ba'b'} + x^1_{ba'b'} $ and  
 $K^2_{abb'} = s^2_{abb'} + v^2_{abb'} + x^2_{abb'}$. 
 The physical two-point function  obtained by fixing all the gauge degrees of freedom in Ref. $\cite{phy}$ is denoted by $P_{bcb'c'}(x,x')$. 
Therefore the covariant  Wightman graviton two-point function is physically equivalent in linearized gravity to the physical Wightman two-point function.

We established the equivalence of the vector-tensor part of the  Wightman graviton two-point function to the physical two-point function $P_{aba'b'}(x,x')$ of Ref.
$\cite{phy}$ in detail using the mode expansion. In Ref. $\cite{15a}$ it was shown that the scalar sector is of pure gauge form for a specific value of 
$\alpha$ and $\beta$. Here we 
have generalized this to arbitrary value of $\alpha$ and $\beta$.  

Now suppose that a covariant Wightman graviton two-point function has an infrared cut-off $\epsilon$ and that it is divergent in the limit $\epsilon\to 0$.  
 As it is physically equivalent in linearized gravity to $P_{aba'b'}(x,x')$, and as $P_{aba'b'}(x,x')$ is not infrared divergent, 
 the two-point function of a local gauge-invariant 
tensor field will not depend on $\epsilon$, i.e. will not be infrared divergent.   This also means that the infrared divergences in the Wightman 
graviton  two-point function 
for a certain gauge choice, for example as noted in Ref. $\cite{7cha5}$, will not appear in the
Wightman two-point function of any local gauge-invariant tensor field. So in this sense these infrared divergences in the Wightman 
graviton  two-point function are  gauge artefacts in the context of linearized gravity.

\chapter{Conclusion}
This thesis consisted of two parts. In the first part we studied the Feynman ghost propagators for Faddeev-Popov ghosts for 
Yang-Mills theories and perturbative quantum gravity in de Sitter
spacetime and  in the second part we analyzed the equivalence of the covariant Wightman graviton 
 two-point function to the physical one obtained in Ref. $\cite{phy}$.

We found that the ghosts and anti-ghosts in Yang-Mills theories satisfy the minimally-coupled massless scalar field equation and as a result 
the Feynman propagator for Yang-Mills theories in de~Sitter spacetime is IR divergent. 
We also found that the Feynman propagator for perturbative quantum gravity is also IR divergent. 

However it was shown  that if we regularize these propagators by adding a small mass then the modes 
responsible for these IR divergences do not contribute to the time-ordered product of the fields.
 We thus found the effective propagators by first subtracting these regularized modes which cause the IR divergences 
and then taking the zero mass limit. This way we obtained the effective IR finite propagators 
for Yang-Mills theories and perturbative quantum gravity. 
 
It  seems  likely that the use of these effective propagators will lead to  consistent perturbative theories. 
However the consistency of these theories is not obvious because it is not clear how the removal of  some of the  modes from the Faddeev-Popov ghosts will 
affect the BRST symmetry of these theories. We can not naively 
remove the  constant  or Killing modes from these theories because if we do so, they will be brought back as BRST transformation of some other modes. 
This point needs further investigation and we would  have to modify the original BRST transformations for consistency.  
It will be interesting to find the modified BRST transformations for these theories after constant or the Killing modes are removed.

 We have dealt only with integer-spins 
and it will also be interesting to derive the  propagators for the half-integer spin ghost fields in supergravity.
 At present it is not clear if these propagators will be IR finite or IR divergent. It is also not clear that if they are
IR divergent, then  we can still use the method we used for integer spin fields to remove the IR divergences. 

We   analyzed  the Wightman graviton two-point function with arbitrary gauge parameters 
in the second part of this thesis. We found  that the covariant Wightman graviton 
two-point function  is physically equivalent to the physical one obtained in Ref. $\cite{phy}$, in the sense that they result
 in the same two-point function of any local gauge-invariant quantity. We then argued that as the
  physical Wightman  graviton two-point function obtained in Ref. $\cite{phy}$ 
 is known to be IR finite and  the covariant Wightman graviton  two-point  function is  physically equivalent to it,
 so any IR divergences in the Wightman graviton  two-point  function, for example as noted in Ref. $\cite{7cha5}$, 
has to be a gauge artefacts at least in the context of linearized gravity. 

We should  emphasize that we have only studied the free theory and   at present have nothing to say about Wightman two-point functions
 in the interacting theories. It will be interesting to analyze what happens when  interactions are introduced. 

 It will also be interesting to analyze the  equivalence of the covariant Wightman graviton two-point function to the physical one in anti-de~Sitter 
spacetime in arbitrary 
dimensions. In fact we can also analyze  this equivalence  for  Wightman graviton two-point functions  in four dimensional 
flat spacetime. This is widely believed to hold but 
 has not been done yet and it will be interesting to see if a result 
similar to the one obtained in case of de~Sitter spacetime  holds for flat spacetime also.

\appendix

\chapter{Symplectic Product}
\section{ General Procedure}
We want to relate the symplectic product  for scalar and vector  modes to the Klein-Gordon type product for them. For this purpose the following
 formula is useful 
\begin{equation}
  \nabla^a\nabla_b \phi \nabla^b \phi' -(\phi\leftrightarrow\phi') \cong - \nabla^a \phi \Box \phi'  + (\phi\leftrightarrow\phi'), \label{lpo}
\end{equation}
where  $(\phi\leftrightarrow\phi')$ is defined to  mean that the preceding expression is repeated but with  
 $\phi$ replaced by $\phi'$ and $\phi'$ replaced by $\phi$, 
and  $\cong$ indicates equality up to terms that vanish when they are integrated over the space. 
To show this we first note that 
\begin{equation}
 \nabla^a\nabla_b \phi \nabla^b \phi' -(\phi\leftrightarrow\phi') = Q^a  - [\nabla^a \phi \Box \phi'  -(\phi\leftrightarrow\phi')],
 \end{equation}
where $Q^a$ is a total divergence of an antisymmetric tensor:
\begin{equation}
 Q^a = \nabla_b [\nabla^a \phi \nabla^b \phi' - \nabla^b \phi \nabla^a \phi' ].
\end{equation}
This total divergence of an antisymmetric tensor can be converted into a surface term by Gauss's divergence theorem on the spatial section $S^3$ because 
\begin{eqnarray}
 \int d \Sigma_b \nabla_a F^{ab} &=& \int d^3 {\bf{x}} \sqrt{-g} \nabla_a F^{a0} \nonumber \\ &=& \int d^3{\bf{x}} \partial_i (\sqrt{-g} F^{i0}). 
\end{eqnarray}
 As there is no boundary on $S^3$, we just drop such terms. 
We have used Eq. $(\ref{d3x})$ here. 
\section{Symplectic Product for the Scalar Modes}

We shall  calculate the contribution to the symplectic product coming from scalar sector here.
Let $B^{\ell \sigma}, B^{\ell' \sigma'}, h^{\ell \sigma}, h^{\ell' \sigma'}$ be solutions to Eqs.  $(\ref{B12})$ 
and  $(\ref{h12})$ and let 
\begin{eqnarray}
C^{\ell \sigma}_{bc} & =& \frac{1}{4}g_{bc} h^{\ell \sigma} + \frac{1}{4} B^{\ell \sigma}_{bc}, \nonumber \\
C^{\ell' \sigma'}_{bc} & = & \frac{1}{4}g_{bc} h^{\ell' \sigma'} +\frac{1}{4}B^{\ell'\sigma'}_{bc},
\end{eqnarray}
where 
\begin{eqnarray}
B^{\ell \sigma}_{bc}& =& \left(4\nabla_b \nabla_c - g_{bc}\Box \right)B^{\ell \sigma}, \nonumber \\
B^{\ell'\sigma'}_{bc} & = &  \left(4\nabla_{b} \nabla_{c} -g_{bc}\Box \right)B^{\ell'\sigma'}.
\end{eqnarray}
We denote $C^{\ell \sigma}_{bc}$, $B^{\ell \sigma}_{bc}$, $B^{\ell \sigma}$, $h^{\ell \sigma}$ by $C_{ab}$, $B_{bc}$, $B$, $h$ and
 $C^{*\ell' \sigma'}_{bc}$, $B^{*\ell' \sigma'}_{bc}$, $B^{*\ell' \sigma'}$, $h^{*\ell' \sigma'}$ by $C'_{ab}$, $B'_{bc}$ $B'$, $h'$ here for simplicity of notation. 
However it may be noted that they are modes and not fields which will be promoted to operators. 

Momentum current conjugate to  $C_{bc}$ can be found by substituting $h_{cb} = C_{bc}$  in  Eq. $(\ref{p2ia})$. Now if the 
 momentum current conjugate to  $C_{bc}$ is $\pi^{abc}_{(C)}$, then we have 
\begin{equation}
  \pi^{abc}_{(C)} = \pi^{abc}_{(B)} + \pi^{abc}_{(h)},
\end{equation}
where  we define $\pi^{abc}_{(h)}$ to be the contribution coming from the pure trace part of the scalar modes 
 which is obtained by substituting $h_{bc} =  g_{bc}h/4$
 in Eq. $(\ref{p2ia})$ and $\pi^{abc}_{(B)}$ as the contribution coming from the traceless part of the scalar
 modes which is obtained by substituting $h_{bc} =  B_{bc}/4$ in  Eq. $(\ref{p2ia})$.
Thus we get 
\begin{eqnarray}
\sqrt{-g}\pi^{abc}_{(h)} &=& \frac{4+3\beta - \alpha\beta}{8\alpha\beta}(g^{ab}\nabla^c h + g^{ac}\nabla^b h)\nonumber \\ 
&& + \left(\frac{1}{4} + \frac{1+\beta}{4\alpha\beta} - \frac{(1+\beta)^2}{\alpha\beta^2}\right)g^{bc}\nabla^a h,
\\
\sqrt{-g} \pi^{abc}_{(B)}&=&- \frac{1}{2}\nabla^a \left[ \nabla^c \nabla^b B - \frac{1}{4} g^{cb}\Box B\right] +\left( \frac{1}{2}-\frac{1}{2\alpha}\right) \times
 \nonumber \\ &&  \left[g^{ac}\nabla_d \left[ \nabla^d \nabla^b B - \frac{1}{4} g^{db}\Box B\right]  + g^{ab}\nabla_d \left[ \nabla^d \nabla^c B -
 \frac{1}{4} g^{dc}\Box B\right]\right]   \nonumber \\ && + \left(- \frac{1}{2} +\frac{\beta +1}{\beta \alpha}\right) 
\left[ g^{bc}\nabla_d \left[ \nabla^d \nabla^a B - \frac{1}{4} g^{da}\Box B\right] \right]. 
\end{eqnarray}

We can simplify $\pi^{abc}_{(B)}$ by using 
\begin{equation}
\nabla^d \left(\nabla_d\nabla_a - \frac{1}{4}g_{da}\Box \right)B = \frac{3}{4}\nabla^a(\Box + 4)B,
\end{equation}
as follows,  
\begin{eqnarray}
\sqrt{-g} \pi^{abc}_{(B)}
& = & \frac{3}{8}\left(1-\frac{1}{\alpha}\right)(g^{ab}\nabla^c + g^{ac}\nabla^b)(\Box + 4)B\nonumber \\
&& + \frac{3}{4}\left(-\frac{1}{2}+\frac{1+\beta}{\alpha\beta}\right)g^{bc}\nabla^a(\Box + 4)B\nonumber \\
&&- \frac{1}{2}\nabla^a\left(\nabla^b\nabla^c - \frac{1}{4}g^{bc}\Box \right)B.
\end{eqnarray}
If $J^a_C$ is the  contribution to $J^a$ defined in Eq. $(\ref{jj})$, from the scalar modes, then we have 
\begin{eqnarray}
 J^a_C &=&  - \frac{i}{\sqrt{-g} }[C_{cb} \pi'^{abc}_{(C)} - C'_{cb}\pi^{abc}_{(C)}] \nonumber \\ &=& 
-i J^a_{(1)} -i J^a_{(2)} -i J^a_{(3)}, 
\end{eqnarray}
where 
\begin{eqnarray}
 J^a_{(1)} &=&     \frac{1}{4 \sqrt{-g} }[hg_{cb} \pi'^{abc}_{(h)} - h'g_{cb}\pi^{abc}_{(h)}],  \\
 J^a_{(2)} &=&  \frac{1}{4\sqrt{-g} }[hg_{cb} \pi'^{abc}_{(B)} - h'g_{cb}\pi^{abc}_{(B)}\nonumber   \\ && 
 \, \, \, \, \,\, \, \, \, \, \,\, \, \, \, \, \,  +  B_{bc}   \pi'^{abc}_{(h)}  -  B'_{bc} \pi^{abc}_{(h)}],\\
 J^a_{(3)} &=&   \frac{1}{4\sqrt{-g} }[ B_{bc}   \pi'^{abc}_{(B)}  -  B'_{bc} \pi^{abc}_{(B)}]. 
\end{eqnarray}
Thus we have 
\begin{eqnarray}
 J^{a}_{(1)} & =& \left(\frac{4+3\beta-\alpha\beta}{16\alpha\beta} + 
\frac{1}{4} + \frac{1+\beta}{4\alpha\beta} - \frac{(1+\beta)^2}{\alpha\beta^2}\right) \nonumber\\ &&
  \times ( h\nabla^a h' - h'\nabla^a h), \\ 
 J^a_{(2)}& =& \frac{4+3\beta-\alpha\beta}{4\alpha\beta} (\nabla_a \nabla_b B \nabla^b h'
- \nabla_a \nabla_b B' \nabla^b h) \nonumber \\
&& - \frac{4+3\beta-\alpha\beta}{16\alpha\beta}(\Box B\nabla^a h' - \Box B'\nabla^a h)\nonumber \\
&& - \frac{3}{16}\frac{4+3\beta-\alpha\beta}{\alpha\beta}
(h'\nabla^a(\Box + 4)B\nonumber \\ && \,\,\,\,\,\,\,\,\,\,\,\,\,\,\,\,\,\,\,\,\,\,\,\,\,\,\,\,\,\,\,\,\,\,\,\,\, - h\nabla^a(\Box +4)B'),\\
J^a_{(3)} &= &\frac{3}{4}\left(1-\frac{1}{\alpha}\right)(\nabla^a\nabla_b B\cdot\nabla^b(\Box + 4)B' \nonumber \\ && 
 \,\,\,\,\,\,\,\,\,\,\,\,\,\,\,\,\,\,\,\,\,\,\,\,\,\,\,\,\,\,\,\,\,\,\,\,\,- \nabla^a\nabla_b B'\cdot\nabla^b(\Box + 4)B)\nonumber \\
&& - \frac{3}{16}\left(1-\frac{1}{\alpha}\right)( \Box B\nabla^a(\Box + 4)B' \nonumber \\ &&
\,\,\,\,\, \,\,\,\,\,\,\,\,\,\,\,\,\,\,\,\,\,\,\,\,\,\,\,\,\,\,\,\,\,\,\,\,
 -  \Box B'\nabla^a(\Box + 4)B) 
\nonumber \\&& -\frac{1}{2}(\nabla_b\nabla_c B\cdot\nabla^a\nabla^b\nabla^c B' 
 -\nabla_b\nabla_c B'\cdot\nabla^a\nabla^b\nabla^c B)
 \nonumber \\ && + \frac{1}{8}(\Box B\cdot \nabla^a \Box B'-\Box B'\cdot \nabla^a \Box B). \nonumber \\ &&
\end{eqnarray}
 To simplify the term $\nabla_b\nabla_c B\nabla^a\nabla^b \nabla^c B' - (B\leftrightarrow B')$  in $J^a_{(3)}$, we first note that 
\begin{eqnarray}
&& \nabla_b\nabla_c B\cdot \nabla^a\nabla^b\nabla^c B' \nonumber \\ 
 &= &\nabla_b\nabla_c B\cdot \left[(g^{ca}\nabla^b B' - g^{cb}\nabla^a B')
+ \nabla^b\nabla^a \nabla^c B'\right]\nonumber \\
 &= & \nabla_a\nabla_b B\cdot \nabla^b B' - \Box B\nabla^a B' 
 \nonumber  \\&& + \nabla_b\nabla_c B\cdot \nabla^b \nabla^a \nabla^c B'. \label{wqe1}
\end{eqnarray}
Then by considering this term together with the term obtained by interchanging $B$ and $B'$, we have 
\begin{eqnarray}
&& \nabla_b\nabla_c B\nabla^b\nabla^a\nabla^c B'- \nabla_b\nabla^a \nabla_c B \nabla^b\nabla^c B'\nonumber \\
& = & \nabla_b(\nabla^b\nabla_c B\nabla^a\nabla^c B' - \nabla^a\nabla^c B \nabla^b \nabla_c B')\nonumber \\
&&  - \Box \nabla_b B\nabla^a\nabla^b B' + \nabla^a\nabla^b B\Box \nabla_b B'\nonumber \\
&\cong& - \nabla_b \Box B \nabla^a\nabla^b B' + \nabla^a \nabla^b B\nabla^b \Box  B' \nonumber \\
&& - 3\nabla_b B \nabla^a\nabla^b B' + 3\nabla^a\nabla^bB\nabla_b B'. \label{wqe2}
\end{eqnarray}
From Eqs. $(\ref{wqe1})$ and $(\ref{wqe2})$, we have 
\begin{eqnarray}
&&\nabla_b\nabla_c B\nabla^a\nabla^b \nabla^c B' - (B\leftrightarrow B')\nonumber \\ 
&\cong& \nabla^a\nabla_b B \nabla^b B' - \Box B\nabla^a B'
- \nabla_b\Box B\nabla^a\nabla^b B' \nonumber \\ && + 3\nabla^a\nabla^b B\nabla_b B'- (B\leftrightarrow B')\nonumber \\
& =& -  \Box B \nabla^a B' + \nabla^a\nabla_b B\nabla^b(\Box + 4)B'- (B\leftrightarrow B').
\end{eqnarray}
 So we can now write $J^a_{(3)}$ as
\begin{eqnarray}
J^a_{(3)}& \cong &  \frac{1}{4}\left(1-\frac{3}{\alpha}\right) \nabla^a\nabla^b B\nabla_b(\Box + 4)B' 
 \nonumber \\  && - \frac{3}{16}\left(1-\frac{1}{\alpha}\right) \Box B \nabla^a(\Box + 4)B' \nonumber \\
&& + \frac{1}{8}\Box B\nabla^a(\Box + 4)B' - (B\leftrightarrow B')\nonumber \\
&=& \frac{1}{4}\left(1-\frac{3}{\alpha}\right)\nabla^a \nabla^b B\nabla_b(\Box + 4)B' \nonumber \\
&& - \frac{1}{16}\left( 1- \frac{3}{\alpha}\right)\Box B \nabla^a(\Box + 4)B' \nonumber \\
&& - (B\leftrightarrow B').
\end{eqnarray}
Now we note that using Eq. $(\ref{B12})$, we have 
\begin{equation}
(\Box + 4)B = \frac{2\alpha}{\alpha-3}(2-M^2)B + \frac{\alpha\beta - 3\beta -4}{\beta(\alpha-3)}h.
\end{equation}
So the terms of the form $\nabla_a\nabla_b B \nabla^b B'$ or $\nabla_a \nabla_b B \nabla^b h'$ or these 
with the primed and unprimed functions swapped in $J^a_{(2)}$ and
 $J^a_{(3)}$ are 
\begin{eqnarray}
&& \frac{4+3\beta-\alpha\beta}{4\alpha\beta}(\nabla^a\nabla^b B\nabla_b h' - \nabla^a\nabla^b B' \nabla_b h) \nonumber \\
&&+ \frac{1}{4}\left(1-\frac{3}{\alpha}\right) \nabla^b\nabla^a B
\nabla_b\left[\frac{2\alpha}{\alpha-3}(2-M^2)B'+ \frac{\alpha\beta-3\beta-4}{\beta(\alpha-3)}h'\right]\nonumber \\
&& - \frac{1}{4}\left(1-\frac{3}{\alpha}\right) \nabla^b\nabla^a B'
\nabla_b\left[\frac{2\alpha}{\alpha-3}(2-M^2)B + \frac{\alpha\beta-3\beta-4}{\beta(\alpha-3)}h\right]\nonumber \\
&=& \frac{2-M^2}{2}(\nabla^a\nabla^b B\nabla_b B' - \nabla^a\nabla^b B'\nabla_b B).
\end{eqnarray}

Now from Eq. $(\ref{lpo})$, we have 
\begin{equation}
\nabla^a \nabla^b B\nabla_b B' - \nabla^a\nabla^b B'\nabla_a B\cong - \nabla^a B \Box B' + \nabla^a B'\Box B,
\end{equation}
so we can write 
\begin{equation}
 J^a_{(2)} + J^a_{(3)} \cong  J^a_{(BB)} + J^a_{(Bh)} + J^a_{(hh)},
\end{equation}
where 
\begin{eqnarray}
 J^a_{(hh)} &=&
- \frac{3}{16}\frac{(\alpha\beta-3\beta-4)^2}{\alpha \beta^2(\alpha-3)}
(h\nabla^a h' - h'\nabla^a h),\\
J^a_{(BB)} &=& 
\frac{3}{4}\frac{(2-M^2)(6-\alpha M^2)}{\alpha-3}(B\nabla^a B' - B'\nabla^a B),\\
J^a_{(Bh)} & = &
- \frac{4+3\beta-\alpha\beta}{16\alpha\beta} \frac{12-2\alpha M^2}{\alpha-3}
(B\nabla^a h' - B'\nabla^a h)\nonumber \\
&& - \frac{3}{16}\frac{4+3\beta-\alpha\beta}{\alpha\beta}
\frac{2\alpha}{\alpha-3}(2-M^2)(h'\nabla^a B - h\nabla^a B')\nonumber \\
&& + \frac{2-M^2}{2}\frac{\alpha\beta-3\beta-4}{\beta(\alpha-3)}(-\nabla^aB h' + \nabla^a B' h)\nonumber \\&& - \frac{1}{16}\left(1-\frac{3}{\alpha}\right)
\left[ \frac{12-2\alpha M^2}{\alpha-3}\frac{\alpha\beta-3\beta-4}{\beta(\alpha-3)}
(B\nabla^a h' - B'\nabla^a h) \right.\nonumber \\
&& \left. \frac{2\alpha}{\alpha-3}(2-M^2)\frac{\alpha\beta-3\beta-4}{\beta(\alpha-3)}
(h\nabla^a B' -h'\nabla^aB)\right].
\end{eqnarray}
Now we have 
 \begin{eqnarray}
J^a_{(Bh)} & = &
\left[ - \frac{4+3\beta-\alpha\beta}{8\alpha\beta}
\frac{6-\alpha M^2}{\alpha-3}- \frac{1}{8}\frac{\alpha-3}{\alpha}\frac{\alpha-3\beta-4}{\beta(\alpha-3)^2}(6-\alpha M^2)
\right]\nonumber \\ && \times  (B\nabla^a h'- B'\nabla^a h)\nonumber \\
&& + \left[\frac{2-M^2}{2}\frac{\alpha \beta-3\beta-4}{\beta(\alpha-3)} + \frac{3}{8}.\frac{4+3\beta-\alpha\beta}{\beta(\alpha-3)}(2-M^2)\right. \nonumber \\
&& \left.
- \frac{1}{8}\frac{\alpha-3}{\alpha}\frac{\alpha(2-M^2)(\alpha\beta-3\beta-4)}{(\alpha-3)\beta}\right]
\nonumber \\
&& \times (h\nabla^a B' - h'\nabla^a B)\nonumber \\
& =& 0.
\end{eqnarray}
So the $B$-$h$ terms cancel out. The total $B$-$B$ contribution is given by $J^a_{(BB)}$ 
and the total $h$-$h$ contribution is given by $J^a_{(hh)} + J^a_{(1)}$, which is given by 
\begin{eqnarray}
J^a_{(1)} + J^a_{(hh)} &=&
\frac{1}{16}\left[ 3-\frac{(3\beta+4)^2}{\alpha\beta^2} - \frac{3(\alpha\beta-3\beta-4)^2}{\alpha\beta^2(\alpha-3)}\right](h\nabla^a h' - h'\nabla^a h)\nonumber \\ 
&=&  - \frac{1}{\beta^2(\alpha-3)}(h\nabla^a h' - h'\nabla^a h).
\end{eqnarray}

So if 
\begin{eqnarray}
C_{ab} & =& \frac{1}{4}h + \left(\nabla_a \nabla_b - \frac{1}{4}g_{ab}\Box \right)B\,, \nonumber \\
C'_{ab} & = & \frac{1}{4}h' + \left(\nabla_{a} \nabla_{b} - \frac{1}{4}g_{ab}\Box \right)B'\,,\nonumber
\end{eqnarray}
then we can write, 
\begin{equation}
 (C, C') = - K_1 \langle B, B' \rangle - K_2\langle h, h'\rangle,
\end{equation}
where
\begin{equation}
 K_1 = \frac{3}{4(\alpha - 3)}(2- M^2)(6 - \alpha M^2 ),
\end{equation}
\begin{equation}
 K_2 = -\frac{1}{\beta^2 (\alpha - 3)}.
\end{equation}
This is the symplectic  product between two scalar modes. 

\section{Symplectic Product for the Vector and Tensor Modes}
The equation of motion for the vector  modes, Eq. $(\ref{aeq1a})$, can be written as 
\begin{equation}
 [\Box +3 - \alpha M^2 ] A^b = 0. \label{refg}
\end{equation}
Let $A^{n\ell\sigma}_b$ and $A^{n'\ell'\sigma'}_b$ be two solutions to this equation. 
We will  denote $A^{n\ell\sigma}_b$ by $A_b$ and $A^{*n'\ell'\sigma'}_b$ by $A'_b$ for simplicity of notation. 
However it may be noted that they are modes and not fields which will be promoted to operators.   
The  momentum current conjugate to $A_{bc} = \nabla_c A_b + \nabla_b A_c$, can be obtained by   substituting  $h_{bc} = A_{ac}$  in  Eq. $(\ref{p2ia})$. 
Now if the 
 momentum current conjugate to  $A_{bc}$ is $\pi^{abc}_{(A)}$, then we have 
\begin{equation}
  \pi^{abc}_{(A)} = \pi^{abc}_{(A4)} + \pi^{abc}_{(A5)},
\end{equation}
where, 
\begin{eqnarray}
 \sqrt{-g}\pi^{abc}_{(A4)} &=& - \frac{1}{2}\nabla^a (\nabla^b A^c + \nabla^c A^b )
\nonumber \\&&  +\frac{1}{2} [g^{ac}\nabla_d (\nabla^d A^b + \nabla^b A^d)  \nonumber \\ &&
 \,\,\,\,\,\,\,\, + g^{ab}\nabla_d (\nabla^d A^c  + \nabla^c A^d)]   \nonumber \\  && - \frac{1}{2} 
 [ g^{bc}\nabla_d (\nabla^a A^d + \nabla^d A^a) ], \label{piiii}\\
\sqrt{-g}\pi^{abc}_{(A5)} &=& 
-\frac{1}{2\alpha} [g^{ac}\nabla_d (\nabla^d A^b + \nabla^b A^d)  \nonumber \\ &&
   \,\,\,\,\,\,\,\, \,\, + g^{ab}\nabla_d (\nabla^d A^c  + \nabla^c A^d)]   \nonumber \\ &&  +\frac{\beta +1}{\beta \alpha} 
  [ g^{bc}\nabla_d (\nabla^a A^d + \nabla^d A^a) ]. \label{piiiia}\\
\end{eqnarray}
If $J^a_A$ is the  contribution to $J^a$ defined in Eq. $(\ref{jj})$, from the vector modes, then we have 
\begin{eqnarray}
 J^a_A &=&  - \frac{i}{\sqrt{-g} }[A_{cb} \pi'^{abc}_{(A)} - A'_{cb}\pi^{abc}_{(A)}] \nonumber \\ &=& 
-i J^a_{(4)} -i J^a_{(5)},
\end{eqnarray}
where 
\begin{eqnarray}
 J_{(4)}^a = \frac{1}{\sqrt{-g}}[A_{bc}  \pi'^{abc}_{(A4)} - A'_{bc}  \pi^{abc}_{(A4)}], \\
 J_{(5)}^a = \frac{1}{\sqrt{-g}}[A_{bc}  \pi'^{abc}_{(A5)} - A'_{bc}  \pi^{abc}_{(A5)}]. \label{piab}
\end{eqnarray}
It is well known that the symplectic product vanishes for $h_{ab} = A_{ab}$ in the theory without the gauge-fixing   term  
(see Ref. $\cite{vecsyms}$). This implies that 
\begin{equation}
 -i \int d \Sigma_a J_{(4)}^a =0. \label{vecq1}
\end{equation}
So the only contribution to the symplectic product comes from $J^a_{(5)}$. Now we can simplify $\sqrt{-g}\pi^{abc}_{(A2)}$ by using
 Eqs. $(\ref{refg})$ and $(\ref{jkla})$
as
\begin{eqnarray}
\sqrt{-g}\pi^{abc}_{(A5)} &=& 
-\frac{1}{2\alpha} [g^{ac} (\Box  + 3)A^b  + g^{ab} (\Box   + 3)A^c]   \nonumber \\ &&  +\frac{\beta +1}{\beta \alpha} 
  [ g^{bc} (\Box + 3)A^a ] \nonumber \\ &=&
M^2 \left[- \frac{1}{2}(g^{ac}A^b + g^{ab}A^c) + \frac{\beta +1}{\beta } g^{bc} A^a \right]. 
 \label{piiiiab}
\end{eqnarray}
Substituting Eq. $(\ref{piiiiab})$ into Eq. $(\ref{piab})$, we get 
\begin{equation}
 J_{(5)}^a = -M^2 [A_b \nabla^a A'^b -A'_b \nabla^a A^b].  \label{vecq10}
\end{equation}
Thus we have from Eqs.  $(\ref{vecq1})$ and  $(\ref{vecq10})$,
\begin{eqnarray}
 \int d \Sigma_a J_A^a &=&  -i \int d \Sigma_a J_{(4)}^a    -i \int d \Sigma_a J_{(5)}^a \nonumber \\ &=&
i M^2 \int d \Sigma_a [A_b \nabla^a A'^b -A'_b \nabla^a A^b].
\end{eqnarray}
So we have 
\begin{equation}
 (A, A') = -M^2 \langle A, A' \rangle.
\end{equation}
 We can write the equation of motion for tensor modes, Eq. $(\ref{eeq1a})$, as 
 \begin{equation}
 [\Box - 2 -M^2] E_{ab} =0.
\end{equation}
 Let $E^{n\ell\sigma}_{bc}$ and $E^{n\ell'\sigma'}_{bc}$ be two solutions to this equation. 
We will  denote $E^{n\ell\sigma}_{bc}$ by $E_{bc}$ and $E^{*n\ell'\sigma'}_{bc}$ by $E'_{bc}$ for simplicity of notation. 
However it may be noted again that they are modes and not fields which will be promoted to operators.
The  momentum current  conjugate to $E_{bc}$, can be obtained by   substituting  $h_{bc} = E_{ac}$  in  Eq. $(\ref{p2ia})$. Now if the
 momentum current conjugate to  $E_{bc}$ is $\pi^{abc}_{(E)}$, then we have 
\begin{equation}
 \sqrt{-g} \pi^{abc}_{(E)} = - \frac{1}{2}\nabla^a E^{bc}.
\end{equation}
If $J^a_E$ is the  contribution to $J^a$ defined in Eq. $(\ref{jj})$, from the tensor modes, then we have 
\begin{equation}
 J^a_E = -\frac{i}{2} [E_{bc}\nabla^a E'^{bc} - E'_{bc}\nabla^a E^{bc} ].
\end{equation}
So we have 
\begin{equation}
 (E,E') =  \frac{1}{2}\langle E, E' \rangle.  
\end{equation}

\section{Summary}
In summary,
\begin{eqnarray}
 (A^{n \ell \sigma}, A^{n' \ell' \sigma'})& = & -M^2 \langle A^{n \ell \sigma}, A^{n' \ell' \sigma'}\rangle,  \\
 (E^{n \ell \sigma},E^{n' \ell' \sigma'}) &=&  \frac{1}{2}\langle E^{n \ell \sigma}, E^{n' \ell' \sigma'} \rangle,  
\end{eqnarray}
  and if
\begin{eqnarray}
C^{\ell \sigma}_{ab} & =& \frac{1}{4}g_{ab}h^{\ell \sigma} + \left(\nabla_a \nabla_b - \frac{1}{4}g_{ab}\Box \right)B^{\ell \sigma}\,, \nonumber \\
C^{\ell' \sigma'}_{ab} & = & \frac{1}{4}g_{ab}h^{\ell' \sigma'} + \left(\nabla_{a} \nabla_{b} - \frac{1}{4}g_{ab}\Box \right)B^{\ell'\sigma'}\,,\nonumber
\end{eqnarray}
then we can write, 
\begin{equation}
 (C^{\ell \sigma}, C^{\ell'\sigma'}) = - K_1 \langle B^{\ell\sigma}, B^{\ell'\sigma'} \rangle - K_2\langle h^{\ell\sigma}, h^{\ell' \sigma'}\rangle,
\end{equation}
where
\begin{eqnarray}
 K_1& = &\frac{3}{4(\alpha - 3)}(2- M^2)(6 - \alpha M^2 ),\\ \nonumber
 K_2 &=& -\frac{1}{\beta^2 (\alpha - 3)}.
\end{eqnarray}
\chapter{Explicit Form for the Tensor $k_{ab}^{n\ell \sigma}$, for $n =0,1$ }
\section{General Procedure}
We shall  calculate explicitly $k_{ab}^{n\ell \sigma}$, for $n =0,1$ here, for completeness.
It may however, be noted that the explicitly form of $k_{ab}^{n\ell \sigma}$ is not necessary for the main results of this 
thesis. 

Recalling that 
\begin{equation}
 A^{n\ell \sigma}_{ab} = \nabla_a A^{n\ell \sigma}_b + \nabla_b A^{n \ell \sigma}_a,
\end{equation}
we have
\begin{eqnarray}
 A_{\chi \chi}^{n\ell \sigma} &=& 2 D_0 A_\chi^{n\ell \sigma},\\
A_{\chi i}^{n\ell \sigma} &=&  D_{-2} A_i^{n\ell \sigma} + \tilde{\nabla}_i A_\chi^{n\ell \sigma}, \\
A_{ij}^{n\ell \sigma} &=&  \tilde{\nabla}_i  A_j^{n\ell \sigma} +\tilde{\nabla}_j  A_i^{n\ell \sigma}. \\
\end{eqnarray}
So we have 
\begin{eqnarray}
  A_{\chi \chi}^{1\ell \sigma} &=& 0, \label{a5}\\
A_{\chi i}^{1\ell \sigma} &=&  D_{-2} P^{-(\ell +1)}_{L+1}(\cos\chi)Y_i^{\ell \sigma} (\bf{x}), \\
A_{ij}^{1\ell \sigma} &=& P^{-(\ell +1)}_{L+1}(\cos\chi)\left[\tilde{\nabla}_i Y_j^{\ell \sigma}({\bf{x}}) + \tilde{\nabla}_j Y_i^{\ell \sigma}(\bf{x}) \right], \\
 A_{\chi \chi}^{0\ell \sigma} &=& 2 (\sin\chi)^{-2} D_{-2}P^{-(\ell +1)}_{L+1}(\cos\chi)Y^{\ell \sigma}(\bf{x}),\\
A_{\chi i}^{0\ell \sigma} &=& \left[ \frac{1}{\ell(\ell+2)} D_{-2}D_1 + \frac{1}{\sin^2 \chi} \right]P^{-(\ell +1)}_{L+1}(\cos\chi)
   \tilde{\nabla}_i Y^{\ell \sigma}(\bf{x}) , \\
A_{ij}^{0\ell \sigma} &=& \frac{2}{\ell(\ell+2)} D_1  P^{-(\ell +1)}_{L+1}(\cos\chi) \tilde{\nabla}_i\tilde{\nabla}_jY^{\ell \sigma} (\bf{x}) \nonumber \\ 
&& + 2 \cot\chi P^{-(\ell +1)}_{L+1}(\cos\chi) \eta_{ij}Y^{\ell \sigma} (\bf{x}). \label{a51} 
\end{eqnarray}
From Eqs. $(\ref{sp1})$-$(\ref{sp6})$ and   $(\ref{a5})$-$(\ref{a51})$
 we see that the $S^3$ tensor structure for $E^{n\ell \sigma}_{ab}$ is similar to $A^{n\ell \sigma}_{ab}$ for $n =0,1$. 

Now  $k_{ab}^{n\ell \sigma}$ is defined  as: 
\begin{eqnarray}
 k_{ab}^{0\ell \sigma} &=& M^{-2}\left[ 2(\ell-1)E^{0\ell \sigma}_{ab} - A^{0\ell \sigma}_{ab}\right] ,\\
k_{ab}^{1\ell \sigma} &=& M^{-2}\left[ (\ell-1)E^{1\ell \sigma}_{ab} -  A^{1\ell \sigma}_{ab}\right].
\end{eqnarray}
  We first recall that from  Eq. $(\ref{D})$, we have  
\begin{equation}
 D_n = \frac{d}{d\chi} + n\cot\chi, 
\end{equation}
and from Eq. $(\ref{D1})$, we have  
\begin{eqnarray}
-\sin\chi D_{-\nu}P^{-\mu}_{\nu} &=& (\nu-\mu)P_{\nu-1}^{-\mu}(x),\nonumber \\ 
-\sin\chi D_{\nu + 1}P^{-\mu}_{\nu} &=& -(\nu+\mu+1)P_{\nu+1}^{-\mu}(x). 
\end{eqnarray}

\section{Explicit Form for the Tensor $k_{ab}^{1\ell \sigma}$}

We can now calculate $k_{\chi \chi}^{1\ell \sigma} $ as follows: 
\begin{equation}
k_{\chi \chi}^{1\ell \sigma}  =
\frac{1}{M^2}\left[(\ell-1)E^{1\ell \sigma}_{\chi\chi}- A^{1\ell \sigma}_{\chi\chi}\right] \\
 =0,
\end{equation}
To calculate $k_{\chi i}^{1\ell \sigma}$, we note that
\begin{eqnarray}
&& D_{-2} P_{L+1}^{-(\ell+1)}(\cos\chi)\nonumber\\
 & & =
\frac{\sin\chi}{\sin\chi}D_{-(L+1)} P_{L+1}^{-(\ell+1)}(\cos\chi)
 + (L-1)\cot\chi P_{L+1}^{-(\ell+1)}(\cos\chi) \nonumber \\
& & = - \frac{\left[(L-1)-(\ell-1)\right]}{\sin\chi} P_L^{-(\ell+1)}(\cos\chi)
+ (L-1)\cot\chi P_{L+1}^{-(\ell+1)}(\cos\chi) \nonumber \\
& & =\frac{\ell-1}{\sin\chi} P_{\overline{L}+1}^{-(\ell+1)}(\cos\chi)
+ \frac{\ell-1}{\sin\chi}\left[ P_L^{-(\ell+1)}(\cos\chi) -  P_{\overline{L}+1}^{-(\ell+1)}(\cos\chi)\right]\nonumber\\
&& \,\,\,\,
+ \frac{L-1}{\sin\chi}\left[\cos\chi P_{L+1}^{-(\ell+1)}(\cos\chi) -  P_L^{-(\ell+1)}(\cos\chi)\right] \nonumber \\
&&\approx \frac{\ell-1}{\sin\chi} P_{\overline{L}+1}^{-(\ell+1)}(\cos\chi) - \frac{1}{\sin\chi}\left\{ (\ell-1)\frac{3\alpha-5}{15}M^2
\left. \frac{\partial\ }{\partial \nu} P_\nu^{-(\ell+1)}(\cos\chi)\right|_{\nu=1}\right. \nonumber\\
&& \left.
\,\,\,\,\,\,\,\,\,\,\,\,\,\,\,\,\,\,\,\,
- \frac{\alpha M^2}{5}\left[ P_1^{-(\ell+1)}(\cos\chi) - \cos\chi P_2^{-(\ell+1)}(\cos\chi)\right]\right\}.
\label{q1}
\end{eqnarray}
From this we find
\begin{equation}
k_{\chi i}^{1\ell \sigma}  =
\frac{1}{M^2}\left[(\ell-1)E^{1\ell \sigma}_{i\chi}- A^{1\ell \sigma}_{i\chi}\right] \\
 =  \frac{\ell-1}{\sin\chi}q_1(\chi)Y_i^{1\ell \sigma},
\end{equation}
where
\begin{eqnarray}
q_1(\chi) & \equiv & \frac{1}{\ell-1}\left\{ (\ell-1)\frac{3\alpha-5}{15}
\left. \frac{\partial\ }{\partial \nu} P_\nu^{-(\ell+1)}(\cos\chi)\right|_{\nu=1}\right. \nonumber\\
&& \left.
\,\,\,\,\,\,
- \frac{\alpha}{5}\left[ P_1^{-(\ell+1)}(\cos\chi) - \cos\chi P_2^{-(\ell+1)}(\cos\chi)\right]\right\}.
\end{eqnarray}
Similarly to find $k_{ij}^{1\ell \sigma}$, we note that 
\begin{eqnarray}
&& D_2  P_{\overline{L}+1}^{-(\ell+1)}(\cos\chi)\nonumber\\
 & & =
\frac{\sin\chi}{\sin\chi}D_{(\overline{L}+2)} P_{\overline{L}+1}^{-(\ell+1)}(\cos\chi)
 - \overline{L}\cot\chi P_{\overline{L}+1}^{-(\ell+1)}(\cos\chi) \nonumber \\
& & = \frac{(\overline{L}+ \ell+3)}{\sin\chi} P_{\overline{L}+2}^{-(\ell+1)}(\cos\chi)
- \overline{L}\cot\chi P_{\overline{L}+1}^{-(\ell+1)}(\cos\chi) \nonumber \\
& & =\frac{\ell+3}{\sin\chi} P_{L+1}^{-(\ell+1)}(\cos\chi)
+ \frac{\ell+3}{\sin\chi}\left[ P_{\overline{L}+2}^{-(\ell+1)}(\cos\chi) -
 P_{L+1}^{-(\ell+1)}(\cos\chi)\right]\nonumber\\
&& \,\,\,\,
-\overline{L} \frac{1}{\sin\chi}
\left[\cos\chi P_{\overline{L}+1}^{-(\ell+1)}(\cos\chi) -  P_{\overline{L}+2}^{-(\ell+1)}(\cos\chi)\right] \nonumber \\
&&\approx \frac{\ell+3}{\sin\chi} P_{L+1}^{-(\ell+1)}(\cos\chi) + \frac{1}{\sin\chi}\left\{ (\ell+3)\frac{3\alpha-5}{15}M^2
\left. \frac{\partial\ }{\partial \nu} P_\nu^{-(\ell+1)}(\cos\chi)\right|_{\nu=2}\right. \nonumber\\
&& \left.
\,\,\,\,\,\,\,\,\,\,\,\,\,\,\,\,\,\,\,\,
- \frac{M^2}{3}\left[ P_2^{-(\ell+1)}(\cos\chi) - \cos\chi P_1^{-(\ell+1)}(\cos\chi)\right]\right\}.
\end{eqnarray}
Hence we find
\begin{equation}
k_{ij}^{1\ell \sigma}  =
\frac{1}{M^2}\left[(\ell-1)E^{1\ell \sigma}_{ij}- A^{1\ell \sigma}_{ij}\right]
 =  q_2(\chi)\left[\tilde{\nabla}_i Y_j^{1\ell \sigma}+\tilde{\nabla}_j Y_i^{(1\ell\sigma0)}\right],
\end{equation}
where
\begin{eqnarray}
q_2(\chi) & \equiv &
\frac{3\alpha-5}{15}
\left. \frac{\partial\ }{\partial \nu} P_\nu^{-(\ell+1)}(\cos\chi)\right|_{\nu=2} \nonumber\\
&&
\,\,\,\,\,\,
- \frac{1}{3(\ell+3)}\left[ P_2^{-(\ell+1)}(\cos\chi) - \cos\chi P_1^{-(\ell+1)}(\cos\chi)\right].
\end{eqnarray}
\section{Explicit Form for the Tensor $k_{ab}^{0\ell \sigma}$}
Next to find $k_{\chi \chi}^{0\ell \sigma}$ , we use Eq. $(\ref{q1})$,  and get
\begin{eqnarray}
k_{\chi \chi}^{0\ell \sigma} & = &
\frac{1}{M^2}\left[2(\ell-1)E^{0\ell \sigma}_{\chi \chi}- A^{0\ell \sigma}_{\chi\chi}\right]\nonumber \\
& = & \frac{2}{\sin^3\chi}\left\{ (\ell-1)\frac{3\alpha-5}{15}
\left. \frac{\partial\ }{\partial \nu} P_\nu^{-(\ell+1)}(\cos\chi)\right|_{\nu=1}\right. \nonumber\\
&& \left.
\,\,\,\,\,\,
- \frac{\alpha}{5}\left[ P_1^{-(\ell+1)}(\cos\chi) - \cos\chi P_2^{-(\ell+1)}(\cos\chi)\right]\right\}
Y^{0\ell \sigma}\nonumber \\
& = & \frac{2(\ell-1)}{\sin^3\chi}q_1(\chi)Y^{0\ell \sigma}.
\end{eqnarray}
Then we have
\begin{eqnarray}
k_{\chi i}^{0\ell \sigma} & = &
\frac{1}{M^2}\left[2(\ell-1)E^{0\ell \sigma}_{i \chi}- A^{0\ell \sigma}_{i \chi}\right]\nonumber \\
& = & \frac{1}{\ell(\ell+2)} X(\chi)\tilde{\nabla}_i Y^{0\ell \sigma}, \label{kchii}
\end{eqnarray}
where
\begin{eqnarray}
 X(\chi) & = &\lim_{M^2 \to 0}\frac{1}{M^2} \left[ 2(\ell-1)\frac{1}{\sin\chi}D_1 
 P_{\overline{L}+1}^{-(\ell+1)}(\cos\chi) \right.  \nonumber \\
&& \left. - \left[D_{-2}D_1   + \frac{\ell(\ell+2)}{\sin^2\chi}\right] P_{L+1}^{-(\ell+1)}(\cos\chi)\right].
\label{Xchi}
\end{eqnarray}
Noting that
\begin{equation}
q_1(\chi)  = \lim_{M^2 \to 0} \frac{1}{M^2} \frac{\sin\chi}{\ell-1}\left[\frac{\ell-1}{\sin\chi} P_{\overline{L}+1}^{-(\ell+1)}(\cos\chi)
- D_{-2} P_{L+1}^{-(\ell+1)}(\cos\chi)\right],
\end{equation}
we find
\begin{eqnarray}
&& M^2 X(\chi) - \frac{2(\ell-1)}{\sin\chi}M^2 D_1 q_1(\chi)\nonumber \\
&& = \left[\frac{d^2\ }{d\chi^2} + \cot\chi\frac{d\ }{d\chi} + 6 - \frac{(\ell+1)^2}{\sin^2\chi}\right] P_{L+1}^{-(\ell+1)}(\cos\chi)\nonumber \\
&& = \left[\frac{d^2\ }{d\chi^2} + \cot\chi\frac{d\ }{d\chi} + (L+1)(L+2)
- \frac{(\ell+1)^2}{\sin^2\chi}\right] P_{L+1}^{-(\ell+1)}(\cos\chi)\nonumber \\
& &\,\,\, + \left[6-(L+1)(L+2)\right] P_{L+1}^{-(\ell+1)}(\cos\chi)\nonumber \\
&& = \alpha M^2 P_2^{-(\ell+1)}(\cos\chi),
\end{eqnarray}
in the $M\to 0$ limit.  Hence
\begin{equation}
X(\chi) = \frac{2(\ell-1)}{\sin \chi }D_1 q_1(\chi) + \alpha
 P_2^{-(\ell+1)}(\cos\chi).
\end{equation}

Finally to find $k_{ij}^{0\ell \sigma}$,  we write it as 
\begin{equation}
k_{ij}^{0\ell \sigma} = \frac{1}{\ell+3}\left[
\frac{Z(\chi)}{\ell(\ell+2)}\tilde{\nabla}_i \tilde{\nabla}_j 
+ W(\chi)\eta_{ij}\right]Y^{0\ell \sigma},
\end{equation}
where
\begin{eqnarray}
 Z(\chi) & = &\lim_{M^2 \to 0} \frac{1}{M^2} \left[
\sin\chi \left[3 D_2 D_1   - \frac{\ell(\ell+2)}{\sin^2\chi}\right] \right. \nonumber \\ &&  \left.
\,\,\,\,\,\,\,\,\,\times  P_{\overline{L}+1}^{-(\ell+1)(\cos\chi)} 
  - 2(\ell+3)D_1  P_{L+1}^{-(\ell+1)}(\cos\chi) \right] , \label{Z} \\
 W(\chi) & = & \lim_{M^2 \to 0} \frac{1}{M^2} \left[\sin\chi \left[D_2 D_1
  - \frac{\ell(\ell+2)-2}{\sin^2\chi}\right]\right.  \nonumber \\
&& \left.  \,\,\,\,\,\,\,\,\, \times  P_{\overline{L}+1}^{-(\ell+1)}(\cos\chi) - 2(\ell+3)\cot\chi P_{L+1}^{-(\ell+1)}(\cos\chi) \right].\label{W}
\end{eqnarray}
We note
\begin{equation}
q_2(\chi) = \lim_{M^2 \to 0} \frac{1}{M^2} \left[ \frac{\sin\chi}{\ell+3} D_2  P_{\overline{L}+1}^{-(\ell+1)}(\cos\chi)
-  P_{L+1}^{-(\ell+1)}(\cos\chi) \right].
\end{equation}
Then, we have 
\begin{eqnarray}
&& M^2 Z(\chi) - 2(\ell+3)D_1 M^2q_2(\chi)\nonumber \\
&& = \sin\chi\left[\frac{d^2\ }{d\chi^2} + \cot\chi\frac{d\ }{d\chi} + 2 - \frac{(\ell+1)^2}{\sin^2\chi}\right]
 P_{\overline{L}+1}^{-(\ell+1)}(\cos\chi)\nonumber \\
&& = \sin\chi\left[\frac{d^2\ }{d\chi^2} + \cot\chi\frac{d\ }{d\chi} + (\overline{L}+1)(\overline{L}+2) - \frac{(\ell+1)^2}{\sin^2\chi}\right]
 P_{\overline{L}+1}^{-(\ell+1)}(\cos\chi)\nonumber \\
&&\,\,\, + \sin\chi\left[2-(\overline{L}+1)(\overline{L}+2)\right] P_{\overline{L}+1}^{-(\ell+1)}(\cos\chi)\nonumber \\
&& \approx M^2\sin\chi P_1^{-(\ell+1)}(\cos\chi).
\end{eqnarray}
Hence, we get 
\begin{equation}
Z(\chi) = 2(\ell+3)D_1 q_2(\chi) + \sin\chi P_1^{-(\ell+1)}(\cos\chi).
\end{equation}
In a similar manner, in the $M\to 0$ limit, we find
\begin{equation}
M^2 W(\chi) - 2(\ell+3)M^2q_2(\chi)\cot\chi \approx M^2\sin\chi  P_1^{-(\ell+1)}(\cos\chi).
\end{equation}
Hence, we get 
\begin{equation}
W(\chi) = 2(\ell+3)q_2(\chi)\cot\chi + \sin\chi P_1^{-(\ell+1)}(\cos\chi).
\end{equation}

\section{Summary}
In summary, we find
\begin{eqnarray}
k_{\chi\chi}^{1\ell \sigma} & = & 0,\nonumber\\
k_{\chi i}^{1\ell \sigma}
& = & \frac{\ell-1}{\sin\chi}q_1(\chi)Y_i^{\ell \sigma},\nonumber \\
k_{ij}^{1\ell \sigma}
& = & q_2(\chi)\left[\tilde{\nabla}_i Y_j^{\ell \sigma}+\tilde{\nabla}_j Y_i^{(\ell\sigma)}\right],\nonumber\\
k_{\chi\chi}^{0\ell \sigma} & = & \frac{2(\ell-1)}{\sin^3\chi}q_1(\chi)Y^{\ell \sigma},\nonumber \\
k_{\chi i}^{0\ell \sigma}
& = & \frac{1}{\ell(\ell+2)}\left[
\frac{2(\ell-1)}{\sin\chi}\left(\frac{d\ }{d\chi} + \cot\chi\right)q_1(\chi) \right. \nonumber \\ && \left.+ \alpha
 P_2^{-(\ell+1)}(\cos\chi)\right]\tilde{\nabla}_i Y^{\ell \sigma},\nonumber \\
k_{ij}^{0\ell \sigma} & = &
\frac{1}{\ell(\ell+2)}\left[2\left( \frac{d\ }{d\chi} + \cot\chi\right)q_2(\chi)\right. \nonumber \\ && \left. +
\frac{\sin\chi}{\ell+3} P_1^{-(\ell+1)}(\cos\chi)\right] 
\tilde{\nabla}_i \tilde{\nabla}_j Y^{\ell \sigma} \nonumber \\ && +\left[ 2q_2(\chi)\cot\chi + \frac{\sin\chi}{\ell+3}
 P_1^{-(\ell+1)}(\cos\chi)\right]\eta_{ij}Y^{\ell \sigma},
\end{eqnarray}
where
\begin{eqnarray}
q_1(\chi) & \equiv & \frac{1}{\ell-1}\left\{ (\ell-1)\frac{3\alpha-5}{15}
\left. \frac{\partial\ }{\partial \nu} P_\nu^{-(\ell+1)}(\cos\chi)\right|_{\nu=1}\right. \nonumber\\
&& \left.
\,\,\,\,\,\,
- \frac{\alpha}{5}\left[ P_1^{-(\ell+1)}(\cos\chi) - \cos\chi P_2^{-(\ell+1)}(\cos\chi)\right]\right\},\\
q_2(\chi) & \equiv &
\frac{3\alpha-5}{15}
\left. \frac{\partial\ }{\partial \nu} P_\nu^{-(\ell+1)}(\cos\chi)\right|_{\nu=2} \nonumber\\
&&
\,\,\,\,\,\,
- \frac{1}{3(\ell+3)}\left[ P_2^{-(\ell+1)}(\cos\chi) - \cos\chi P_1^{-(\ell+1)}(\cos\chi)\right].
\end{eqnarray}

\end{document}